\documentclass[
  journal=pasa,
  manuscript=research-paper, 
  year=2024,
  volume=00,
]{cup-journal}

\usepackage{graphicx}
\usepackage{lscape}
\usepackage{rotating,graphicx}
\usepackage{color}
\usepackage{booktabs}
\usepackage{amsmath} 

\usepackage{amsmath,amssymb}

\usepackage{caption}
\usepackage{subcaption}
\usepackage{lscape}
\usepackage{aas_macros}
\usepackage{hyperref} 
\hypersetup{colorlinks,citecolor=blue,linkcolor=blue,urlcolor=blue}

\newcommand{\hi}{\mbox{H\,{\sc i}} } 
\newcommand{\hix}{\mbox{H\,{\sc i}}} 

\title{\color{MidnightBlue} VLBI studies of FLASH \hi 21-cm absorbers - I}

\author{J.N.H.S. Aditya}
\affiliation{Shanghai Astronomical Observatory, CAS, 80 Nandan Road, Shanghai 200030, P. R. China}
\alsoaffiliation{Sydney Institute for Astronomy, School of Physics A28, University of Sydney, NSW 2006, Australia}
\alsoaffiliation{ARC Centre of Excellence for All-Sky Astrophysics in 3 Dimensions (ASTRO 3D)}
\email[J.N.H.S. Aditya]{adi.jnhs@shao.ac.cn, adi.jnhs@gmail.com}
\alsoaffiliation{State Key Laboratory of Radio Astronomy and Technology, A20 Datun Road, Chaoyang District, Beijing, P. R. China}

\author{Elaine M. Sadler}
\affiliation{Sydney Institute for Astronomy, School of Physics A28, University of Sydney, NSW 2006, Australia}
\alsoaffiliation{ARC Centre of Excellence for All-Sky Astrophysics in 3 Dimensions (ASTRO 3D)}
\alsoaffiliation{ATNF, CSIRO, Space and Astronomy, PO Box 76, Epping, NSW 1710, Australia}

\author{Raffaella Morganti}
\affiliation{ASTRON, the Netherlands Institute for Radio Astronomy, Oude Hoogeveensedijk 4, NL-7991 PD Dwingeloo, The Netherlands}
\alsoaffiliation{Kapteyn Astronomical Institute, University of Groningen, Postbus 800, NL-9700 AV Groningen, The Netherlands}

\author{L. Y. Petrov}
\affiliation{NASA Goddard Space Flight Center, Code 61A, 8800 Greenbelt Rd, Greenbelt, 20771 MD, USA}

\author{Tao An}
\affiliation{Shanghai Astronomical Observatory, CAS, 80 Nandan Road, Shanghai 200030, P. R. China}
\alsoaffiliation{State Key Laboratory of Radio Astronomy and Technology, A20 Datun Road, Chaoyang District, Beijing, P. R. China}

\author{Emily F. Kerrison}
\affiliation{Sydney Institute for Astronomy, School of Physics A28, University of Sydney, NSW 2006, Australia}
\alsoaffiliation{ARC Centre of Excellence for All-Sky Astrophysics in 3 Dimensions (ASTRO 3D)}
\alsoaffiliation{ATNF, CSIRO, Space and Astronomy, PO Box 76, Epping, NSW 1710, Australia}

\author{Elizabeth K. Mahony}
\affiliation{ATNF, CSIRO, Space and Astronomy, PO Box 76, Epping, NSW 1710, Australia}

\author{Hyein Yoon}
\affiliation{Institute for Data Innovation in Science, Seoul National University, 1 Gwanak-ro, Gwanak-gu, Seoul 08826, Republic of Korea}
\alsoaffiliation{Astronomy Program, Department of Physics and Astronomy, Seoul National University, 1 Gwanak-ro, Gwanak-gu, Seoul 08826, Republic of Korea}

\author{Renzhi Su}
\affiliation{Key Laboratory for Research in Galaxies and Cosmology, Shanghai Astronomical Observatory, 80 Nandan Road, Shanghai 200030, P. R. China}
\alsoaffiliation{University of Chinese Academy of Sciences, 19A Yuquan Road, Beijing 100049, China}

\author{Matthew Whiting}
\affiliation{ATNF, CSIRO, Space and Astronomy, PO Box 76, Epping, NSW 1710, Australia}

\author{Vanessa A. Moss}
\affiliation{ATNF, CSIRO, Space and Astronomy, PO Box 76, Epping, NSW 1710, Australia}
\alsoaffiliation{Sydney Institute for Astronomy, School of Physics A28, University of Sydney, NSW 2006, Australia}

\author{Simon Weng}
\affiliation{Aix Marseille Univ, CNRS, CNES, LAM, Marseille, France.}

\doi{10.1017/pas.\the\year.xxx}

\received {dd Mmm YYYY}
\revised  {dd Mmm YYYY}
\accepted {dd Mmm YYYY}
\published{dd Mmm YYYY}


\begin{document}

\begin{abstract}

We have conducted VLBA 1.4\,GHz (L-band) continuum  observations towards twelve sources with \hi 21-cm absorption detections at redshift $0.4<z<0.7$ in the pilot surveys of FLASH, an ongoing survey with the ASKAP radio telescope. 11 of the 12 targets are resolved in the VLBA observations. Using the parsec scale radio images, we have classified the source morphology and identified the radio core. Six of the twelve targets have core-jet morphology, four have two-sided jet morphology, one has a complex morphology, and one is unresolved.
We describe a methodology to test whether the emission from the core or the total emission detected in the VLBA image has sufficient flux density to cause the entire \hi 21-cm absorption, and we estimate limits on the gas covering factor and velocity-integrated optical depth (VOD). We find that for seven of the twelve sources, the core has sufficient flux density to cause all the \hi 21-cm absorption detected in the ASKAP spectrum.
 For three other targets, with projected sizes in the range $\rm 305-409 \ pc$, a large fraction of the entire emission in the VLBA map could be occulted by the gas. For 0903+010 (NVSS\,J090331+010846), we estimate that at least $\approx 73\%$ of the peak absorption detected in the ASKAP spectrum could arise against the emission detected in the VLBA image. For the target 0023+010 (NVSS\,J002331+010114), we estimate an upper limit on the VOD of $\rm 169 \ km \ s^{-1}$, the highest in our sample. For 0903+010 (NVSS\,J090331+010846) we estimate a lower limit of $\rm 104 \ km \ s^{-1}$ on the VOD. We find that the distribution of \hi 21-cm VODs at $0.4<z<1.0$ could increase by up to a factor of three after correction for the covering factors using our VLBA measurements.

\end{abstract}

\section{Introduction}
\label{sec:intro}

The Australian SKA Pathfinder Telescope (ASKAP) is currently conducting the First Large Absorption Survey in \hi \citep[FLASH,][]{allison2022}, an untargeted survey for \hi 21-cm absorption at redshifts $0.4 < z < 1.0$, occurring against background radio sources. The FLASH observations will span an epoch of 3.3 Gyrs ($\approx 25 \%$ of the history of the Universe) during which the cosmic star-formation rate density declined steadily \citep[e.g.][]{madau2014}. Our survey \citep[][]{allison2022, su2022, aditya2024, yoon2024} will probe both AGN-associated and intervening absorption. In the former case the absorbing clouds are present either in the vicinities of an AGN or in the AGN-host galaxy, while in the latter case the absorbing clouds are present in the interstellar or circumgalactic medium of a foreground galaxy that is intersecting our line of sight towards a background radio AGN. The search for \hi 21-cm absorption will probe the optical depth and kinematics of the \hi gas, essential for understanding the evolution of gas availability and its relation to the cosmic star formation rate density. 

The size of the ASKAP synthesized beam is $\approx 15$ arcsec in the 700-1000 MHz observing band, corresponding to a physical size of over 100 kpc for a radio source at a redshift of 0.7, the median redshift of the FLASH survey. The resolution is significantly larger than the size of a typical galactic gaseous disk, up to $\approx10$ kpc. From our FLASH pilot surveys, reported in \citet[][]{yoon2024}, we see absorption to be predominantly occurring in sources with peaked or inverted radio SEDs (spectral energy distribution) \citep[e.g.][]{kerrison2024}. Radio sources with these SEDs are likely to be young, undeveloped, and are expected to be spatially compact, with sizes $\lesssim 1$ kpc \citep{odea2021}. 

ASKAP observations cannot resolve the sub-kpc structure of the background radio emission or probe the distribution of the \hi gas against the radio source at sub-kpc scales.  A major drawback of this is that, in cases where the foreground \hi gas does not cover the entire background emission, the measured (or apparent) \hi optical depth ($\tau_{\rm app}$) in the ASKAP observations will be lower than the true optical depth ($\tau_{\rm true}$) of the absorber. The measured and true optical depths are related as 
\begin{equation}\label{eqn1}
    \tau_{\rm true} = -{\rm ln}( 1 +  \frac{(e^{-\tau_{\rm app}} - 1.0)}{f} ) 
\end{equation},
where $f$ is the gas covering factor, introduced by \citet[][]{briggs1983}. They defined this factor as the fraction of the QSO emission that we see covered by the absorbing cloud. This is estimated as the ratio of radio emission obscured by the \hi gas versus the total emission in the ASKAP beam.  

The most direct way to overcome this issue would be to conduct spectroscopic Very Large Baseline Interferometric (VLBI) observations at the redshifted \hi 21-cm frequency, to trace the \hi distribution against the resolved radio source and estimate the true optical depth. However, the existing VLBI telescopes do not cover the 700-1000 MHz frequencies at which we are searching for \hi 21-cm absorption. 

An alternative way to estimate an upper limit on the optical depth is to measure the VLBI continuum flux density near the redshifted \hi 21-cm frequency. For example, assuming that \hi absorption arises only against the VLBI core, the ratio between the core and the total flux density (core fraction) yields the covering factor $f$. While the core flux density would be estimated using the VLBI image, total flux density would be inferred from observations with a smaller interferometer \citep[see e.g.][]{kanekar2009}. As we assume here that the foreground gas covers emission only from the AGN core, the core fraction will be a lower limit to the covering factor, and thus, the corrected optical depth will be an upper limit. For an extended gaseous structure that covers an emission larger than that of the core, the optical depth would be lower than the estimated upper limit. 

In the case of AGN-associated systems, \citet[][]{maccagni2017} report that among 66 detections from their WSRT survey, in systems with narrow \hi 21-cm profiles with full width at 20\% intensity (FW20) $\rm \lesssim 250 \ km \ s^{-1}$, a bulk of the gas is probably located in a circumnuclear structure, likely obscuring the core and emission in the close surroundings. B2352+495 ($z=0.2379$) and TXS 1954+513 ($z=1.223$) are typical examples where narrow absorption is detected against the bright radio core of a CSO (compact symmetric object) and an FR-II radio source, respectively \citep[e.g.][]{araya2010, aditya2017}. 

Although absorption against radio jets and hot spots has been detected in multiple cases (see e.g. NGC\,1052, \citealp[][]{vangorkom1986}; 4C12.50, \citealp[][]{morganti2013}; 3C\,84, \citealp[][]{morganti2023}), these absorption profiles are typically wide, have shallow wings, and show complex structure. For narrow absorption profiles, the gas tends to accumulate around the core, whereas for shallow and broad features, the gas is unsettled, possibly because of interactions with the radio source. Although it is possible that systems with narrow absorption could reveal shallow and wide wings upon deeper observations, a bulk of the gas that is causing the narrow component is likely accumulated around the core.


The spatial extent of the so-called core, which is typically a bright unresolved component identified in the radio continuum image, may  depend strongly on the achieved spatial resolution.
In the literature, spectroscopic VLBI observations of associated \hi 21-cm absorbers have been conducted for systems at typical redshifts of $z \lesssim 0.1$ \citep[][]{schulz2021, murthy2021}. These observations achieved a spatial resolution of $\rm \approx few \times 10 \,pc$, probe the central $\rm few \times 100$ pc, and are generally insensitive to diffuse radio emission at kiloparsec (kpc) scales.   For VLBI observations of \hi absorption systems at redshift of $z \approx 0.7$ (probed by FLASH), the spatial resolution at the absorber redshift will be $\approx 4$ times larger than that of $z \approx 0.1$ objects, for the same angular resolution. The size of the unresolved radio core could be $\gtrsim 100$ pc, the scale of the entire region probed by the $z < 0.1$ studies.
Although having a low spatial resolution, these observations would be sensitive to diffuse radio emission from larger spatial scales, up to a few kpc. 

Spectroscopic VLBI studies of associated absorbers reported that \hi velocities up to even $\rm \approx 1000 \ km \ s^{-1}$ have been recovered within the central 100-150 pc \citep[e.g.][]{morganti2013, schulz2021, murthy2021}. This shows that most of the strong interactions between the radio jets and the gas could be occurring at these scales, although cases with wide and shallow features have been detected at kpc scales \citep[e.g. NGC 1052][]{vangorkom1986}. Therefore, conservatively, for associated absorbers with relatively narrow line width where the gas would accumulate near the pc-scale core, we can expect the bulk of the absorption to be detected within the central 100-200 pc scales. 

\citet[][]{briggs1983} and \citet[][]{kanekar2009} estimated the covering factors of the intervening absorption systems and have found that the core fraction is a close approximation to the gas covering factor. For intervening absorption systems at $0.4 \lesssim z \lesssim  3$, \citet[][]{kanekar2009} achieved spatial resolutions of 130 pc to a $\rm few \times 100 $ pc with VLBI observations, implying that a large fraction of the background emission occurs at these spatial scales, resulting in a high covering factor.

The VLBI continuum observations towards $z\approx 0.7$ absorbers would be sensitive to the emission at the scales of a $\rm few \times 100$ pc, up to a few kpc. These scales probe the region where the bulk of the absorption is expected for both associated and intervening absorbers, and, in addition, probe larger scales where trace absorption arising against the diffuse continuum is expected. These maps should therefore yield a close approximation, or at least a reasonable lower limit, to the gas covering factor.   

Estimating upper limits on the true \hi 21-cm optical depth using the covering factor (see equation \ref{eqn1}) is necessary to 1) identify potential targets that could have a high \hi column density, and 2) estimate the distribution of the \hi-21 cm optical depth in a redshift interval. This is critical to test a possible evolution in \hi 21-cm absorption strength as a function of redshift, and in turn, to measure the gas availability at various redshifts to fuel the global star formation rate density \citep[see e.g.][]{madau2014} .

\subsection{This work} \label{sec:this_work}
\cite{yoon2024} presented new detections of 30  radio sources with \hi 21-cm absorption in the FLASH pilot surveys. Of these, 12 targets north of declination -30\,degrees were observed using the 1.4\,GHz (L-band) receiver of the Very Large Baseline Array (VLBA), and the parsec-scale continuum maps were made. The VLBA observations had a typical resolution of $25 \times 5$ mas. In this paper, we discuss the results of these observations. 

The remaining (southern) objects detected by FLASH were observed at 1.4\,GHz with the Australian Long Baseline Array (LBA) and  these results will be published in a subsequent paper. The 1.4\,GHz  band was chosen for both VLBA and LBA observations so that the observing frequency is as close as possible to the redshifted \hi 21-cm frequency of the targets. In addition to the 12 VLBA images at 1.4~GHz from our observations, we also used images of four targets at other frequencies, found in the publicly available astrogeo VLBI FITS image database \citep{petrov2025}\footnote{\url{http://doi.org/10.25966/kyy8-yp57}}, hereafter called astrogeo. To estimate the total flux densities of our targets at L band, we use data from the Rapid ASKAP continuum Survey-mid (RACS; \citealp[][]{duchesne2023}). The RACS-mid survey was conducted at an effective central frequency of 1367.5 MHz, with 144 MHz bandwidth.

We use the VLBI continuum images to:
\newline
1) Classify the source morphology at the VLBI resolution and estimate the number of \hi detections in each morphological subclass.
\newline
2) Describe a methodology to test whether the radio core or the total emission detected in the VLBA image has sufficient flux density to cause the entire \hi 21-cm absorption detected in the ASKAP spectrum.
\newline
3) For targets with narrow \hi 21-cm line width, $\rm FW20 < 250 \ km \ s^{-1}$, we estimate the limits on the covering factor. Here, we adopt $\rm FW20 < 250 \ km \ s^{-1}$ as a definition for narrow line width, based on the results reported in \citet[][]{maccagni2017}. Using the limits on the covering factor, we derive the upper limits to the \hi optical depth.
\newline
4) Test correlations between the \hi 21-cm covering factor, the velocity-integrated optical depth, and the spatial extent of the source at the absorber redshift. We use the velocity-integrated optical depth (VOD; $\equiv \int \tau \, dv$) as our primary measure of absorption strength, following \citet{aditya2016}. This parameter is directly proportional to the \hi column density ($ {\rm N}_{\rm HI} = 1.823 \times 10^{18} \times {\rm T_s} \times \int \tau \ dv $, where $\rm T_s$ is the spin temperature) and allows for easier comparison with observations in the literature.


In the following sections, we discuss the details of the VLBA observations, results, and source characterisation.

\section{VLBA observations and data reduction}
Twelve targets were observed with the Very Large Baseline Array (VLBA) in the L band. The observations were conducted between 29 February and 1 April 2024; the exact date of the observations for each target is given in Table~\ref{vlba_obs}. Nine antennas participated in the observing runs; Brewster (BR), Fort Davis (FD), Hancock (HN), Kitt Peak (KP), Los Alamos (LA), Mauna Kea(MK), North Liberty (NL), Owens Valley (OV), and Pie Town (PT). The observing band covered the range 1220 MHz to 1284 MHz, and is divided into 2 IFs of 32 MHz, each further divided into 128 channels of width 250\,kHz. We observed strong calibrators 0234+285, 4C39.25, and 3C454.3 for fringe finding and bandpass calibration. Interleaved phase calibrator and target scans were repeated a few times for each target, with a 2-3 min scan on the phase-referencing source, followed by a 4-5 min scan on the target. An on-source time of 30-40 mins was achieved for each target, as listed in Table~\ref{vlba_obs}.

We used the VLBA\_RUN pipeline for processing our 1.4 GHz data. This is part of the Astronomical Image Processing Software (AIPS; \citealp[][]{greisen2003}) available at \url{https://www.aips.nrao.edu/}. Major steps in this pipeline involve applying ionospheric total electron content corrections, amplitude calibration, bandpass calibration, correcting the parallactic angle, instrumental delay correction, delay-and-phase calibration by fringe-fitting and imaging. The pipeline produced calibrated visibilities of the phase calibrator and the target. Multiple rounds of imaging and self-calibration were performed on the target assuming a point-source model for the first few rounds. Phase-only self-calibration was performed until the noise in the image did not improve further, followed by one or two rounds of amplitude and phase self-calibration, to produce the final image of the target.

\begin{figure}[!h]
\centering

    \includegraphics[width=1.1\textwidth]{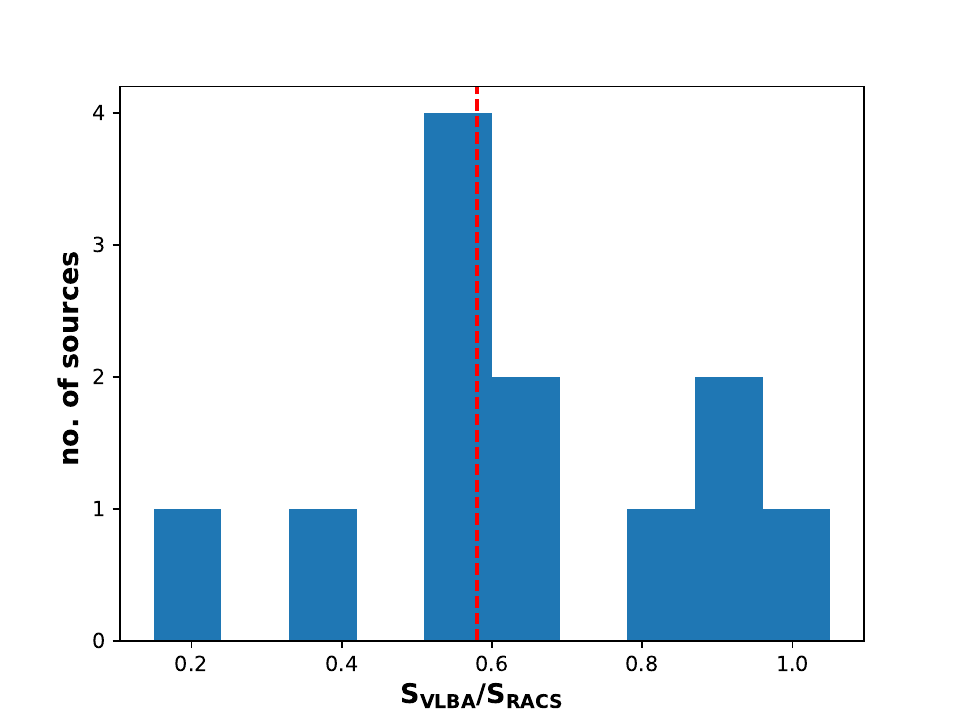}
\caption{Histogram showing the ratios of integrated VLBA and RACS flux densities. The vertical red line shows the median ratio.}\label{vlba_racs}
\end{figure}

Since interferometric visibilities sample a certain range of spatial frequencies, synthesized images recover flux density only until certain
scales and are insensitive to diffuse, extended emission. Depending on frequency, VLBI high resolution are limited to spatial scales 50--100~mas and finer, which corresponds to up to 100 -- 1000 parsecs for systems at $z \approx 0.7$, being insensitive to diffuse emission from extended AGN lobes or diffuse star-formation regions at kiloparsec scales. As a result, flux density integrated over a VLBI image at parsec scales is usually less than the flux density determined with RACS that accounts for both compact and extended emission up to a scale of 10 -- 100 kiloparsecs.
Figure~\ref{vlba_racs} shows the ratios between the integrated VLBA and RACS flux densities of our sources. The red dashed line represents the median ratio. We note that these values are consistent with the results of \citet[see e.g.][]{njeri2024}. 


\section{VLBA source characterisation and covering factor}
\subsection{Methodology for source characterisation}
We produced continuum images for 12 targets in the L band, using our VLBA data (see Figures \ref{fig:0011-023} to \ref{fig:2236-251}). We used the AIPS task JMFIT to identify the peaks of various source components in the image and measure their flux densities. The task fits Gaussian components to a region in an image using the least-squares method. The initial guess for the Gaussians was made by manual inspection.

We morphologically categorize the sources into five classes, \citep[see e.g.][]{devries2009}:
\newline
1. Unresolved or barely resolved (U).
\newline
2. Compact Double (CD). The source has two components with similar flux density.
\newline
3. Core-Jet (CJ). A source with two or more components with significantly different flux densities, where the component with the highest peak flux density is relatively compact and located at an extremity of the source. 
\newline
4. Two-sided jet (TSJ). Sources with three or more components, where the  component with the highest peak flux density not located at an extremity of the source.
\newline
5. Complex (CX). Sources with a complex morphology that does not fall into any of the above categories.

The core component is identified using the source morphology and relative compactness, assessed using the peak-to-integrated flux-density ratio. We note that the source morphology is given first priority, followed by the compactness of the component in cases where the morphology is not clear. We also used spectral indices for identifying the core in a few cases where VLBI maps at other frequencies are available in the literature.

\subsection{Methodology for covering factor estimation}
\subsubsection{Parameters}\label{params}
The parameters used in this analysis are: 

\vspace*{0.1cm}
\noindent
1) {\bf Parameters from FLASH observations}
\vspace*{-0.1cm}
\begin{itemize}
\item 
S$_{\rm ASKAP}$ - the integrated flux density of the source measured by ASKAP   at $\sim 700$ MHz
\item 
\rm $\lvert S_{\rm HI} \rvert$ -
the flux density of the \hi 21-cm absorption line in absolute value.
\item
PF - the \hi peak fraction, defined by $\rm PF = \lvert S_{\rm HI} \rvert / S_{\rm ASKAP} $
\item 
FW20 - the full width of the \hi line at 20\% 
\end{itemize}

 To measure $\rm S_{\rm HI}$, we fitted the \hi profiles with a 1-D Gaussian and estimated the peak flux density in the line. For Gaussian fits to the spectra of all sources, a chi-square test was used to determine the `goodness’ of the fit; the reduced $\chi ^{2}$ value is close to one for all profiles. 
 
 The values $\rm S_{ASKAP}$, PF, and FW20 for each target are listed in Table~\ref{askap_results}. We note that FW20 is less than $\rm 250 \ km \ s^{-1}$ for all absorbers in this sample. 
\newline
\newline
2) {\bf Parameters from VLBA observations}: 
\vspace*{-0.1cm}
\begin{itemize}
\item
$\rm S_{\rm core}$ - the integrated flux density of the core in the VLBA image 
\item 
$\rm S_{\rm tot}$ - the integrated flux density of the total emission detected in the VLBA image 
\item  
$\rm S_{\rm min}$ - an estimate of the minimum flux density arising from \hix-obscured emission (see section \ref{sec:method}) . 
\end{itemize}
These values for each target are listed in Table~\ref{vlba_results}.
\newline
\newline
3) {\bf Other parameters}: 
\vspace*{-0.1cm}
\begin{itemize}
\item
$\rm S_{\rm RACS}$: the 1.4\,GHz flux density from the RACS-mid survey 
\item
$\rm S_{min}/S_{\rm RACS}$: the lower limit on the covering factor $f$ 
\end{itemize}
These values are also listed in Table~\ref{vlba_results}.

\subsubsection{Methodology}
\label{sec:method}
As a general principle, we estimate a minimum flux density $\rm S_{min}$ using either the core or the total emission detected in the VLBA map, such that the flux density is large enough to cause the \hi 21-cm absorption detected in the ASKAP spectrum. We use the technique described in \ref{sec:appendix} to determine $\rm S_{min}$. 

For each target, we estimate the values of PF, $\rm S_{core} / S_{RACS}$, and $\rm S_{tot} / S_{RACS}$, then test the conditions described below. Based on the condition that the values satisfy, we use either $\rm S_{core}$ or $\rm S_{tot}$ as the minimum flux density, $\rm S_{min}$. We do not use a flux density that is intermediate to $\rm S_{core}$ and $\rm S_{tot}$. This is because of the clumpy nature of the \hi gas at parsec scales \citep[e.g.][]{schulz2021, morganti2013}, it is difficult to predict the \hi distribution and identify obscured continuum components.
After estimating $\rm S_{min}$, the ratio $\rm S_{min} / S_{RACS}$ yields a lower limit to the covering factor.
\newline
\newline
1. If PF < $\rm S_{core} / S_{RACS}$, then it implies that the emission from the core has sufficient flux density to cause the complete \hi 21-cm absorption. In this case, we use the core flux density as $\rm S_{min}$, to estimate a lower limit on the covering factor; i.e. $\rm S_{min} = S_{core}$ and $ {\rm S_{min} / S_{RACS}} < f < 1 $.
\newline
\newline
2.  If $\rm S_{core} / S_{RACS} < PF < S_{tot} / S_{RACS}$, this implies that the emission from the core is insufficient, but the total emission in the VLBA image has sufficient flux density to cause the complete \hi 21-cm absorption. In this case, we use $\rm S_{tot}$ to estimate a lower limit to the covering factor, i.e. $\rm S_{min} = S_{tot}$. 
\newline
\newline
3. If $\rm S_{tot} / S_{RACS} < PF$, this implies that the entire emission detected in the VLBA image 
has insufficient flux density to cause the complete \hi 21-cm absorption. In this case, $\rm S_{min} > S_{tot}$; we cannot estimate $\rm S_{min}$, and we use PF as a lower limit to the covering factor, i.e. ${\rm PF} < f  < 1$ (see \ref{sec:appendix} for a description). We cannot place an upper limit on the optical depth.

\subsection{Evaluation of each source}

\begin{figure*}
\caption{0011-023: On left, the VLBA L-band continuum image. Contours are at the levels $\rm 0.6 \ mJy \times (1, 2, 4, 8..)$, where 0.6 mJy is the three times the r.m.s noise ($\sigma$) on the continuum image, as listed in Table \ref{vlba_results}. The peak flux density in the image is 26.6 mJy/bm. The red cross marks represent the peaks of 2D Gaussian fits to various emission components. The core is labelled with the text `C'. On right, the ASKAP \hi 21-cm absorption spectrum. The shaded region represents $5\sigma$ noise on the spectrum. The red dotted line is the 1-D Gaussian fit to the absorption profile.} \label{fig:0011-023}
\centering

\includegraphics[width=0.4\textwidth]{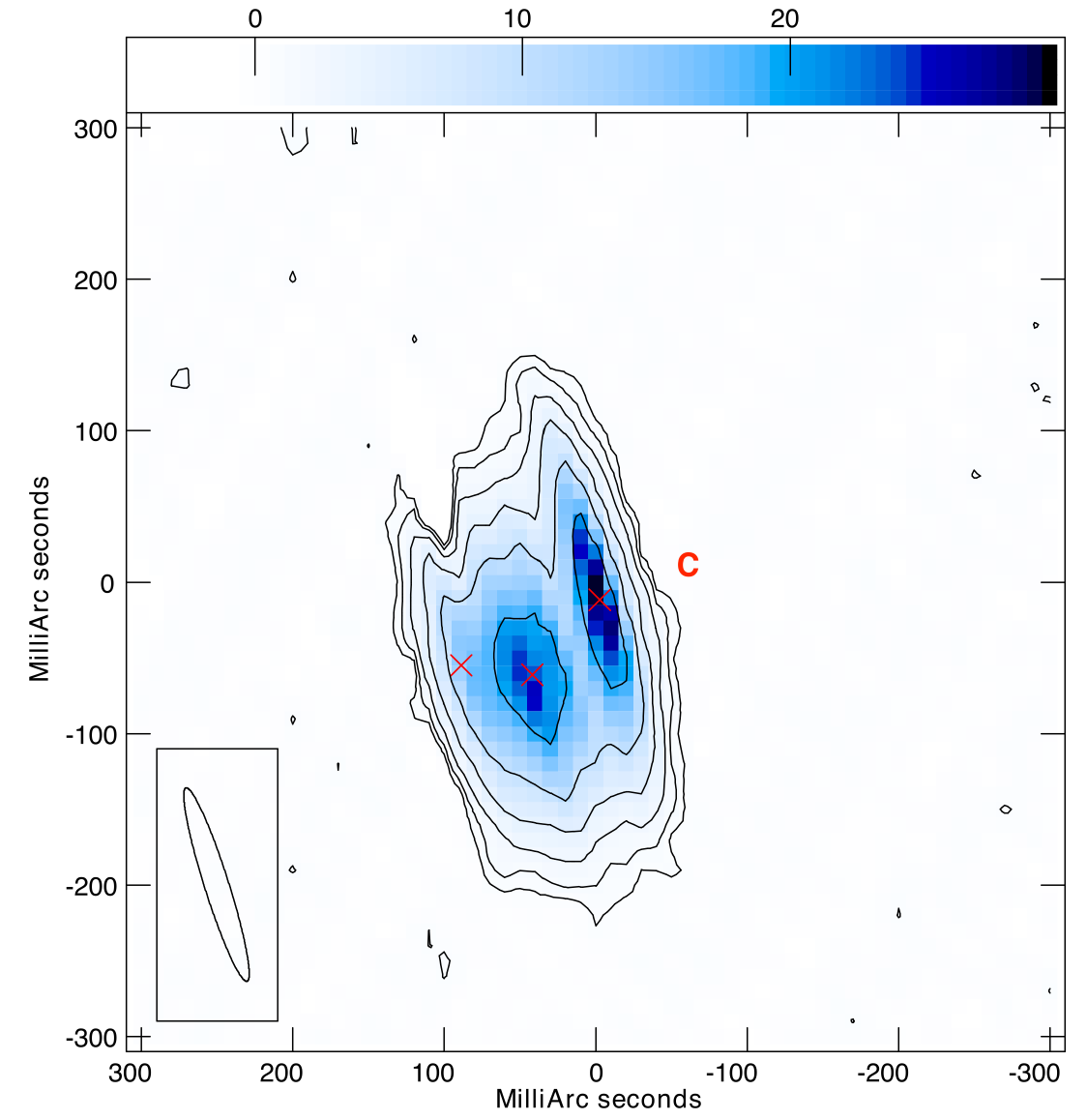}
\includegraphics[width=0.4\textwidth]{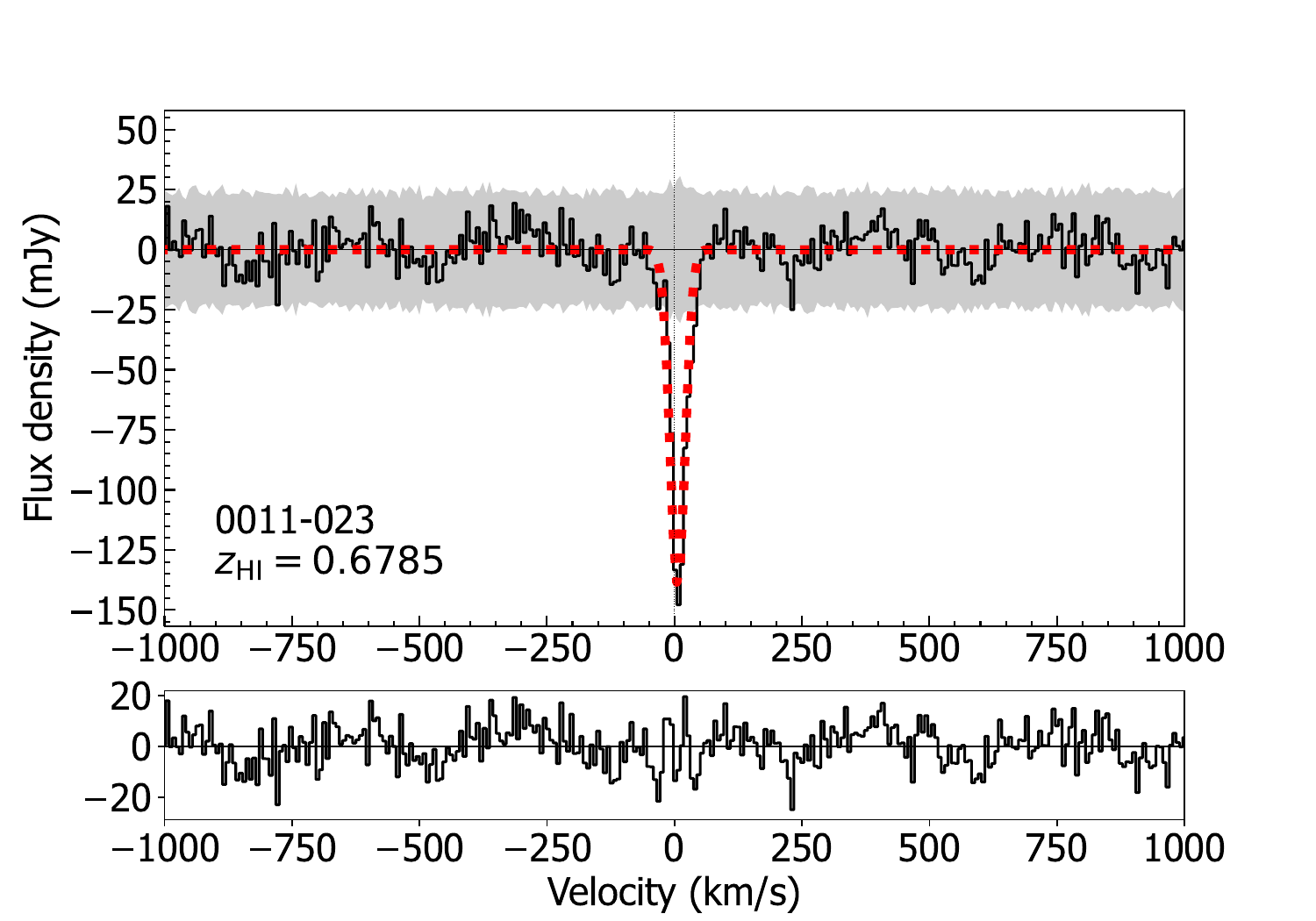}

\end{figure*}

\begin{figure*}
    \includegraphics[width=0.45\textwidth]{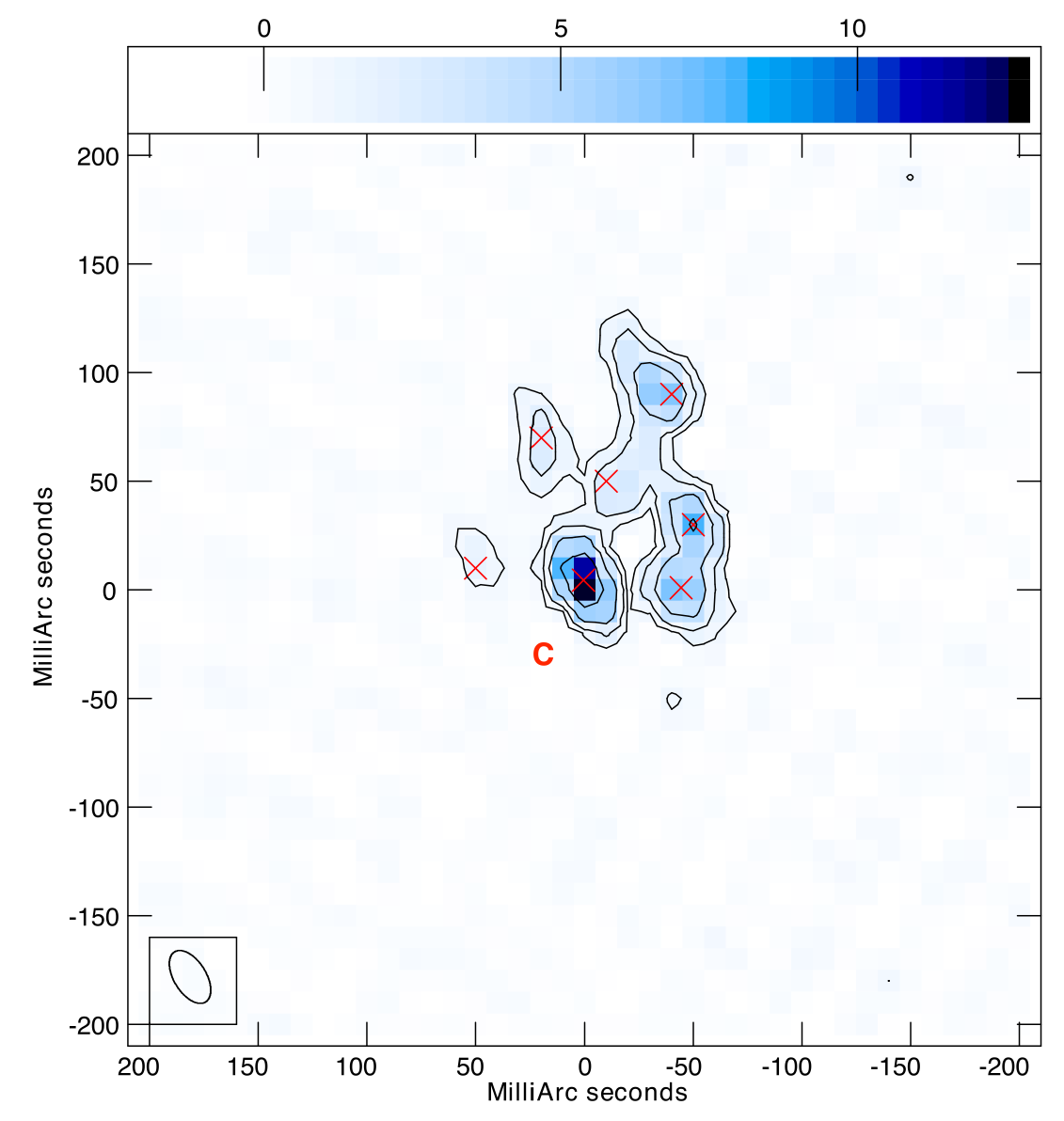}
    \includegraphics[width=0.5\textwidth]{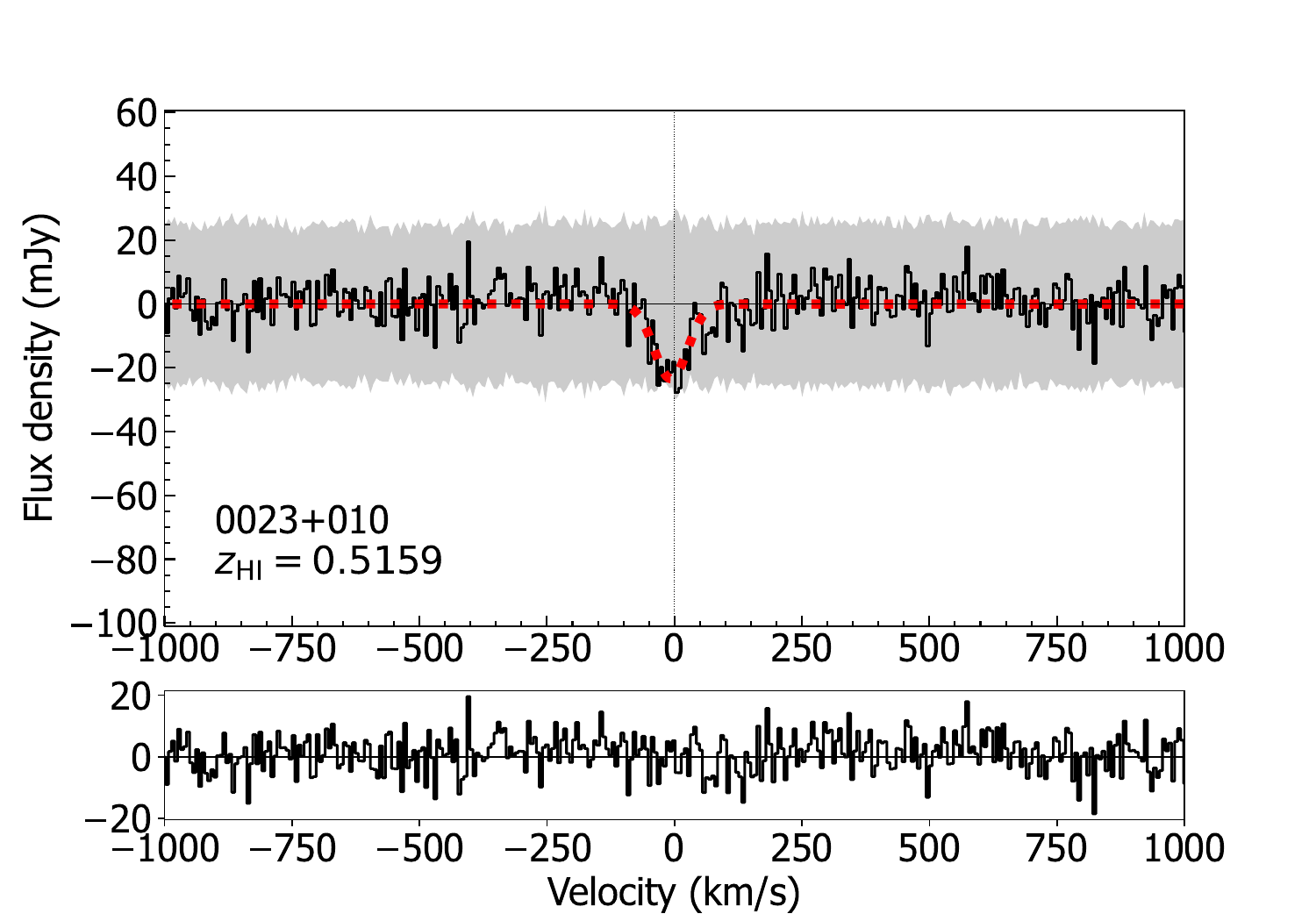} 
        \includegraphics[width=0.4\textwidth]{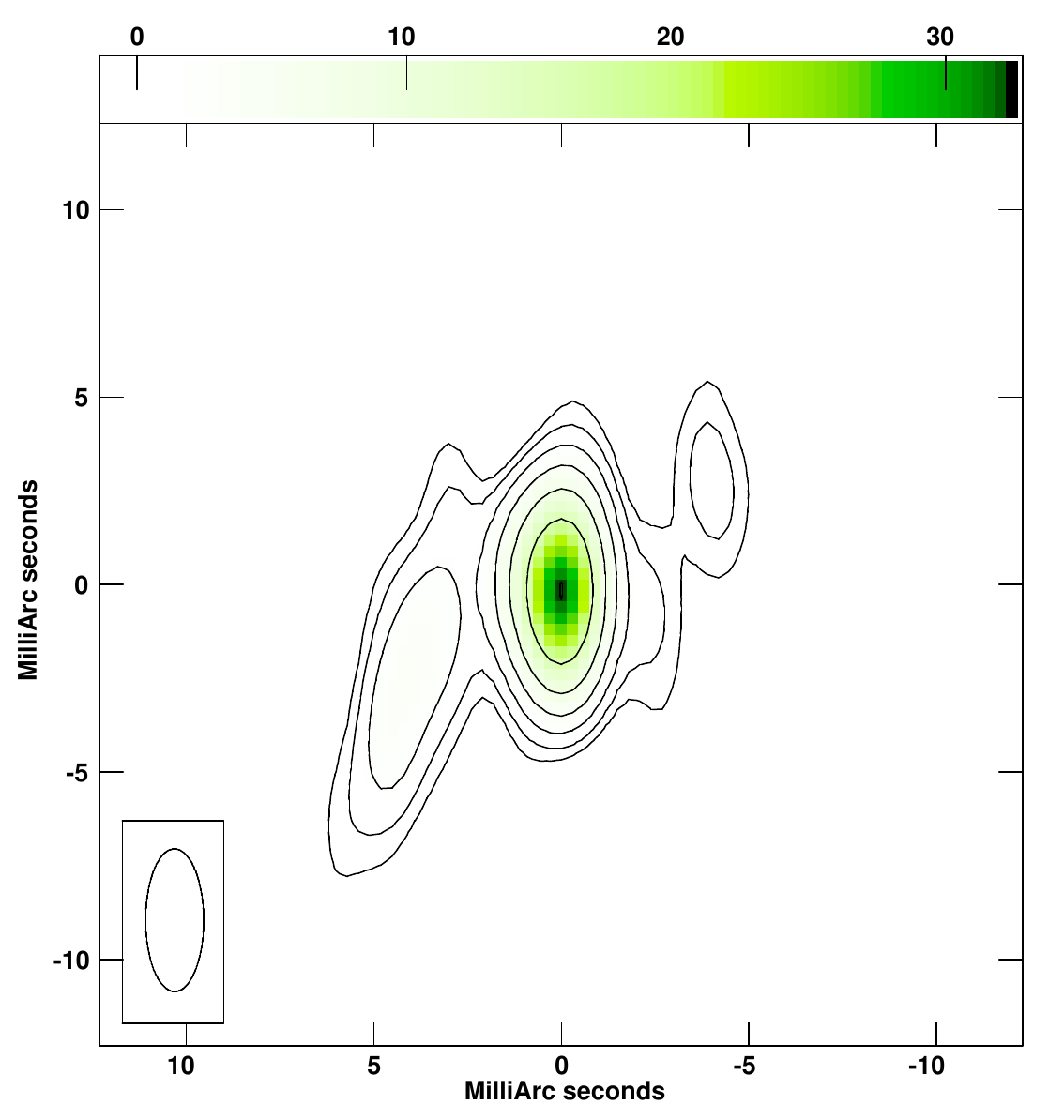}
    \includegraphics[width=0.4\textwidth]{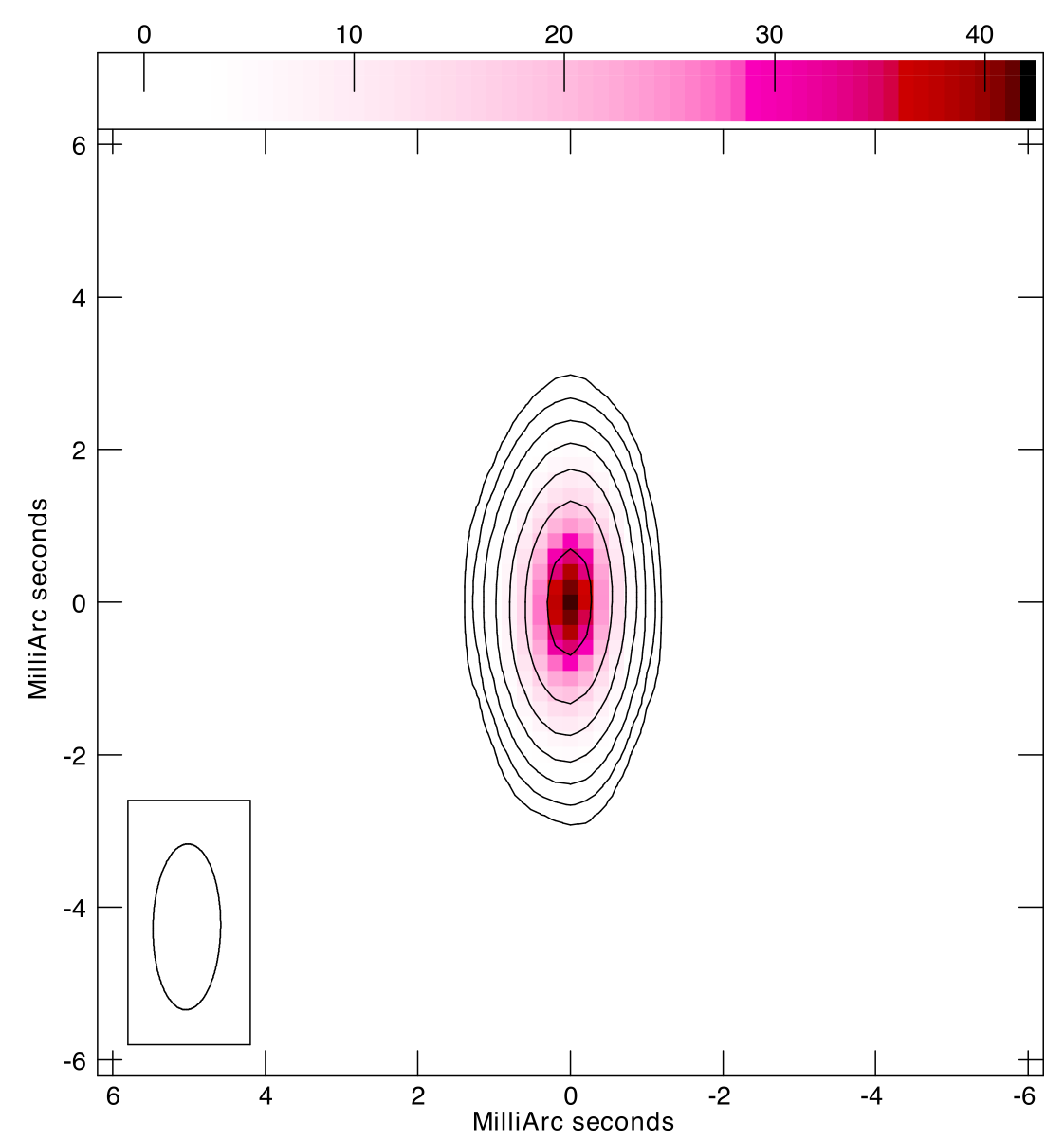} 
    
    \caption{0023+010: Details of the top left and top right images are same as that of Figure~\ref{fig:0011-023}. Contours in the top-left panel are at the levels $\rm 0.9 \ mJy \times (1, 2, 4, 8..)$. The peak flux density in the image is 13.5 mJy/bm. The left and right panels in the bottom are the 4.3 GHz and 7.6 GHz continuum images, where the data was taken from astrogeo. The contours in both images are at $\rm 0.5 \ mJy \times (1, 2, 4, 8..) $. The peak flux densities in the images are 31 mJy/bm and 42 mJy/bm, respectively. } \label{fig:0023+010}
    
\end{figure*}

\begin{figure*}
    \includegraphics[width=0.4\textwidth]{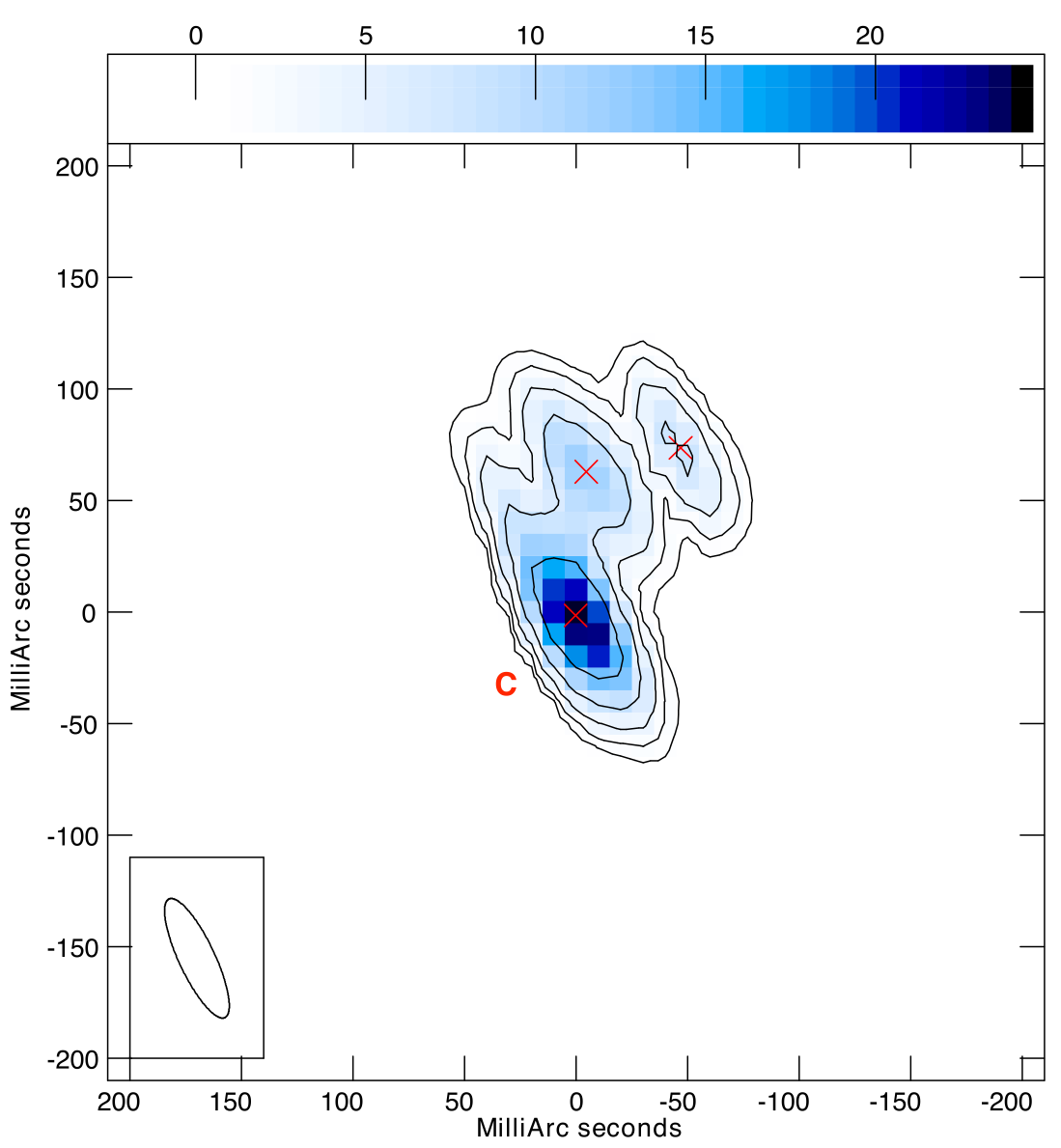}
    \includegraphics[width=0.5\textwidth]{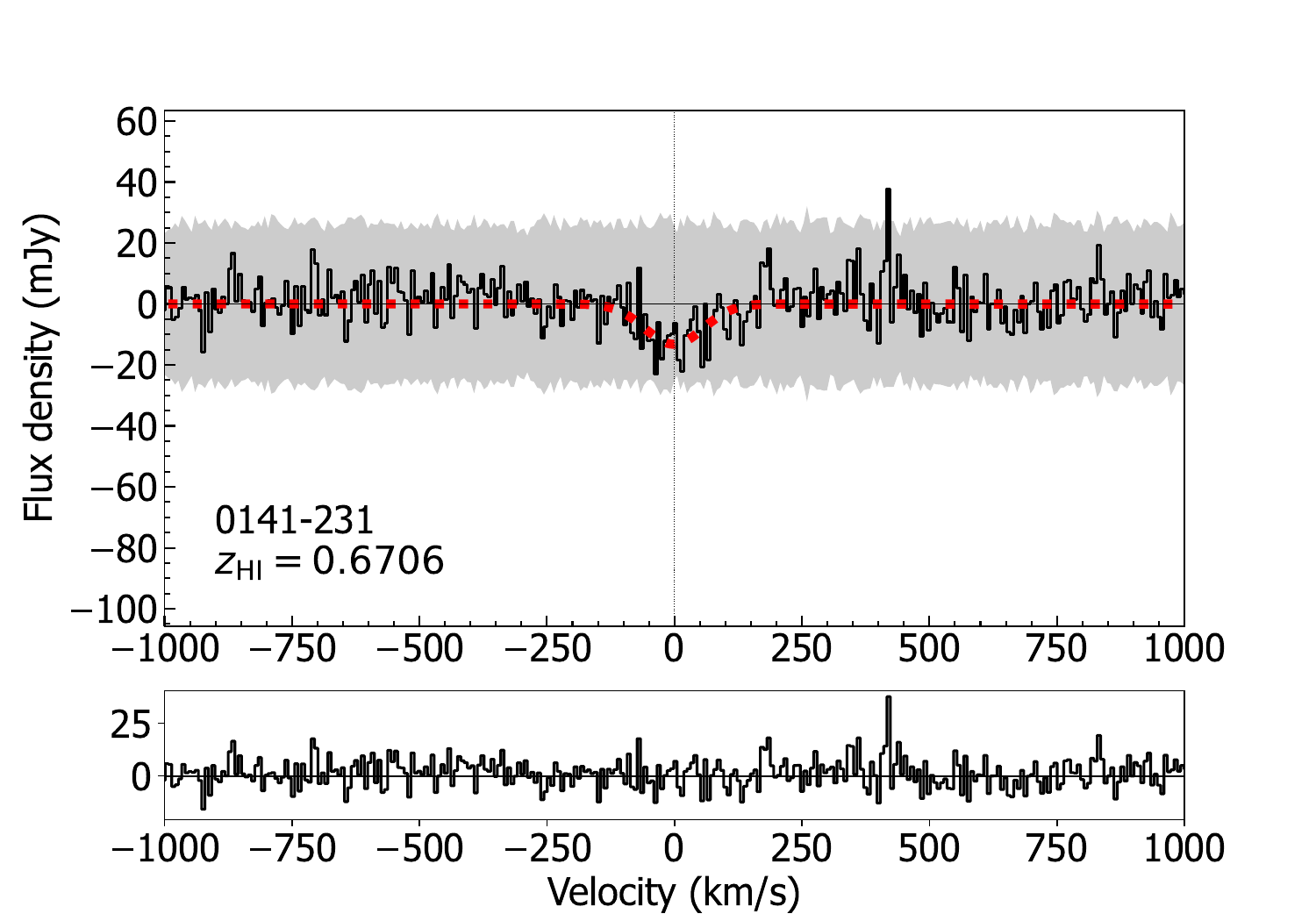}
        \includegraphics[width=0.4\textwidth]{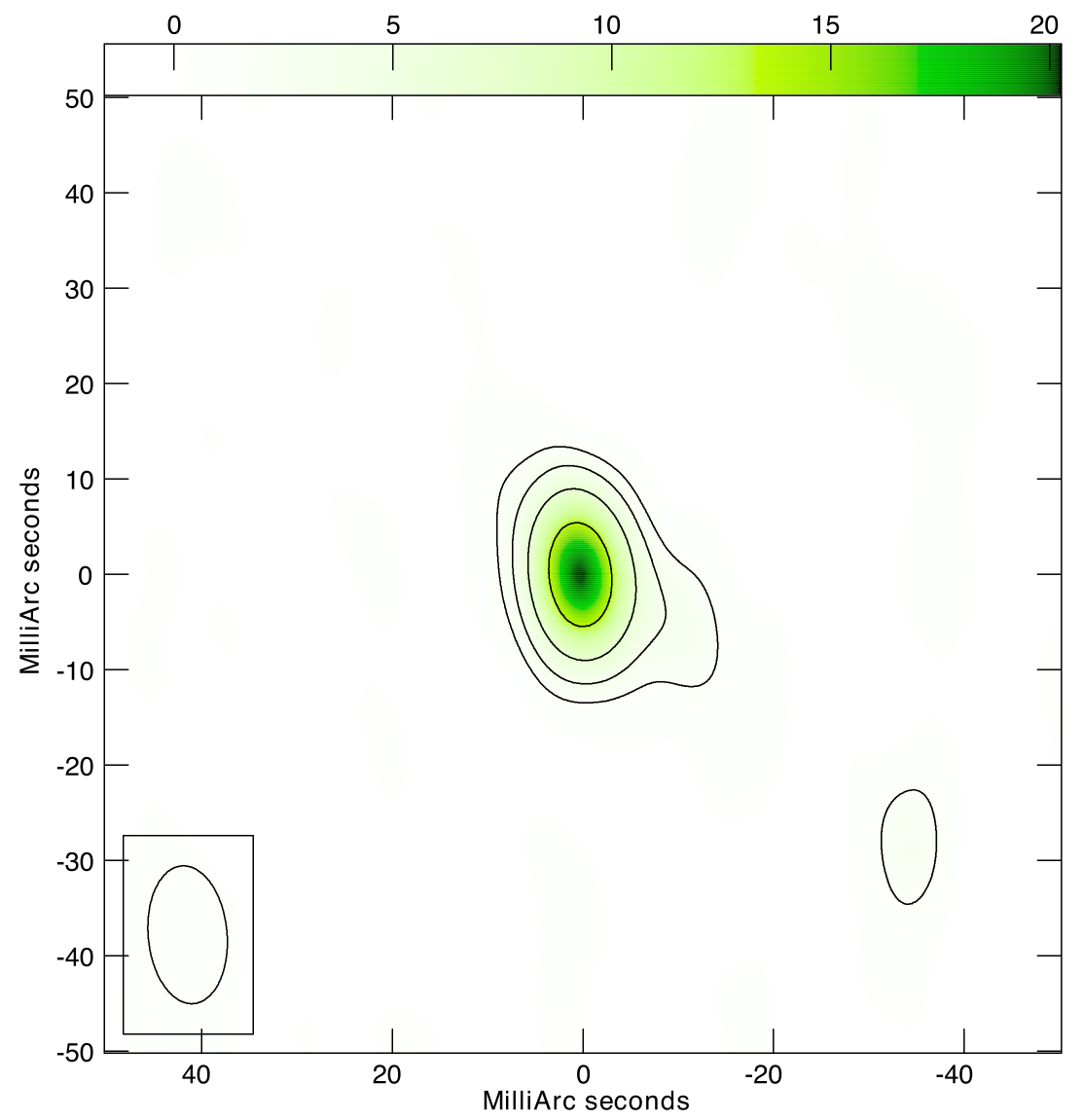}
     \caption{0141-231: Details of the top left and top right images are same as that of Figure~\ref{fig:0011-023}. Contours in the top-left panel are at the levels $\rm 0.9 \ mJy \times (1, 2, 4, 8..)$. The peak flux density in the image is 25.5 mJy/bm. The bottom panel is the 4.3 GHz continuum image, where the data was taken from astrogeo. The contours in the image are at $\rm 1.7 \ mJy \times (1, 2, 4, 8..) $. The peak flux density in the image is 19 mJy/bm.  } \label{fig:0141-231}
\end{figure*}

\begin{figure*}[!h]
\centering

\includegraphics[width=0.4\textwidth]{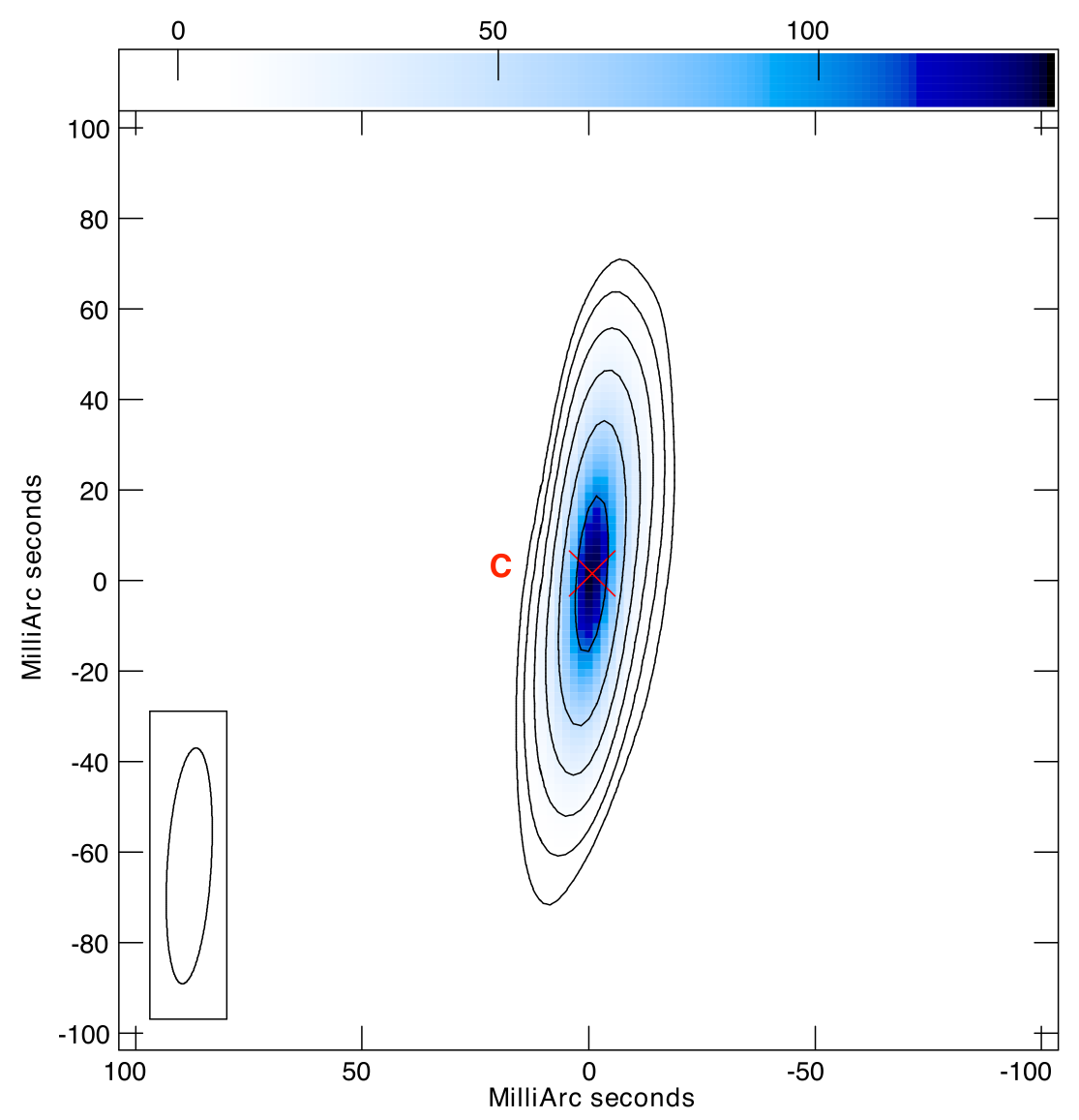}
    \includegraphics[width=0.4\textwidth]{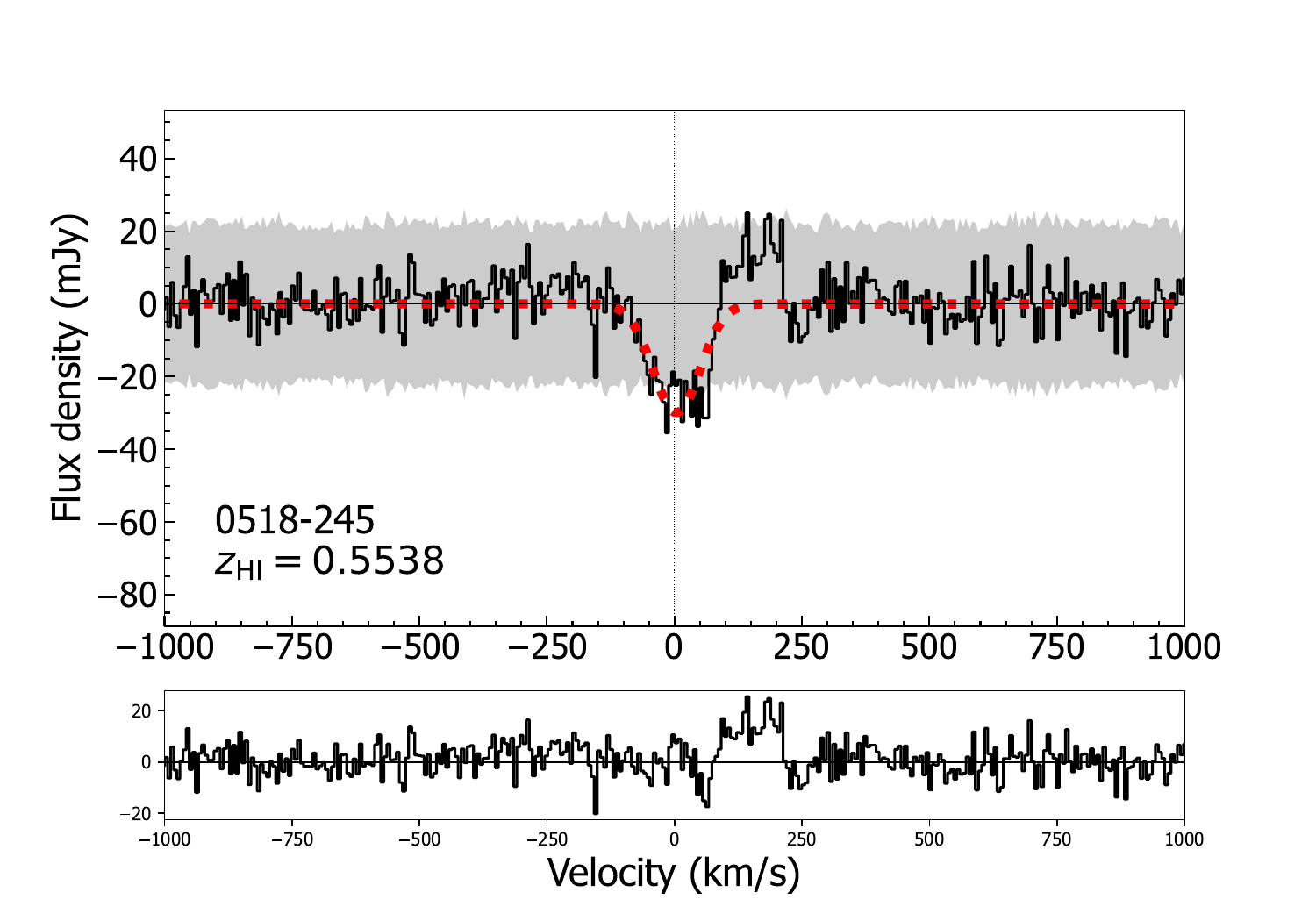}
    \includegraphics[width=0.4\textwidth]{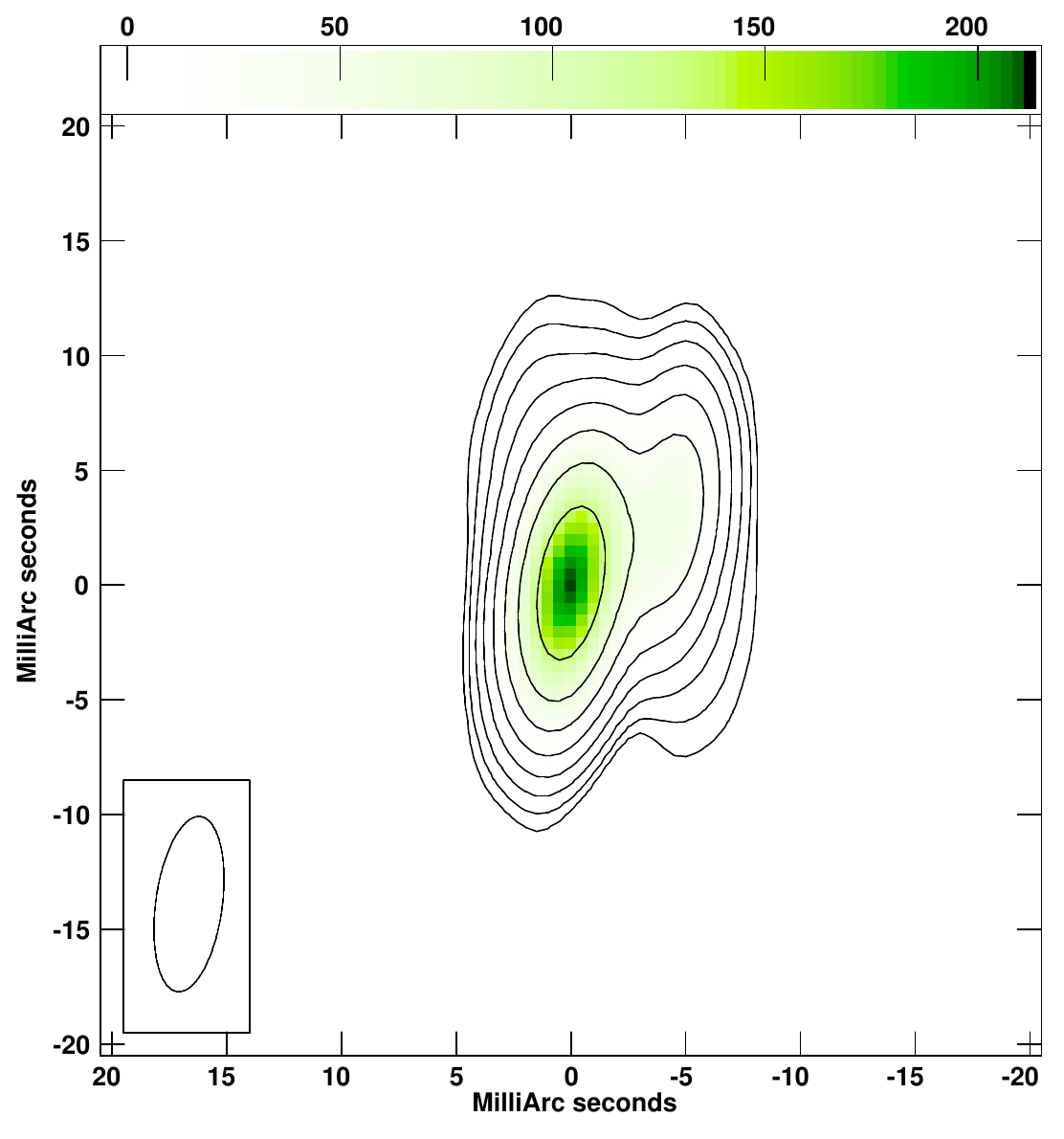}
    \includegraphics[width=0.4\textwidth]{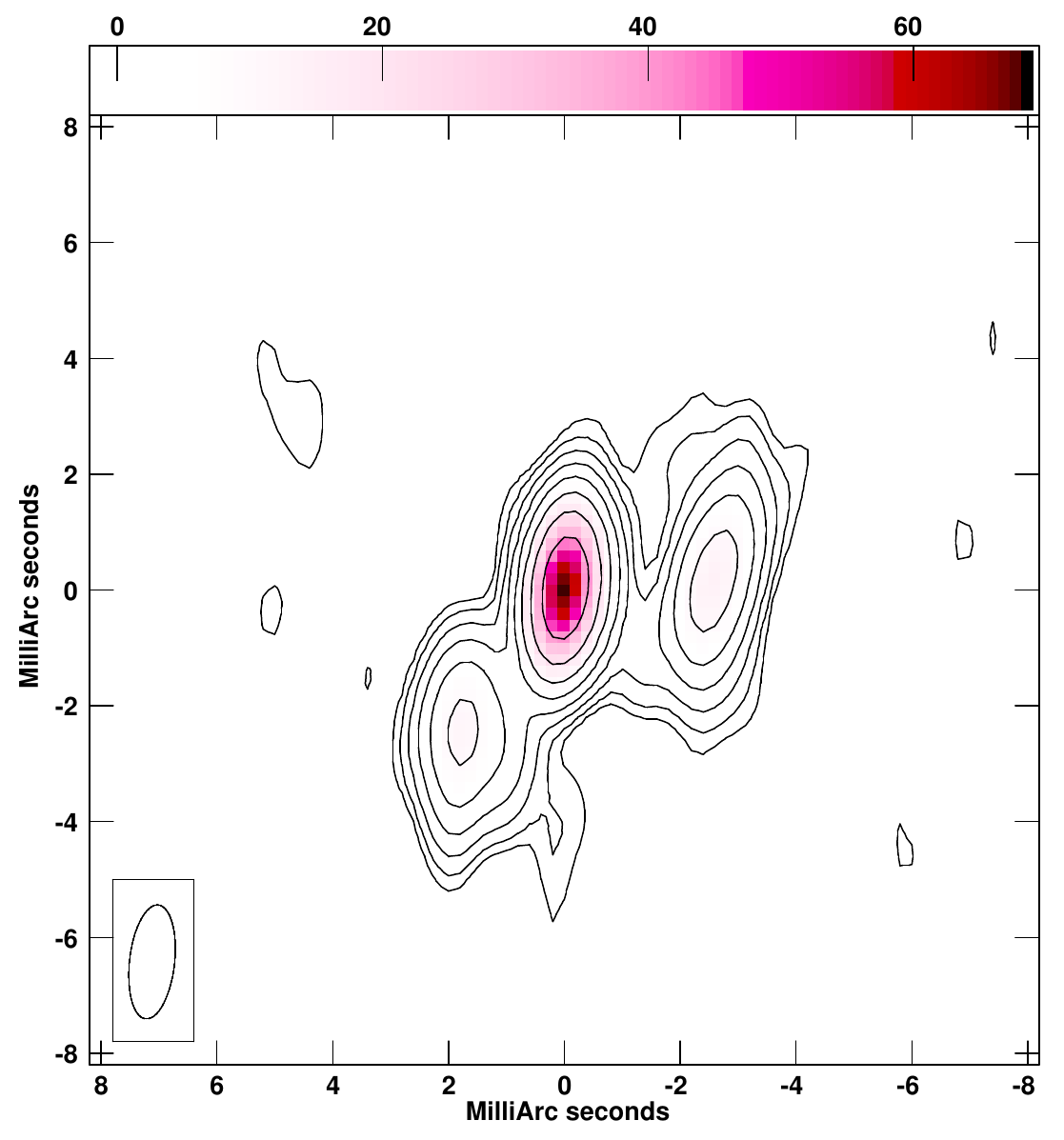}

   \caption{0518-245: Details of the top left and top right images are same as that of Figure~\ref{fig:0011-023}. Contours in the top-left panel are at the levels $\rm 3.3 \ mJy \times (1, 2, 4, 8..)$. The peak flux density in the image is 131.9 mJy/bm. The left and right panels in the bottom are the 2.3 GHz and 8.7 GHz continuum images, where the data was taken from astrogeo. The contours in the 2.3 GHz image are at $\rm 1.0 \ mJy \times (1, 2, 4, 8..) $, and those in the 8.7 GHz image are at $\rm 0.3 \ mJy \times (1, 2, 4, 8..) $. The peak flux densities in the two images are 205 mJy/bm and 64 mJy/bm, respectively.}\label{fig:0518-245}
\end{figure*}

\begin{figure*}[!h]
\centering

 \includegraphics[width=0.4\textwidth]
    {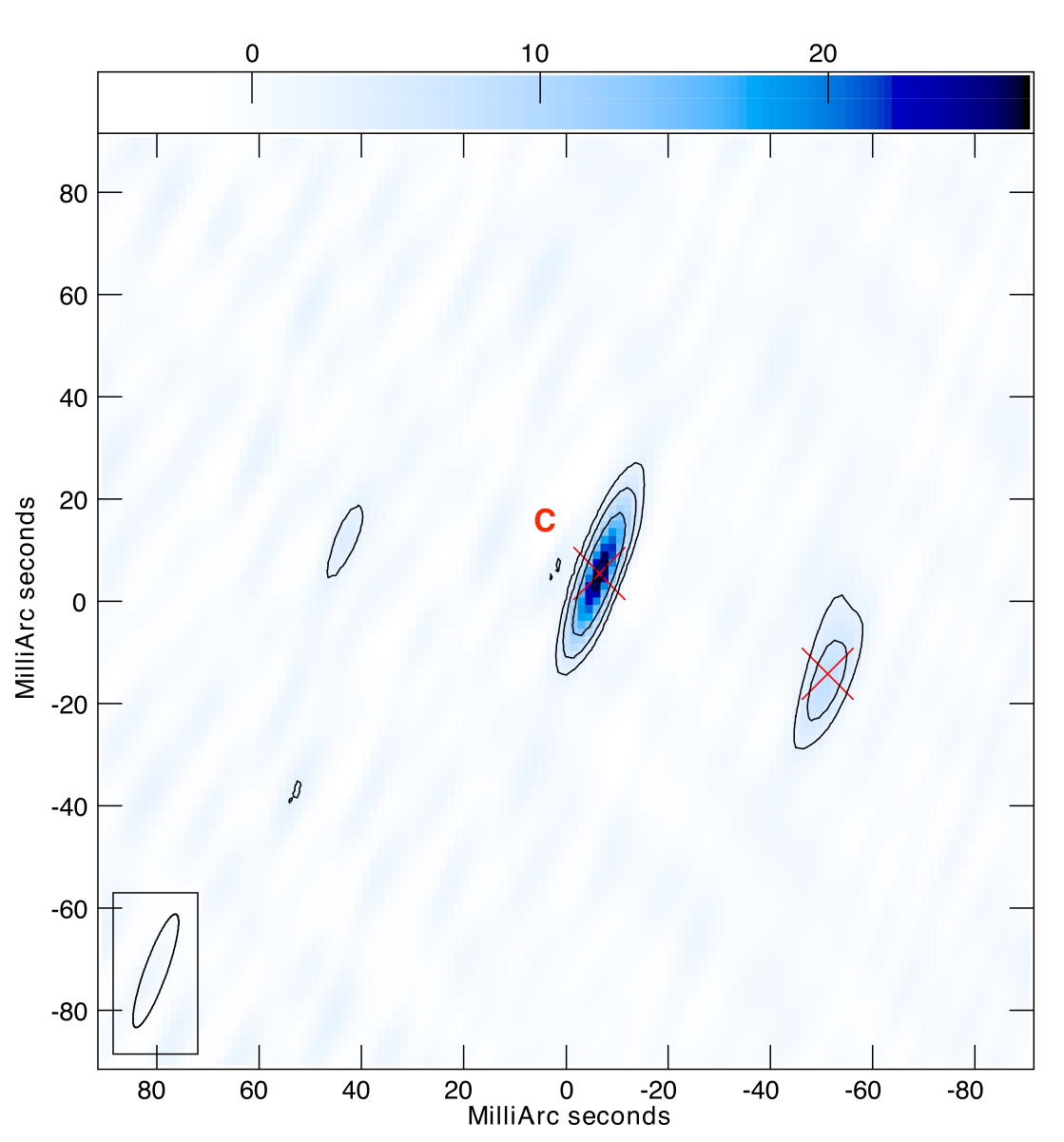}
    \includegraphics[width=0.4\textwidth]{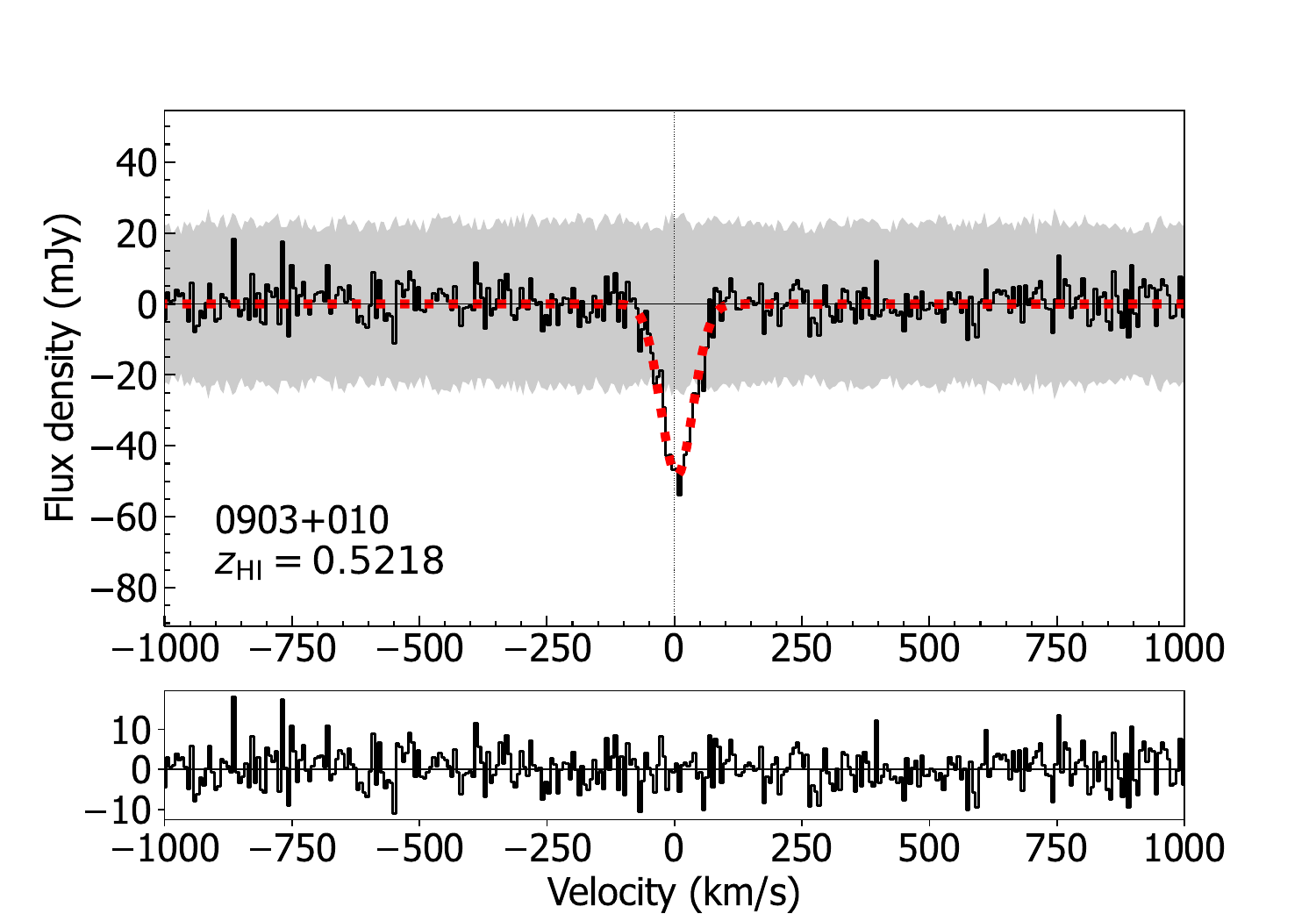} 

\caption{0903+010: Details of the images are same as that of Figure~\ref{fig:0011-023}. Contours in the left panel are at the levels $\rm 2.4 \ mJy \times (1, 2, 4, 8..)$. The peak flux density in the image is 26.5 mJy/bm.}\label{fig:0903+010}
\end{figure*}

\begin{figure*}[!h]
\centering

    \includegraphics[width=0.4\textwidth]{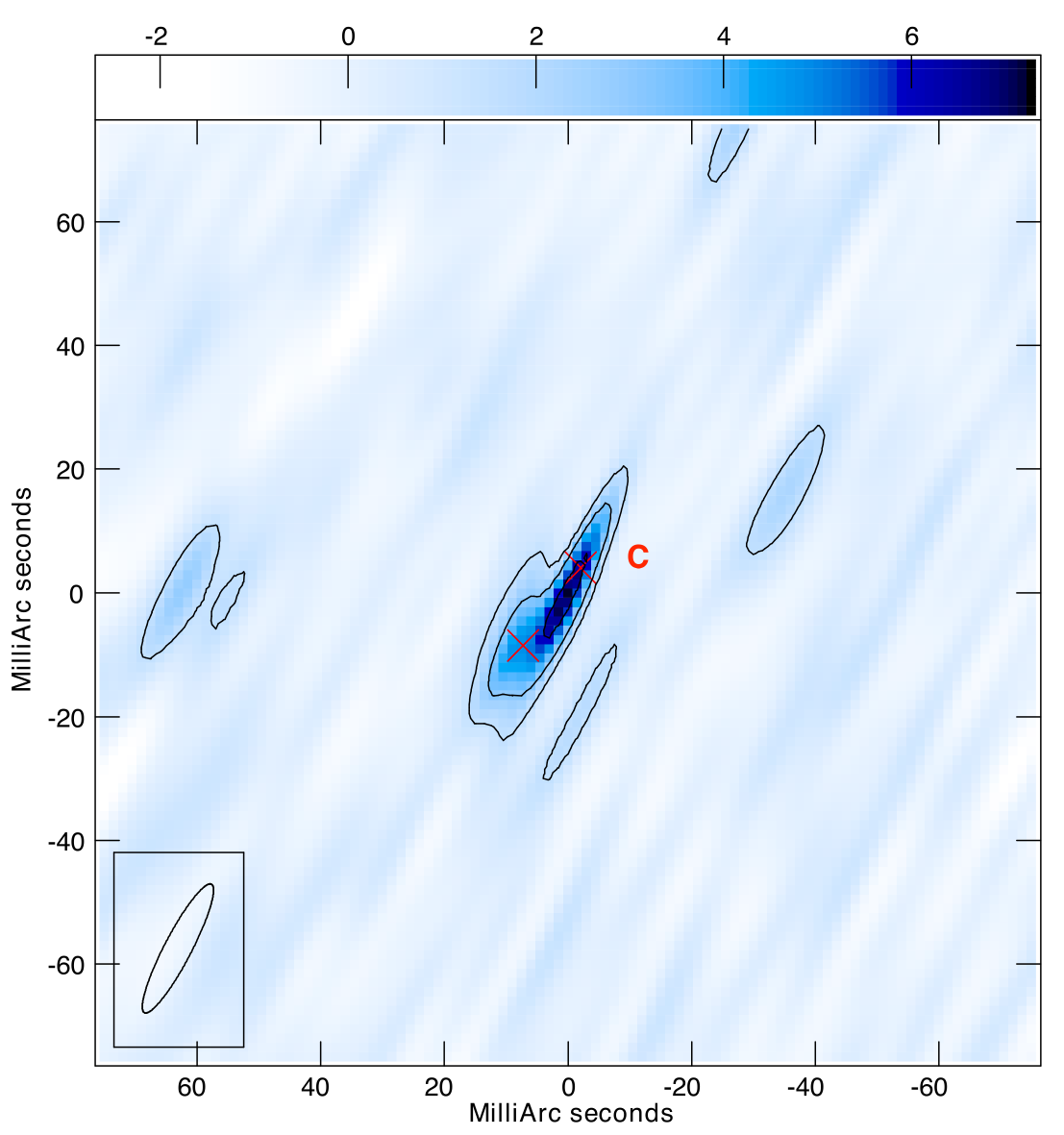}
    \includegraphics[width=0.4\textwidth]{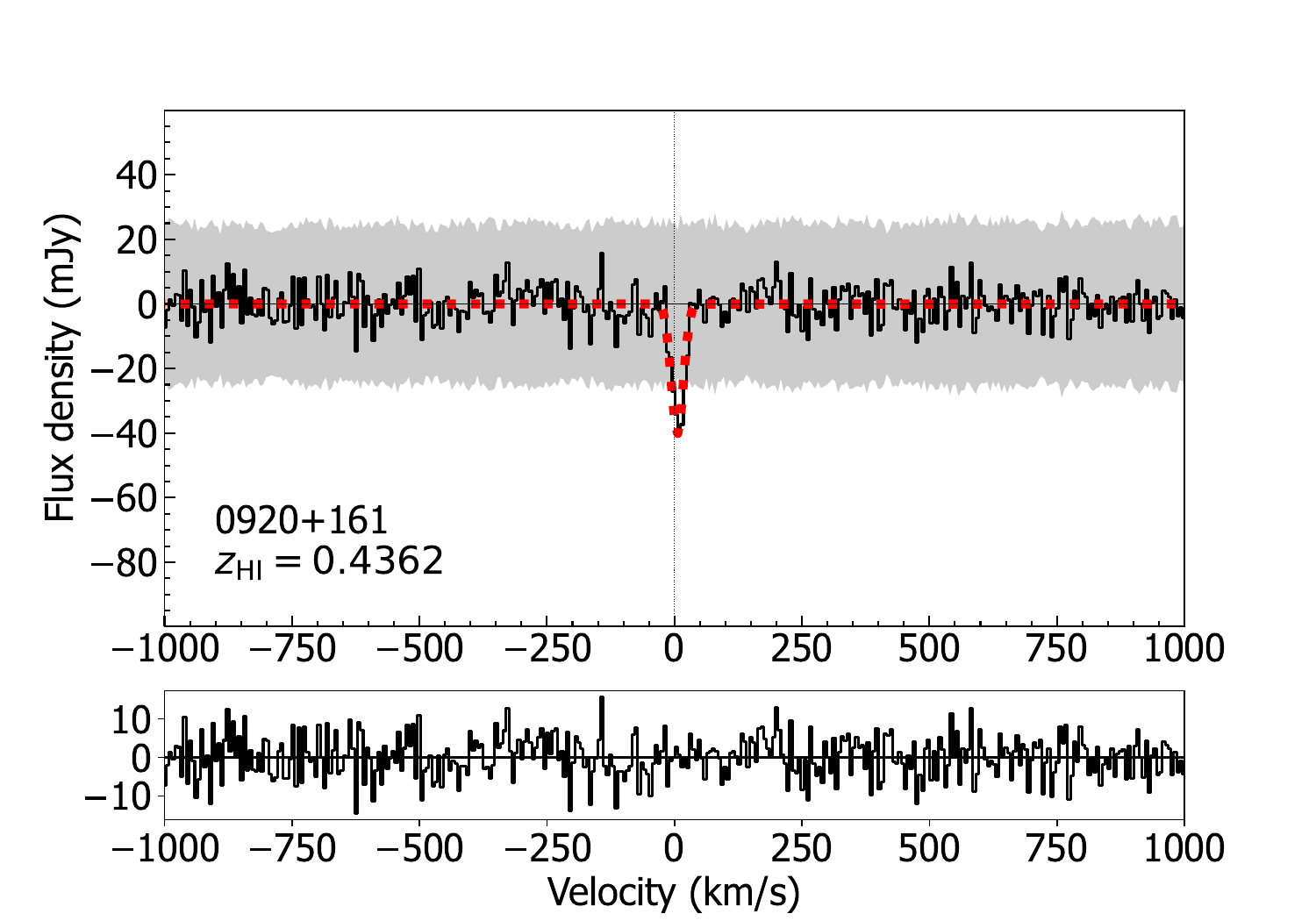}
\caption{0920+161: Details of the images are same as that of Figure~\ref{fig:0011-023}. Contours in the left panel are at the levels $\rm 1.5 \ mJy \times (1, 2, 4, 8..)$. The peak flux density in the image is 5.8 mJy/bm. }\label{fig:0920+161}
\end{figure*}

\begin{figure*}[!h]
\centering
    \includegraphics[width=0.4\textwidth]{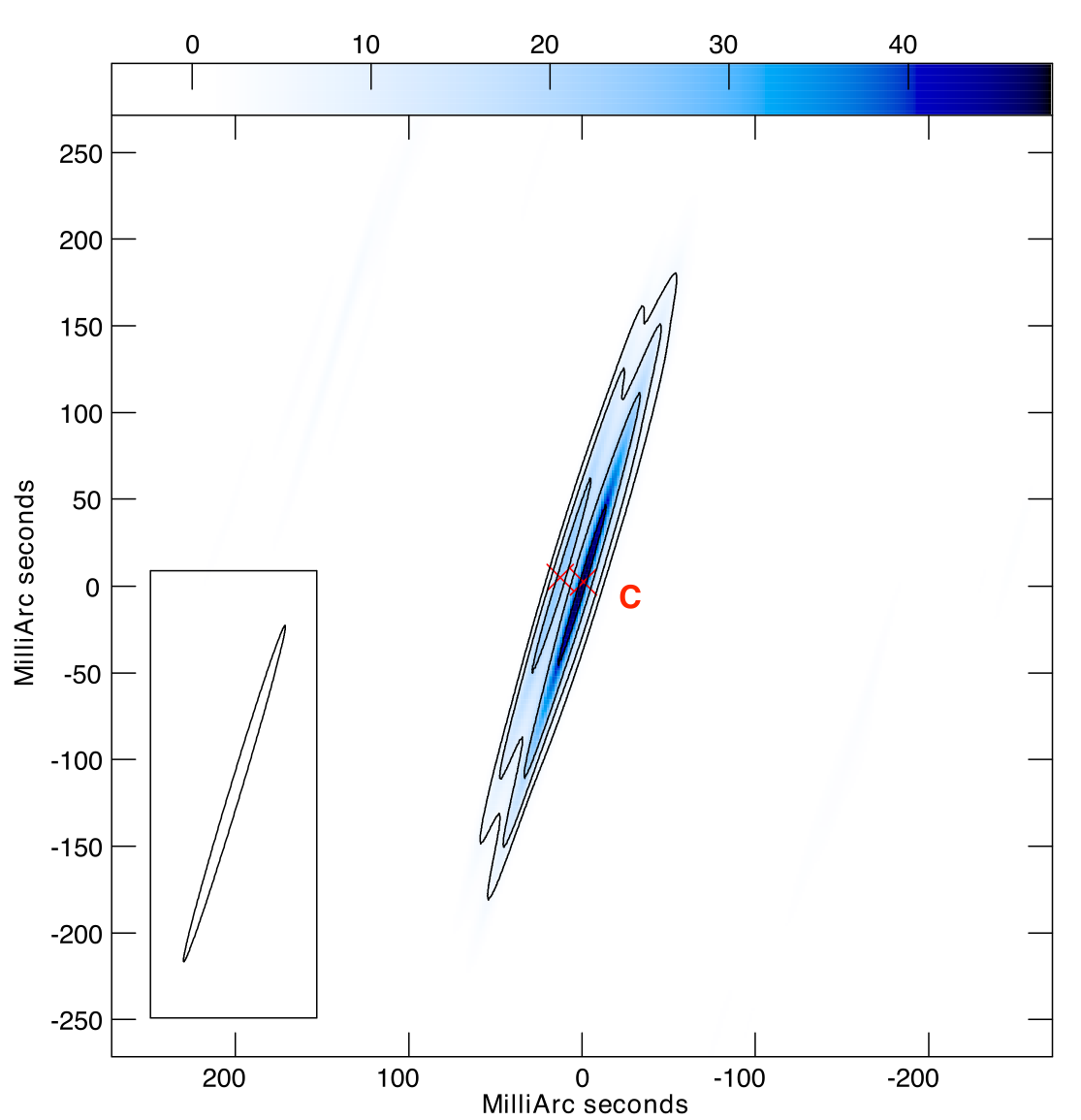}
    \includegraphics[width=0.4\textwidth]{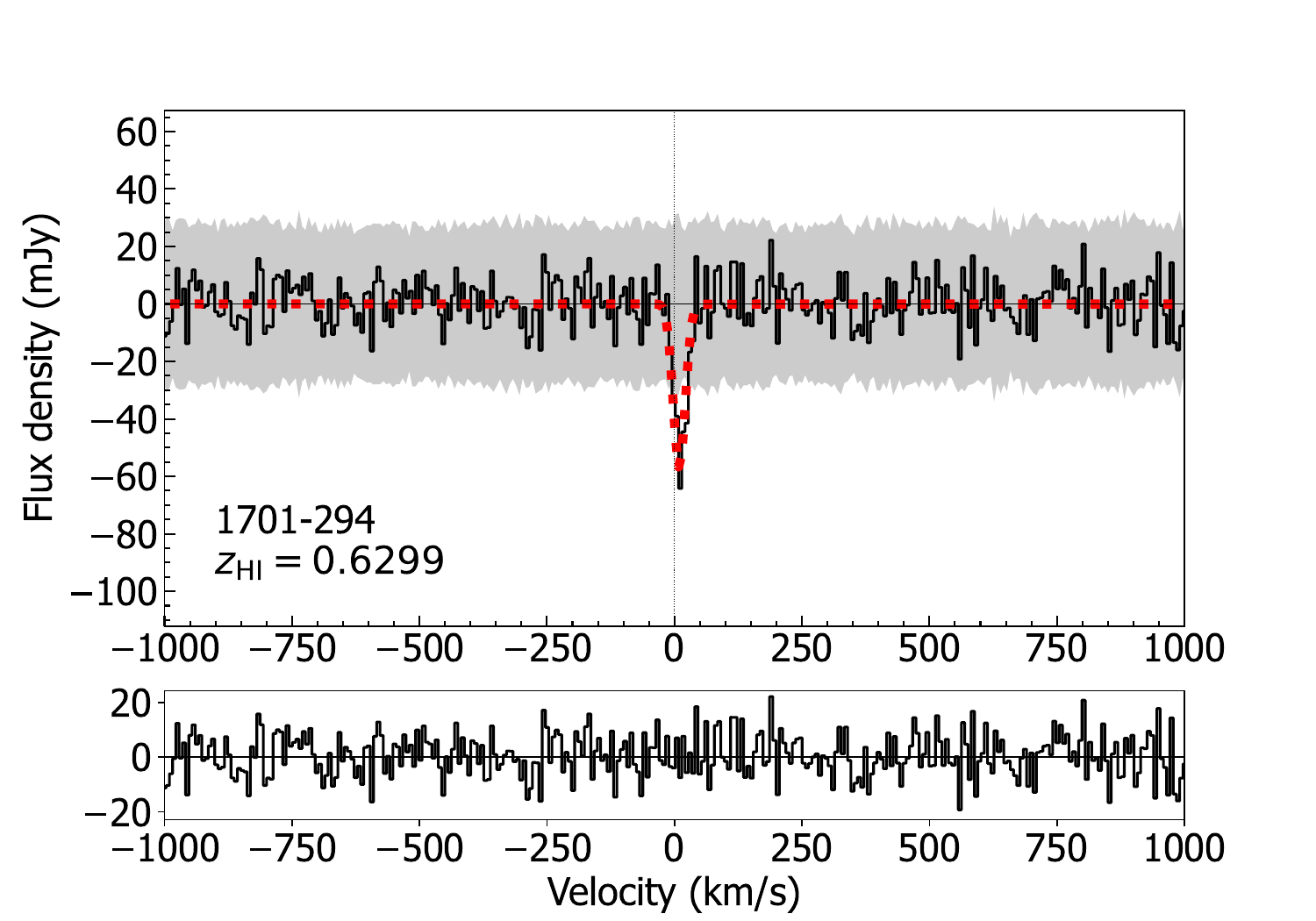}

\caption{1701-294: Details of the images are same as that of Figure~\ref{fig:0011-023}. Contours in the left panel are at the levels $\rm 5.1 \ mJy \times (1, 2, 4, 8..)$. The peak flux density in the image is 43.7 mJy/bm.}\label{fig:1701-294}
\end{figure*}

\begin{figure*}[!h]

\centering  
    \includegraphics[width=0.4\textwidth]{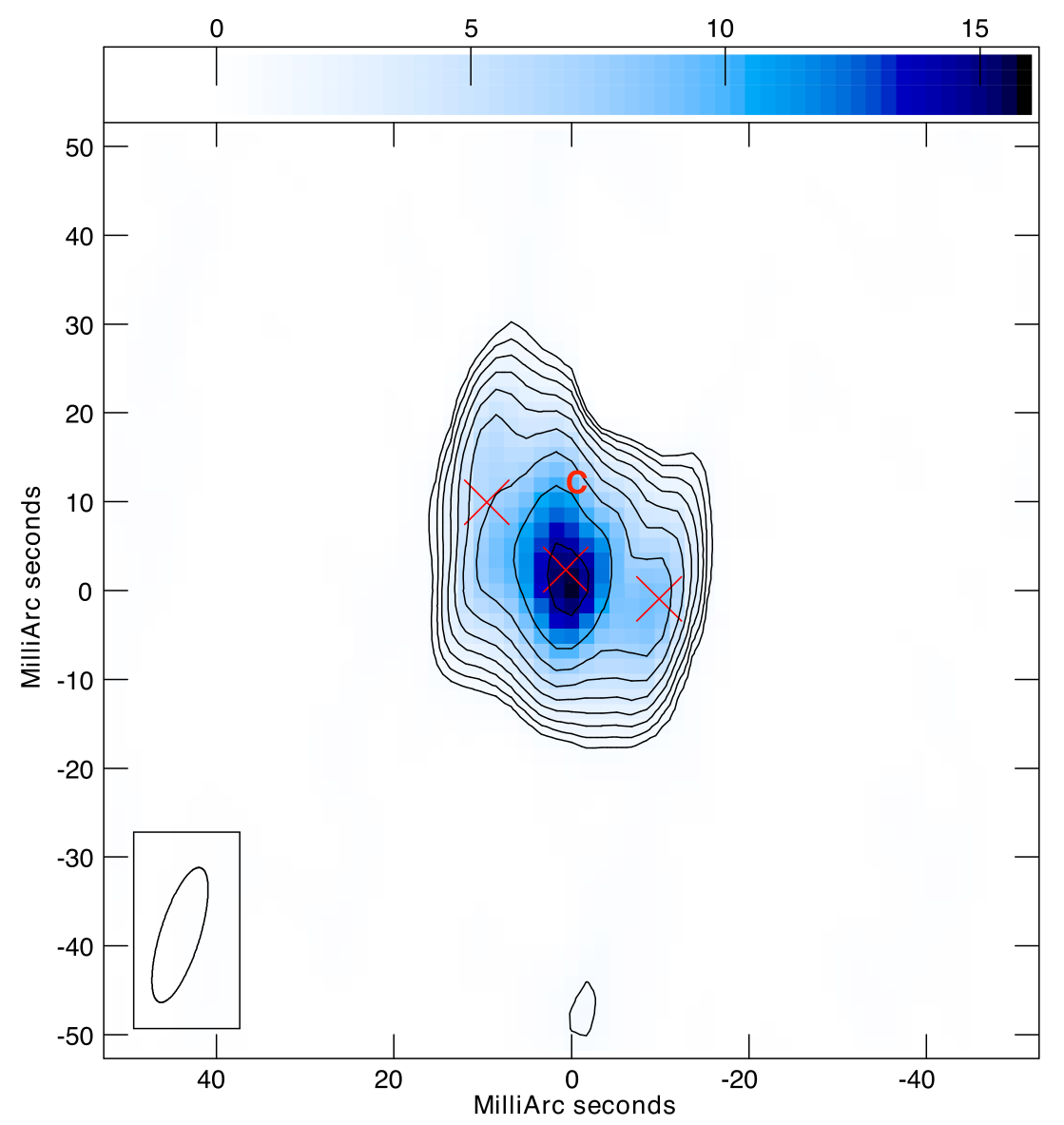}
    \includegraphics[width=0.4\textwidth]{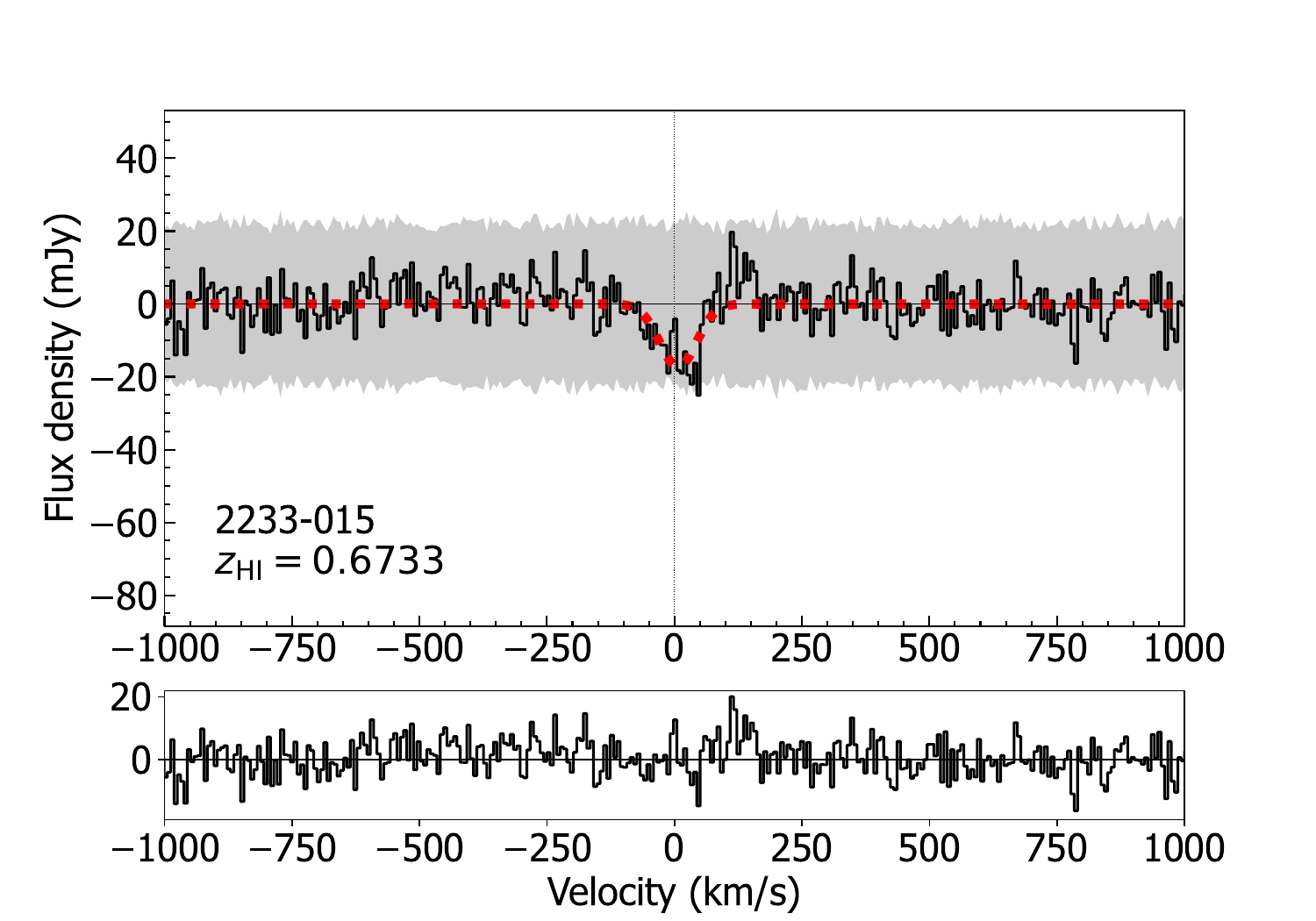} 
  \caption{2233-015: Details of the images are same as that of Figure~\ref{fig:0011-023}. Contours in the left panel are at the levels $\rm 0.9 \ mJy \times (1, 2, 4, 8..)$. The peak flux density in the image is 16.1 mJy/bm. }\label{fig:2233-015}  
\end{figure*}

\begin{figure*}[!h]

    \includegraphics[width=0.4\textwidth]{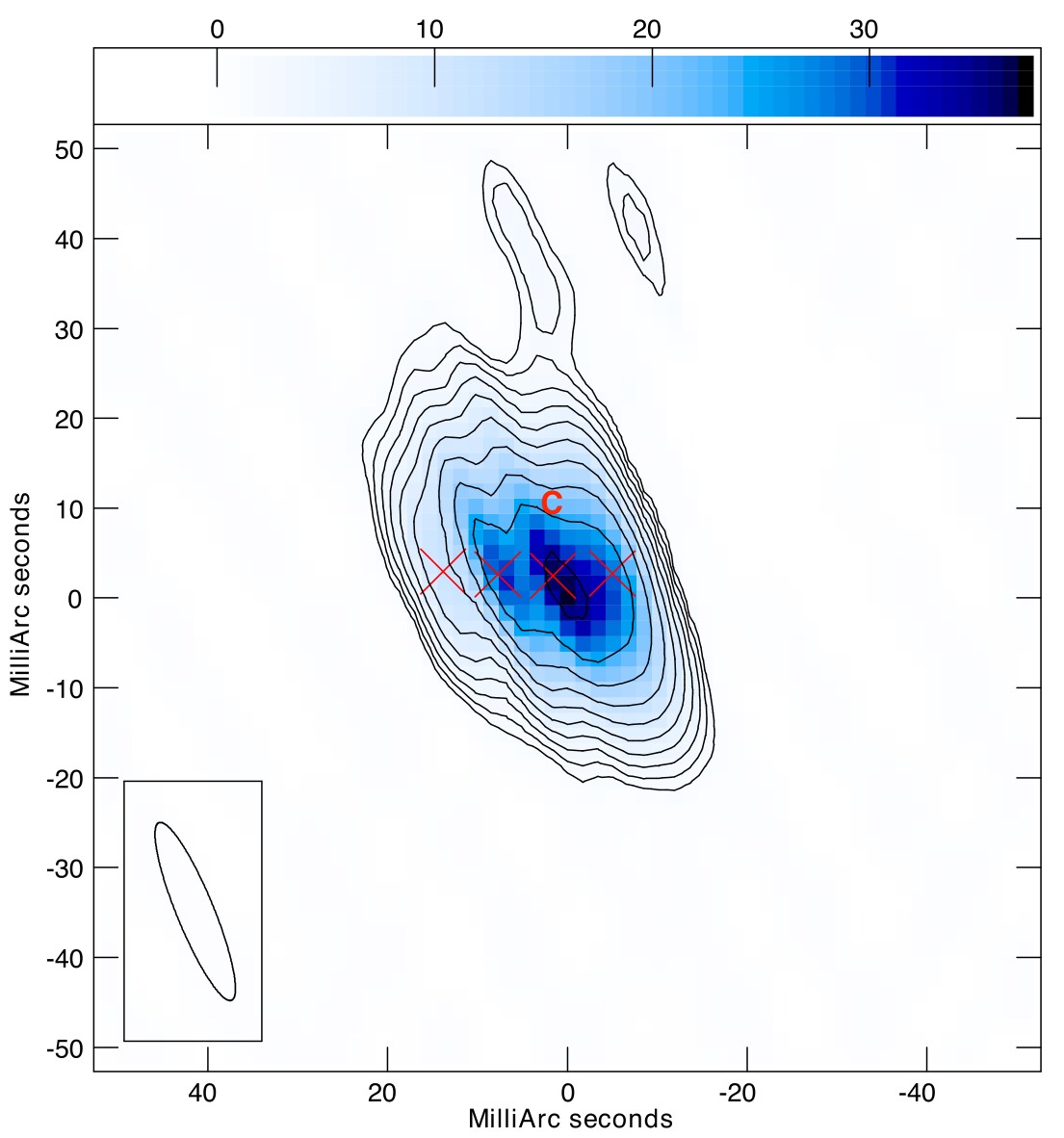}
    \includegraphics[width=0.4\textwidth]{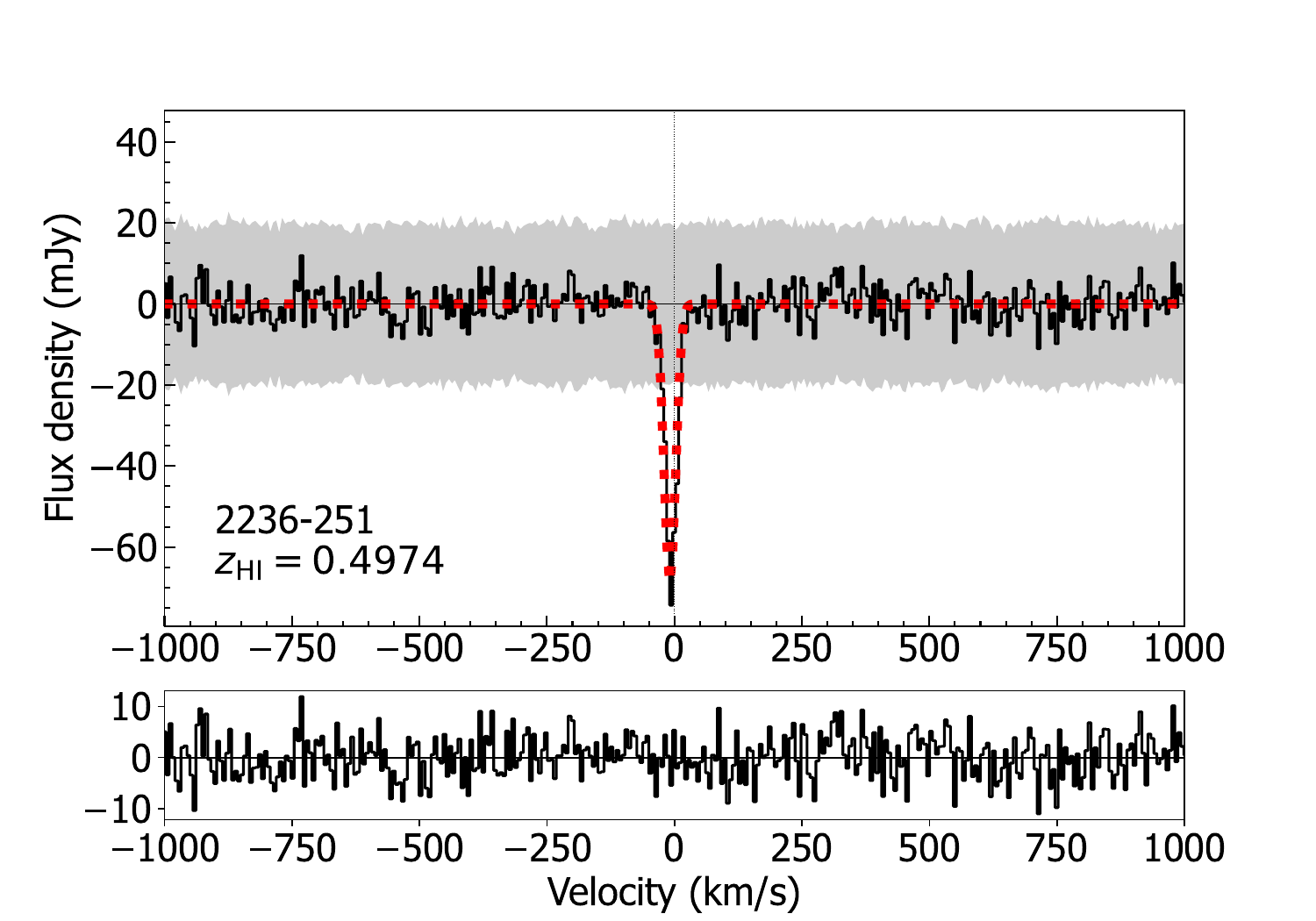}
  \caption{2236-251: Details of the images are same as that of Figure~\ref{fig:0011-023}. Contours in the left panel are at the levels $\rm 1.5 \ mJy \times (1, \sqrt{2}, 2, 2\sqrt{2}, 4,..)$. The peak flxu density in the image is 33.2 mJy/bm. }\label{fig:2236-251}
\end{figure*}

\begin{description}[font=$\bullet$\scshape\bfseries]
    \item \textbf{0011-023:}

    The source in the L-band image is fitted with three two-dimensional (2-D) Gaussian components (see Figure~\ref{fig:0011-023}). The source could either have a two-sided jet or core-jet morphology. However, the westernmost component is the most compact with a peak-to-integrated flux density ratio of 0.55, whereas the component in the middle of the source has a ratio of 0.24. We hence identify the westernmost component as the core, which also has the highest peak flux density. We classify the source with core-jet morphology (CJ), with multiple jet components. 
    \citet[][]{petrov2025} reported that the source was observed at 4.4 GHz and 7.6 GHz but was not detected. The upper limit on the flux density is $\rm \approx 12 \ mJy/bm$.
\newline
\newline
    For this target, $\rm PF=0.21$, $\rm S_{core} / S_{RACS} = 0.11$ and $\rm S_{tot} / S_{RACS} = 0.38$; the values satisfy the relation $\rm S_{core} / S_{RACS} < PF < S_{tot} / S_{RACS}$. 
    The core has insufficient flux density to cause the entire \hi 21-cm absorption. We therefore use $\rm S_{min} = S_{tot}$, i.e. the total VLBA flux density, and estimate a lower limit on the covering factor, $\rm S_{min} / S_{RACS} = 0.38$ 
\newline
\newline
    The core is relatively compact, implying that the gas covering factor could be relatively higher. However, the core does not have sufficient flux density to cause the entire absorption, indicating a larger emission should be occulted by the gas. Assuming that the absorption is less likely to arise against large-scale diffuse emission, the results indicate that a significant fraction of the total emission detected in the VLBA image, including the core, could be covered by gas. However, we note that the value $\rm (S_{RACS} - S_{tot}) / S_{RACS} = 0.62$, i.e. a major fraction of $\rm S_{RACS}$ is not detected in the VLBA image. A trace absorption could be arising against this undetected emission. 
\newline
\newline

    \item \textbf{0023+010:}

    The source in the L-band image has a complex structure (Figure~\ref{fig:0023+010}), with seven Gaussian components. We classify the source as having a complex (CX) morphology. Multiple components are compact with equal peak and integrated flux densities, so it is not clear which of them is the core. We tentatively identify the dominant continuum component as the core. We note that there could be an uncertainty in our identification.
\newline
\newline
    For this target, PF=0.35 and $\rm S_{core}/S_{RACS} = 0.37$. However, we note that total VLBA flux density, $\rm S_{tot} = 51.4$ mJy, is slightly higher than $\rm S_{RACS} = 49.1$ mJy. This could be due to temporal variation of the source brightness or to measurement errors. In this case, we use the VLBA flux density as the total flux density of the source at L band, as it avoids errors due to temporal variations. Therefore we compare PF with $\rm S_{min}/S_{tot} = 0.36$ (instead of $\rm S_{core}/S_{RACS}$), and find that the values satisfy $\rm PF < S_{core}/S_{tot}$. This indicates that the core has sufficient flux density to cause the entire \hi 21-cm absorption, however, the $\rm S_{core}/S_{tot}$ is only slightly larger than PF. This indicates that while a significant fraction of the \hi 21-cm absorption could be arising against the core, some fraction could be arising against emission beyond the core. We estimate a lower limit on the covering factor, $\rm S_{min}/S_{tot} = 0.36$.
\newline
\newline
    The dominant component appears resolved in the 4.3 GHz image (data taken from astrogeo), having a TSJ morphology. The projected size is $\approx 15$ mas, corresponding to a physical extension of 94 pc. The source is unresolved in the 7.6 GHz image. Typical \hi gas clouds in local galaxies, with the highest column densities ($\rm \approx 10^{23} \ cm^{-2}$ ), are found to have sizes of $\rm \approx 100 \ pc$ \citep[][]{braun2012}. Note that such a gas cloud can cover the entire emission detected in the 4.3 GHz map. Using the integrated flux densities of the core identified in the 1.4 GHz image and the source detected in the 4.3 GHz map, we estimate a spectral index of 0.8 (using $\rm S \propto $ $ \nu ^ {\alpha} $ ).
\newline
\newline
   
    \item \textbf{0141-231:}
    The source has been fitted with three Gaussian components in the L-band image (see Figure~\ref{fig:0141-231}). No diffuse emission is detected. The brightest component is located at the southern end of the structure; the structure satisfies the criterion of CJ morphology. The brightest component is identified as the core. 
\newline
\newline
    The PF (0.09) and $\rm S_{core}/S_{RACS}$ (0.57) satisfy the relation $\rm PF < S_{core}/S_{RACS}$, i.e. the core has sufficient flux density to cause the entire \hi 21-cm absorption. However, this does not rule out some absorption occurring against emission beyond the core. We use $\rm S_{min} = S_{core}$, and estimate a lower limit on the covering factor, $\rm S_{min}/S_{RACS} = 0.57$. 
\newline
\newline
    In the 4.3 GHz image, the source appears nearly unresolved, but with a small faint extension to a dominant core, towards the south-west. The spectral index of the core in the 1.4 GHz image, estimated by using the integrated flux density of the total emission in the 4.3 GHz image, is -0.76.
\newline
\item \textbf{0518-245:}
    The source is unresolved (U) in the L-band image (see Figure~\ref{fig:0518-245}), and the single component is identified as the core.
    The PF (0.16) and $\rm S_{core}/S_{RACS}$ (0.65) satisfy the relation $\rm PF < S_{core}/S_{RACS}$.
    So we use $\rm S_{min}=S_{core}$, and estimate a lower limit on the covering factor, $\rm S_{min} / S_{RACS} = 0.65$. 
\newline
\newline 
    The source is resolved into multiple components in both 2.3-GHz and 8.7-GHz images. The 8.7 GHz image shows three dominant continuum components, resembling a TSJ morphology. The largest projected extent of the emission in the 2.3 GHz and 8.7 GHz maps is 162 pc and 52 pc, respectively. Note that either one or two \hi clouds of 100 pc size could completely obscure the 2.3 GHz and 8.7 GHz radio sources. 
\newline
\newline
    \item \textbf{0903+010:}
    The source has been fitted with two Gaussian components with similar fit sizes, the components are separated by $\approx 55$ mas (projected spatial extent of 347 pc, at a redshift of 0.5218, see Table~\ref{vlba_results}). The brighter component is more compact with a peak-to-integrated flux density ratio of 0.8, compared to 0.55 for the weaker component. The source is classified as a core-jet (CJ) system, where the compact component is identified as the core. 
\newline
\newline
    For this target, PF (0.82) and $\rm S_{tot}/S_{RACS}$ (0.78) satisfy the relation $\rm S_{tot}/S_{RACS} < PF$. This indicates that the absorbed emission in \hi 21-cm line is greater than the total emission at parsec scales found in the VLBA image. This implies that a large fraction of the emission detected in the VLBA image is likely covered by gas, and some fraction would arise at scales larger than 347 pc, the largest scale of the emission in the VLBA image. 
\newline
\newline
     At the same time, the fraction of the flux density undetected in the VLBA image is just 0.22, which is 3.7 times lower than PF. This implies that atleast $\approx 73\%$ of the peak absorption (of the \hi profile) cannot arise against the large-scale undetected emission, and should arise against the emission detected in the VLBA image. 
\newline
\newline
    Therefore, the two radio lobes detected in the VLBA image could be probably obscured, either by separate gas clouds or by a single larger cloud of size $\gtrsim 347$ pc. We use PF as a lower limit to the covering factor, and we cannot place an upper limit on the optical depth.
\newline
\newline
     \item \textbf{0920+161:}
     
     The source has a compact component along with a diffuse extension to the south-west. Two Gaussians are fitted to the emission region. The compact component has similar peak and integrated flux densities ($\approx 5.8$ mJy) whereas the diffused component has peak and integrated flux densities of 4.2 mJy and 11.3 mJy, respectively. The integrated flux density of the diffused component is significantly higher than that of the compact component. We identify the compact component as the core and classify the source as a core-jet (CJ) system.
\newline
\newline
     Again for this target, PF (0.23) and $\rm S_{tot}/S_{RACS}$ (0.15) satisfy the relation $\rm S_{tot}/S_{RACS} < PF$, implying that the emission detected in the VLBA image is not sufficient to cause the entire \hi absorption. The largest extent of the VLBA source is 305 pc (see Table~\ref{vlba_results}). Assuming that a bulk of the absorption is expected at the $\approx 300$ pc scales, a significant fraction of the emission in the VLBA image, including the core, could be obscured by the gas. Since the $\rm S_{tot}$ is insufficient to cause the entire absorption, a fraction of the absorption could arise at scales larger than 305 pc. Note that a major fraction of total RACS emission is not detected in the VLBA image, i.e. $\rm (S_{RACS} - S_{tot}) / S_{RACS} = 0.85$.  
\newline
\newline
\item \textbf{1002-195:}
     The source in the L-band VLBA image has a connected structure, and is fitted with three Gaussian components (see Figure~\ref{fig:1002-195}). The central and most compact component, having the highest peak-to-integrated flux density ratio of 0.9, is identified as the core. All three components have similar integrated flux densities. We classify the source as a TSJ system.
\newline
\newline
     Here, the values PF (0.41), $\rm S_{core}/S_{RACS}$ (0.17) and $\rm S_{tot}/S_{RACS}$ (0.53), satisfy the relation $\rm S_{core}/S_{RACS} < PF < S_{tot}/S_{RACS}$.
     While the flux density of the core in insufficient to cause the entire absorption, the flux density of the total emission in the VLBA image is sufficient. The values imply that while a large fraction of the emission from the core and the total emission in the VLBA image (largest extent of 409 pc) is probably obscured, we cannot rule out some fraction of absorption arising against large-scale undetected emission.
\newline
\newline
     We use $\rm S_{min} = S_{tot}$, i.e. the flux density of the total emission in the VLBA image, to estimate a lower limit to the covering factor, $\rm S_{min} / S_{RACS} = 0.53$. 
\newline
\newline
\item \textbf{1136+004:}
    Three linearly aligned components are detected in the L-band image (see Figure~\ref{fig:1136+004}). The first and second components are connected, while the separation between the second and third components is $\approx 35$ pc. The structure resembles an asymmetric TSJ system. The central brightest component is identified as the core.
\newline
\newline
   For this target, the PF (0.18) and $\rm S_{core}/S_{RACS}$ (0.27) satisfy the relation $\rm PF < S_{core}/S_{RACS}$. The core has sufficient flux density to cause the entire \hi absorption. We therefore use $\rm S_{min}=S_{core}$, and estimate a lower limit to the lower factor, $\rm S_{min} / S_{RACS} = 0.27$.
\newline
\newline
\item \textbf{1701-294:}
    The beam is highly elongated along the North-East to South-West direction. The source appears to be resolved into two components and resembles a core-jet (CJ) morphology. Both components have similar peak-to-integrated flux density ratios, 0.6. We identify the dominant component as the core.    Due to the large beam size, we note that the uncertainty on our measurements of the flux densities and the projected sizes could be large.
\newline
\newline
    For this target, the values PF(0.14) and $\rm S_{core}/S_{RACS}$ (0.41) satisfy the relation $\rm PF < S_{core}/S_{RACS}$, implying the core has sufficient flux density to cause the entire absorption. We therefore use the core flux density as $\rm S_{min}$ to estimate the lower limit on the covering factor, $\rm S_{min}/S_{RACS} = 0.41$.
\newline
\newline
\item \textbf{2007-245:}
     The source in the L-band image has two dominant lobes whose peaks are separated by $\approx 40$ mas (see Figure~\ref{fig:2007-245}). The western lobe is resolved into two Gaussian components, while the eastern lobe is fitted with a single component. Both 4.4 GHz and 7.6 GHz images show a dominant lobe located in the centre of the image, along with multiple weaker lobes to the east, which clearly resemble a core-jet (CJ) system. Based on the relative locations of various components in the L-band map, we identify their counterparts in the 4.4 GHz and 7.6 GHz images. Using the 4.4 GHz map, we estimate the spectral indices ($\alpha$) of the three components identified in the 1.4 GHz image. From west to east, the spectral indices are -0.34, -1.7, and -0.95. The westernmost component in the 1.4 GHz map has a flat spectral shape, while the other two components have relatively steeper shapes. We hence identify the westernmost component as the core. 
\newline
\newline
     Here, the absorption is fitted with two Gaussians, with PF values PF 0.05 and 0.01 (see Table~\ref{askap_results}). The value of $\rm S_{core}/S_{RACS}$ is 0.09, which satisfies $\rm PF < S_{core}/S_{RACS}$ for both the absorption components. Therefore the core has sufficient flux density to cause all of the absorption.  
\newline
\newline
     However, the core fraction, $\rm S_{core}/S_{RACS}$ is quite low (0.09), and the fraction of the continuum emission other than the core, i.e. $\rm (S_{RACS} -S_{core})/S_{RACS}$ is 0.91. Note that the fraction of the total emission in the VLBA image is $\rm S_{tot} / S_{RACS} = 0.54$.
     If the absorber is associated with the source, then the chances of the gas obscuring the core would be high (see Section~\ref{sec:intro}). Whereas if the absorber is an intervening system, there is a high chance that it could arise against emission in the VLBA image other than the core. So we treat this source as an exception, and use $\rm S_{min} = S_{tot}$, i.e. the total VLBA flux density, instead of the core flux density, to estimate a lower limit to the covering factor, $\rm S_{min}/S_{RACS} = 0.54$.
\newline
\newline
\item \textbf{2233-015:}
    The structure resembles a TSJ morphology, where the component with highest peak flux density is located in the middle of the structure. Three continuum Gaussian components of similar sizes are fitted to the emission, and the brightest component in the centre is identified as the core. 
    \newline
    \newline
    For this source, the values PF (0.07) and $\rm S_{core}/S_{RACS}$ (0.36) satisfy the relation $\rm PF < S_{core}/S_{RACS}$. The values imply that while the core has sufficient flux density to cause the entire \hi absorption.
    We therefore use $\rm S_{min} = S_{core}$ to estimate the lower limit on the covering factor, $\rm S_{min}/S_{RACS} = 0.36$.  
 \newline
 \newline
\item \textbf{2236-251:}
     Again, the morphology resembles that of a TSJ system. Four continuum Gaussian components with very similar sizes are fitted to the emission structure, and the brightest among them, second from the right, is identified as the core.
     \newline
     \newline
     The values PF (0.32) and $\rm S_{core}/S_{RACS}$ (0.33) satisfy the relation $\rm PF < S_{core}/S_{RACS}$. We therefore use $\rm S_{min} = S_{core}$ to estimate the lower limit on the covering factor, $\rm S_{min} / S_{RACS} = 0.33$. However, the core flux density is barely sufficient to cause the entire \hi absorption, implying that a fraction of the absorption could be arising at scales larger than that of the core.
\newline
\newline
     
\end{description}

\subsection{Optical depth limits and source extension}


In the ASKAP measurements, it is assumed that the \hi gas obscures the entire emission detected in the beam, i.e. $f = 1$. Since $f = 1$ is the upper limit to the covering factor, the estimated  VOD ($\int \tau_{\rm ASKAP}\,dv$) is a lower limit on the VOD. These values for each target are listed in Table~\ref{askap_results}. The upper limit on the velocity-integrated optical depth ($\int \tau_{\rm VLBI}\,dv$) is estimated by applying the lower limit on the covering factor ($ f = \rm S_{min}/S_{RACS}$) to $\int \tau_{\rm ASKAP} dv$, using equation~\ref{eqn1}. Here, the covering factor is applied to the optical depth in each channel of the \hi 21-cm absorption profile and then the values are added.

We also estimated the largest projected spatial extent of the total source and the core, at the \hi 21-cm absorption redshift. For the core components, we used the major axis of the Gaussian fit to estimate the spatial extent. The spatial extents of the complete source lie in the range of 305 pc to 2713 pc, with a median size of 508 pc. The spatial extents of the core lie in the range 143 pc to 1472 pc, with a median value of 188 pc. 

For all targets, we provide the following parameters in Table~\ref{vlba_results}: the source name, size of the synthesized beam, the r.m.s. noise in the continuum image, peak flux density of the Gaussian emission component in the VLBA image ($\rm S_{peak}$), integrated flux density of the component ($\rm S_{int}$), total flux density of all components ($\rm S_{tot}$), estimate of minimum flux density arising from occulted emission ($\rm S_{min}$), the RACS flux density of the source ($\rm S_{RACS}$), the lower limit on the covering factor ($\rm S_{min}/S_{RACS}$), largest angular size of the source in milliarcseconds (mas), largest spatial extent of the source in parsecs, at the redshift of the \hi 21-cm line, largest angular extent of the core in mas, largest spatial extent of the core, at the \hi 21-cm redshift,  and the upper limit to the velocity-integrated optical depth ($\int \tau_{\rm VLBA} dv$).

\begin{table*} 
\centering
\footnotesize
\tabcolsep 5pt
\caption{VLBA observations. The columns are: 1) the source name that is used in this paper, 2) the NED name, 3) \& 4)  the ASKAP right ascension and declinations of the source, 5) the date of observation, 6) the phase-reference calibrator, and 7) the on-source time.
\newline $^{a}$ Here, the target is bright enough for self-calibration. Phase-referencing was not used.   } \label{vlba_obs}
\begin{tabular}{c l c c l c c}
\hline	
\hline
\multicolumn{1}{c}{Source name}	&	\multicolumn{1}{c}{NED name}	& \multicolumn{1}{c}{RA}	& \multicolumn{1}{c}{DEC} &	\multicolumn{1}{c}{Obs. date}	&	\multicolumn{1}{c}{Phase-ref}  & \multicolumn{1}{c}{On-source time}	\\

\multicolumn{1}{c}{}	&	\multicolumn{1}{c}{}		&	\multicolumn{1}{c}{ [h:m:s]}	&	\multicolumn{1}{c}{[d:m:s]}	&	\multicolumn{1}{c}{}	&	\multicolumn{1}{c}{}	& 	\multicolumn{1}{c}{[mins]} \\

\multicolumn{1}{c}{(1)}	&	\multicolumn{1}{c}{(2)}		&	\multicolumn{1}{c}{(3)}	&	\multicolumn{1}{c}{(4)}	&	\multicolumn{1}{c}{(5)}	&	\multicolumn{1}{c}{(6)}	&	\multicolumn{1}{c}{(7)}	  \\
\hline
\hline

        0011-023 & PKS 0011-023 & 00h14m25.5710s & -02d05m55.620s & 2-Mar-24 & 0016-001 & 36 \\ 
        0023+010 & NVSS J002331+010114 & 00h23m31.5900s & +01d01m15.200s & 2-Mar-24 & 0022+001 & 42 \\ 
        0141-231  & NVSS J014141-231511 & 01h41m41.7300s & -23d15m11.100s & 2-Mar-24 & 0138-225 & 42 \\ 
        0518-245 & NVSS J051806-245502 & 05h18m06.0500s & -24d55m01.900s & 2-Mar-24 & --$^{a}$ & 55 \\ 
        0903+010 & NVSS J090331+010846 & 09h03m31.6000s & +01d08m47.500s & 29-Feb-24 & 0909+012 & 35 \\ 
        0920+161  & NVSS J092012+161238 & 09h20m12.5200s & +16d12m38.200s & 29-Feb-24 & 0908+160 & 42 \\ 
        1002-195  & NVSS J100238-195917 & 10h02m38.5700s & -19d59m17.100s & 29-Feb-24 & 1006-215 & 42 \\ 
        1136+004 & NVSS J113622+004850 & 11h36m22.0400s & +00d48m52.600s & 29-Feb-24 & 1133+004 & 42 \\ 
        1701-294  & NVSS J170135-294918 & 17h01m35.6400s & -29d49m18.700s & 29-Feb-24 & 1650-294 & 42 \\ 
        2007-245 & PKS 2007-245 & 20h10m45.1500s & -24d25m45.600s & 1-Apr-24 & 2005-231 & 36 \\ 
        2233-015 & NVSS J223317-015739 & 22h33m17.2800s & -01d57m39.600s & 1-Apr-24 & 2226+005 & 42 \\ 
        2236-351 & NVSS J223605-251919  & 22h36m05.8700s & -25d19m19.800s & 1-Apr-24 & 2237-250 & 42 \\ 

 \hline
\hline
\end{tabular}
\end{table*}

\begin{table*} 
\centering
\small
\tabcolsep 5pt
\caption{\hi 21-cm line properties estimated from ASKAP observations, reported in \citet[][]{yoon2024}. The columns are: 1) the source name, 2) redshift of the \hi 21-cm line, 3) the flux density measured in ASKAP observations, at $\approx 700$ MHz, 4) the ratio of the peak absorption flux density and continuum flux density in absolute values,  5) the velocity-integrated optical depth; a lower limit, measured in the ASKAP observations, 6) the line width (full width at 20\% intensity)} \label{askap_results}
\begin{tabular}{c c c c c c }
\hline	
\hline
\multicolumn{1}{c}{Source name}	&	\multicolumn{1}{c}{redshift (\hi)} & $\rm S_{ASKAP}$ & $\rm PF$	& \multicolumn{1}{c}{$\int \tau_{\rm ASKAP} \ dv$}	& \multicolumn{1}{c}{FW20}  \\

\multicolumn{1}{c}{}	&	\multicolumn{1}{c}{}		& [mJy] & &	\multicolumn{1}{c}{ [$\rm km \ s^{-1}$] }	&	\multicolumn{1}{c}{[$\rm km \ s^{-1}$] }		\\

\multicolumn{1}{c}{(1)}	&	\multicolumn{1}{c}{(2)}		&
\multicolumn{1}{c}{(3)}	&	\multicolumn{1}{c}{(4)}	&	\multicolumn{1}{c}{(5)}	&	\multicolumn{1}{c}{(6)}	 \\
\hline
\hline

        0011-023 & 0.6785 & 694 & 0.21  & 7.9 & 52.7  \\ 
        0023+010 & 0.6745 & 67  & 0.35 & 32.9 & 116.7  \\ 
        0141-231 & 0.6707 & 137 & 0.09 & 14.3 & 204.9   \\ 
        0518-245 & 0.5538 & 186 & 0.16 & 19.7 & 165.4  \\ 
        0903+010 & 0.5218 & 59 & 0.82 & 103.9 & 112.6   \\ 
        0920+161 & 0.4362 & 171 & 0.23 & 7.6 & 42.4  \\ 
        1002-195 & 0.4815 & 57 & 0.41 & 17.1 & 50.0  \\ 
        1136+004 & 0.5632 & 154 & 0.18 & 4.7 & 35.9   \\ 
        1701-294 & 0.6299 & 403 & 0.14 & 4.4 & 42.3  \\ 
        2007-245 & 0.6778 & 1827 & 0.05 & 0.8 & 80.6   \\ 
              &        &      &   0.01   & 0.6 & 15.7 \\ 
        2233-015 & 0.6734 & 235 & 0.07 & 6.8 & 129.8   \\ 
        2236-251 & 0.4974 & 215 & 0.32 & 10.3 & 40.4   \\ 

 \hline
\hline
\end{tabular}
\end{table*}

\begin{landscape}
\begin{table} 

\centering
\footnotesize
\tabcolsep 5pt
\caption{VLBA continuum and \hi 21-cm absorption properties. The columns are: 1) the source name, 2) the beam size in the VLBA continuum image, 3) the r.m.s. noise on the continuum image, 4) the peak flux density of the 2-D Gaussian fit to a continuum component in the image. Note that the first component of a source corresponds to the core. For this and the next column, the listed uncertainty corresponds to the 2-D Gaussian fit in the VLBA images. However, the uncertainty due to calibration errors would be $\approx 10 \%$. 5) the integrated flux density of the continuum component, 6) the total flux density, 7) the minimum flux density arising from emission occulted by \hi gas, 8) the RACS integrated flux density, 9) the lower limit on the covering factor, 10) the largest angular size measured in the continuum image, in milliarcseconds, 11) the spatial extent of the source at the redshift of the \hi 21-cm line (Here, we used the redshift of the \hi 21-cm line peak to estimate the sizes), 12) the largest angular size of the Gaussian fit to the core, 13) the largest spatial extent of the core at the \hi 21-cm redshift, 14) the upper limit to the velocity-integrated \hi 21-cm optical depth ($\int \tau_{\rm VLBA} dv$), estimated by applying the covering factor to the optical depth measured in ASKAP observations (see Table~\ref{askap_results}).    } \label{vlba_results}
\begin{tabular}{l l l l l l l l l l l l  l l}
\hline	
\hline
\multicolumn{1}{c}{Source name}	&	\multicolumn{1}{c}{beam size}	& \multicolumn{1}{c}{r.m.s. cont}	& \multicolumn{1}{c}{S$_{\rm peak}$} &	\multicolumn{1}{c}{S$_{\rm int}$}	&	
\multicolumn{1}{c}{S$_{\rm tot}$}	&
\multicolumn{1}{c}{S$_{\rm min}$}  & 
\multicolumn{1}{c}{S$_{\rm RACS}$}  &
\multicolumn{1}{c}{$\rm S_{min} / S_{RACS}$}	& \multicolumn{1}{c}{source size}	& 
\multicolumn{1}{c}{source size } & 
\multicolumn{1}{c}{core  size}	& 
\multicolumn{1}{c}{core size } & 
\multicolumn{1}{c}{$\int \tau _{\rm VLBA} dv $ } \\

\multicolumn{1}{c}{}	&	\multicolumn{1}{c}{[mas X mas]}		&	\multicolumn{1}{c}{ [mJy $\rm bm^{-1}$]}	&	\multicolumn{1}{c}{[mJy $\rm bm^{-1}$]}	&	\multicolumn{1}{c}{[mJy]} &
\multicolumn{1}{c}{[mJy]}	&
\multicolumn{1}{c}{[mJy]}	& 
\multicolumn{1}{c}{[mJy]}	&
\multicolumn{1}{c}{[low. limit on $\rm f$]} &	
\multicolumn{1}{c}{[mas]} &	
\multicolumn{1}{c}{[pc]} &
\multicolumn{1}{c}{[mas]} &	
\multicolumn{1}{c}{[pc]} &
\multicolumn{1}{c}{[$\rm km \ s^{-1}$]}  
\\

\multicolumn{1}{c}{(1)}	&	\multicolumn{1}{c}{(2)}		&	\multicolumn{1}{c}{(3)}	&	\multicolumn{1}{c}{(4)}	&	\multicolumn{1}{c}{(5)}	&	\multicolumn{1}{c}{(6)}	&	\multicolumn{1}{c}{(7)}	&	\multicolumn{1}{c}{(8)}	&	\multicolumn{1}{c}{(9)}	&	\multicolumn{1}{c}{(10)}  &
\multicolumn{1}{c}{(11)} &
\multicolumn{1}{c}{(12)} &
\multicolumn{1}{c}{(13)} &
\multicolumn{1}{c}{(14)}
\\
\hline
\hline
 
        \hline
        0011-023 & 134.1 X 16.1 & 0.2 & $26.6 \pm 0.3$ & $48.1\pm 0.8$ & $162.2\pm1.8$ & $162.2\pm1.8$ & $431.3\pm25.9$ & $0.38\pm0.06$ & $346\pm5$ & $2471\pm36$ & $156\pm2$ & $1113\pm14$& 24.7 \\ 
        ~ & ~ & ~ & $24.6 \pm 0.3$ & $102.9\pm1.5$ & ~ & ~ & ~ & ~ & ~ \\ 
        ~ & ~ & ~ & $8.8\pm 0.3$ & $11.2\pm0.6$ & ~ & ~ & ~ & ~ & ~ \\ \hline
        0023+010 & 27.1 X 14.8 & 0.3 & $13.5\pm0.3$ & $18.3\pm0.6$ & $51.4\pm1.4$ &
        $18.3\pm0.6$ & $ 49.1 \pm 3.0$ & $0.36\pm0.05$ & $172\pm7$ & $1225\pm50$ & $31\pm1$ & $225\pm7$ & 169.0 \\ 
        ~ & ~ & ~ & $7.4\pm0.3$ & $11.4\pm0.7$ & ~ & ~ & ~ & ~ & ~ \\ 
        ~ & ~ & ~ & $7.9\pm0.3$ & $7.9\pm0.5$ & ~ & ~ & ~ & ~ & ~ \\ 
        ~ & ~ & ~ & $2.8\pm0.3$ & $2.8\pm0.5$ & ~ & ~ & ~ & ~ & ~ \\ 
        ~ & ~ & ~ & $6.7\pm0.3$ & $6.7\pm0.5$ & ~ & ~ & ~ & ~ & ~ \\ 
        ~ & ~ & ~ & $2.6\pm0.3$ & $2.6\pm0.5$ & ~ & ~ & ~ & ~ & ~ \\ 
        ~ & ~ & ~ & $1.7\pm0.3$ & $1.7\pm0.5$ & ~ & ~ & ~ & ~ & ~ \\ \hline
        0141-231 & 58.8 X 16.1 & 0.3 & $25.5\pm0.3$ & $56.3\pm0.9$ & $91.6\pm1.3$& $56.3\pm0.9$ & $99.6 \pm 6.0$ & $0.57\pm0.03$ & $218\pm3$ & $1549\pm21$ & $73\pm1$ & $519\pm7$ & 25.9 \\ 
        ~ & ~ & ~ & $12.4\pm0.3$ & $24.6\pm0.8$ & ~ & ~ & ~ & ~ & ~ \\ 
        ~ & ~ & ~ & $8.5\pm0.3$ & $10.7\pm0.6$ & ~ & ~ & ~ & ~ & ~ \\ \hline
        0518-245 & 52.2 X 9.7 & 1.1 & $131.9\pm0.8$ & $186.9\pm1.7$ & $186.9 \pm 1.7$ &$186.9\pm1.7$ & $293.2 \pm 17.6$   & $0.64 \pm0.03$ & $152\pm1$ & $989\pm7$ & $60\pm1$ & $390\pm7$ & 32.0 \\ \hline
        0903+010 & 23.5 X 4.4 & 0.8 & $26.5\pm0.2$ & $33.1\pm0.4$ &$48.3\pm0.7$ &$>48.3$ & $61.9\pm3.7$ & $>0.82$  & $55\pm1$ & $347\pm6$ & $25\pm1$ & $158\pm6$ & -- \\ 
        ~ & ~ & ~ & $8.4\pm0.2$ & $15.2\pm0.6$ & ~ &     &  & ~ & ~ \\ \hline
        0920+161 & 23.5 X 4.4 & 0.5 & $5.8\pm0.2$ & $5.8\pm0.2$ & $17.1\pm0.7$ &$>17.1$ & $117.2\pm7.0$ & $>0.23$ & $53\pm2$ & $305\pm11$ & $25\pm1$& $143\pm6$ & -- \\ 
        ~ & ~ & ~ & $4.2\pm0.2$ & $11.3\pm0.7$ & ~ & ~ &  & ~ & ~ \\ \hline
        1002-195 & 25.9 X 4.9 & 0.3 & $4.6\pm0.2$ & $5.2\pm0.5$ &$16.1\pm1.2$ &$16.1\pm1.2$ & $30.4\pm1.8$ & $0.53\pm0.08$  & $68\pm5$ & $409\pm30$ & $27\pm1$ & $165\pm6$ & 41.4 \\ 
        ~ & ~ & ~ & $2.7\pm0.2$ & $5.4\pm0.7$ & ~ &  &  & ~ & ~ \\ 
        ~ & ~ & ~ & $2.3\pm0.2$ & $5.5\pm0.8$ & ~ & ~ & ~ & ~ & ~ \\ \hline
        1136+004 & 22.9 X 4.3 & 0.7 & $24.0\pm0.2$ & $45.2\pm0.5$ &$92.7\pm0.8$ &$45.2\pm0.5$ & $164.5\pm10.2$  & $0.27\pm0.03$ & $71\pm1$ & $464\pm7$ & $32\pm1$& $210\pm7$ & 22.9 \\ 
        ~ & ~ & ~ & $16.0\pm0.2$ & $24.1\pm0.4$ & ~ & ~ & ~ & ~ & ~ \\ 
        ~ & ~ & ~ & $18.7\pm0.2$ & $23.4\pm0.4$ & ~ & ~ & ~ & ~ & ~ \\ \hline
        1701-294 & 202.4 X 6.6 & 1.7 & $43.7\pm0.4$ & $75.9\pm1.0$ &$111.6\pm1.3$ &$75.9\pm1.0$ & $185.4\pm11.1$ & $0.41\pm0.03$ & $392\pm6$ & $2713\pm41$ & $213\pm1$ & $1472\pm7$ & 11.7 \\ 
        ~ & ~ & ~ & $23.7\pm0.4$ & $35.7\pm0.9$ & ~ & ~ & ~ & ~ & ~ \\ \hline
        2007-245 & 18.5 X 5.2 & 1.1 & $69.9\pm1.8$ & $111.4\pm4.4$ &$609.3\pm9.2$ &$609.3\pm9.2$ &  $1137.4\pm68.3$  & $0.54\pm0.05$ & $77\pm1$ & $551\pm7$ & $23\pm1$ & $161\pm7$ & 2.7 \\ 
        ~ & ~ & ~ & $120.7\pm1.8$ & $172.7\pm4.1$ & ~ & ~ & ~ & ~ &  \\ 
        ~ & ~ & ~ & $105.5\pm1.8$ & $325.2\pm6.9$ & ~ & ~ & ~ & ~ & ~ \\ \hline

 \hline
\hline
\end{tabular}
\end{table}
\end{landscape}

\begin{landscape}
\begin{table} 

\centering
\footnotesize
\tabcolsep 5pt

 \setcounter{table}{3}
\ContinuedFloat

\caption{ continued} 
\begin{tabular}{l l l l  l l l l l l l  l l l}
\hline	
\hline
\multicolumn{1}{c}{Source name}	&	\multicolumn{1}{c}{beam size}	& \multicolumn{1}{c}{r.m.s. cont}	& \multicolumn{1}{c}{S$_{\rm peak}$} &	\multicolumn{1}{c}{S$_{\rm int}$}	&	
\multicolumn{1}{c}{S$_{\rm tot}$}	&
\multicolumn{1}{c}{S$_{\rm min}$}  & 
\multicolumn{1}{c}{S$_{\rm RACS}$}  &
\multicolumn{1}{c}{$\rm S_{min} / S_{RACS}$}	& \multicolumn{1}{c}{source size}	& 
\multicolumn{1}{c}{source size } & 
\multicolumn{1}{c}{core size}	& 
\multicolumn{1}{c}{core size } &
\multicolumn{1}{c}{$\int \tau _{\rm VLBA} dv $ } \\

\multicolumn{1}{c}{}	&	\multicolumn{1}{c}{[mas X mas]}		&	\multicolumn{1}{c}{ [mJy $\rm bm^{-1}$]}	&	\multicolumn{1}{c}{[mJy $\rm bm^{-1}$]}	&	\multicolumn{1}{c}{[mJy ]} &
\multicolumn{1}{c}{[mJy]}	&
\multicolumn{1}{c}{[mJy]}	& 
\multicolumn{1}{c}{[mJy]}	&
\multicolumn{1}{c}{[low. limit on $\rm f$]} &	
\multicolumn{1}{c}{[mas]} &	
\multicolumn{1}{c}{[pc]} &
\multicolumn{1}{c}{[mas]} &	
\multicolumn{1}{c}{[pc]} &
\multicolumn{1}{c}{[$\rm km \ s^{-1}$]}  
\\

\multicolumn{1}{c}{(1)}	&	\multicolumn{1}{c}{(2)}		&	\multicolumn{1}{c}{(3)}	&	\multicolumn{1}{c}{(4)}	&	\multicolumn{1}{c}{(5)}	&	\multicolumn{1}{c}{(6)}	&	\multicolumn{1}{c}{(7)}	&	\multicolumn{1}{c}{(8)}	&	\multicolumn{1}{c}{(9)}	&	\multicolumn{1}{c}{(10)}  &
\multicolumn{1}{c}{(11)} &
\multicolumn{1}{c}{(12)} &
\multicolumn{1}{c}{(13)} &
\multicolumn{1}{c}{(14)}
\\
\hline
\hline

       2233-015 & 15.8 X 4.5 & 0.3 & $16.1\pm0.2$ & $54.9\pm0.9$ &$80.7\pm1.3$ &$54.9\pm0.9$ &  $152.7\pm9.2$ & $0.36\pm0.03$ & $54\pm1$ & $387\pm7$ & $21\pm1$ & $146\pm7$ &  20.1 \\ 
        ~ & ~ & ~ & $7.2\pm0.2$ & $11.3\pm0.5$ & ~ & ~ & ~ & ~ & ~ \\ 
        ~ & ~ & ~ & $5.3\pm0.2$ & $14.5\pm0.8$ & ~ & ~ & ~ & ~ & ~ \\ \hline
        2236-251 & 21.4 X 4.1 & 0.5 & $33.2\pm0.2$ & $58.6\pm0.4$ &$158.2\pm0.9$ &$58.6\pm0.4$ &  $180.9\pm10.9$ & $0.33\pm0.03$ & $56\pm1$ & $344\pm6$ & $25\pm1$ & $152\pm6$ &  58.4 \\ 
        ~ & ~ & ~ & $25.2\pm0.2$ & $30.6\pm0.4$ & ~ & ~ & ~ & ~ & ~ \\ 
        ~ & ~ & ~ & $11.2\pm0.2$ & $17.6\pm0.4$ & ~ & ~ & ~ & ~ & ~ \\ 
        ~ & ~ & ~ & $26.8\pm0.2$ & $51.4\pm0.5$ & ~ & ~ & ~ & ~ & ~ \\ \hline

 \hline
\hline
\end{tabular}
\end{table}
\end{landscape}

\begin{figure*}[!h]
\centering

\includegraphics[width=0.4\textwidth]{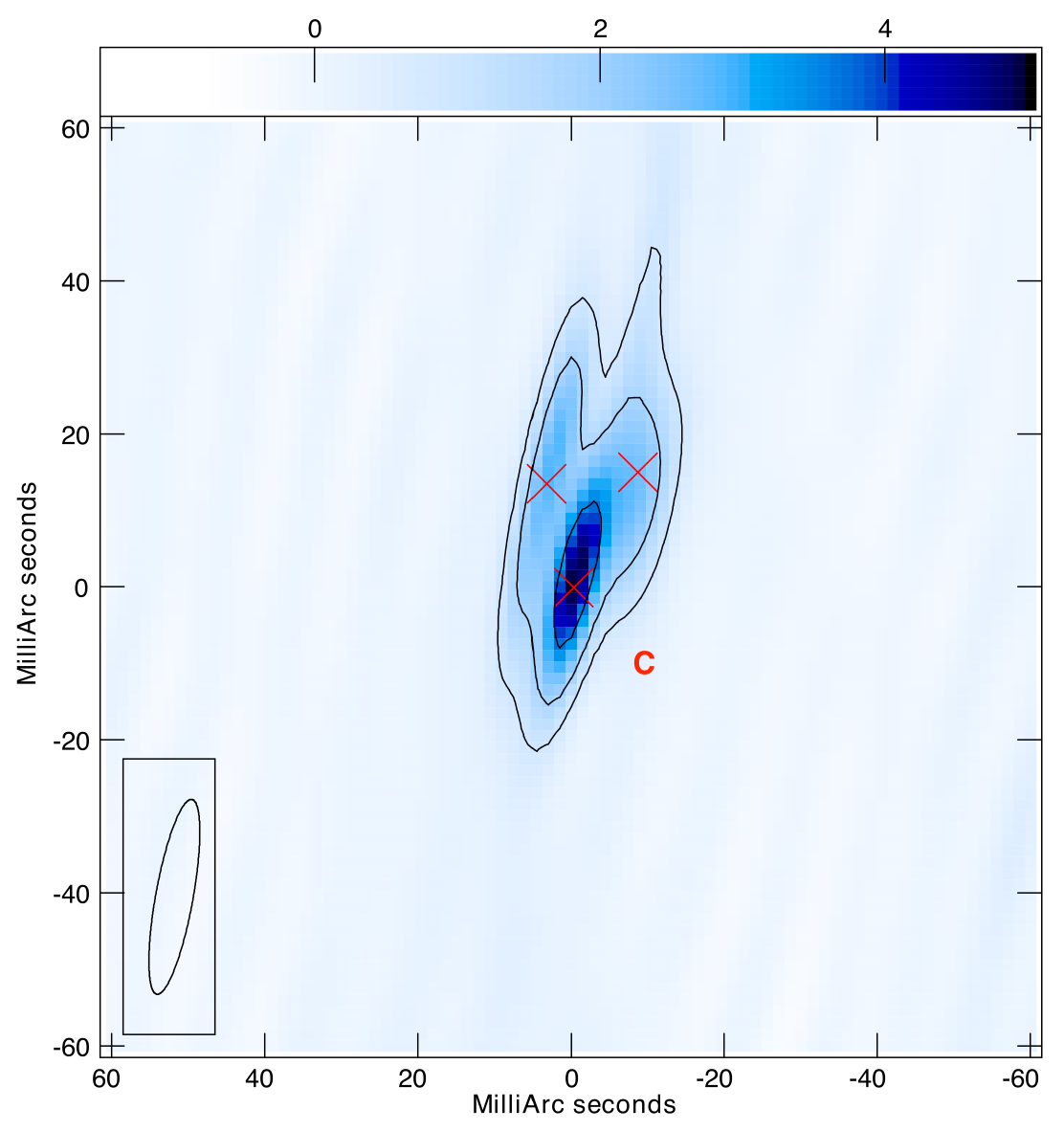}
    \includegraphics[width=0.4\textwidth]{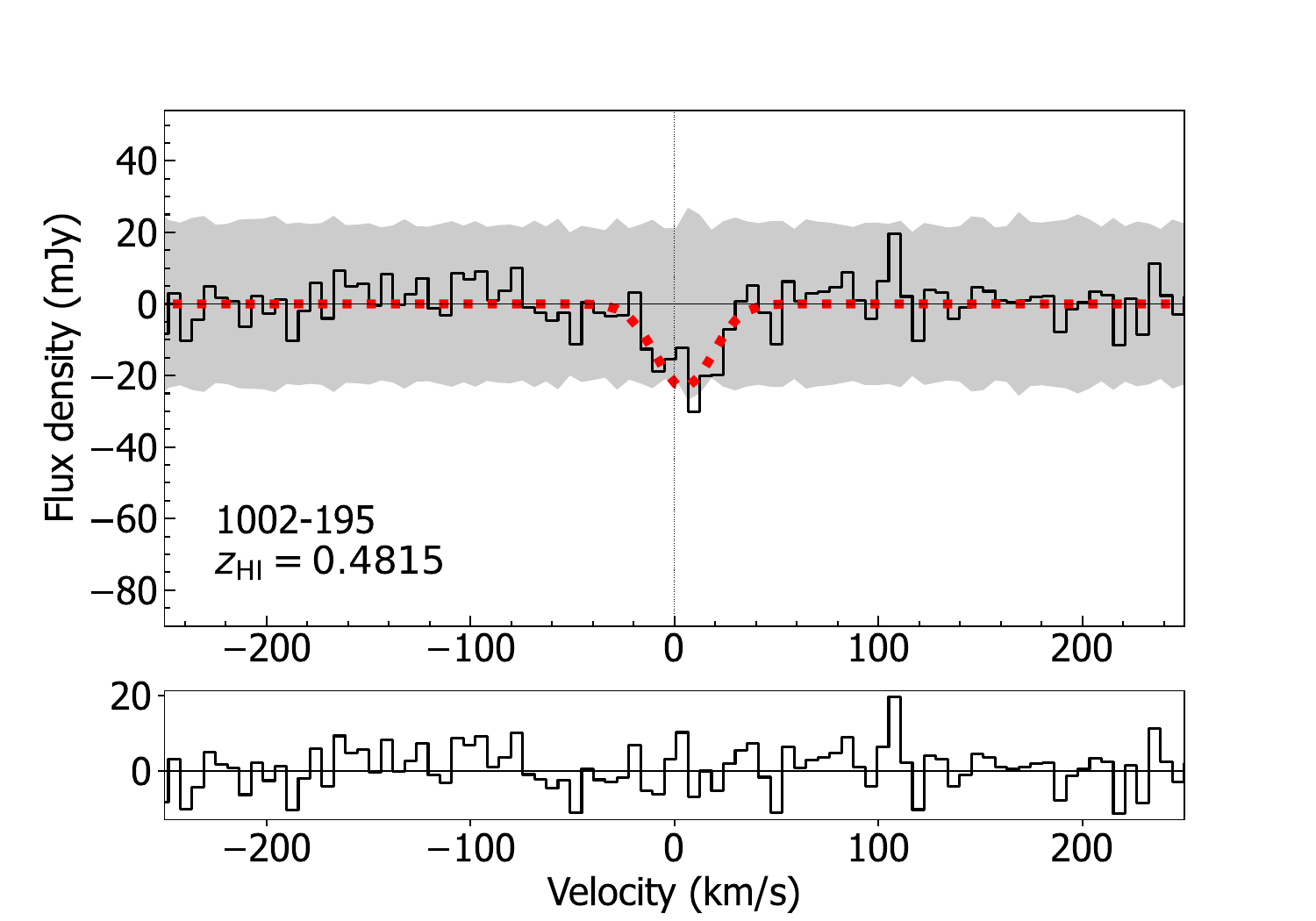}

\caption{1002-195: Details of the images are same as that of Figure~\ref{fig:0011-023}. Contours in the left panel are at the levels $\rm 0.9 \ mJy \times (1, 2, 4, 8..)$. The peak flux density in the image is 4.6 mJy/bm. }\label{fig:1002-195}
\end{figure*}

\begin{figure*}[!h]
\centering    
    \includegraphics[width=0.4\textwidth]{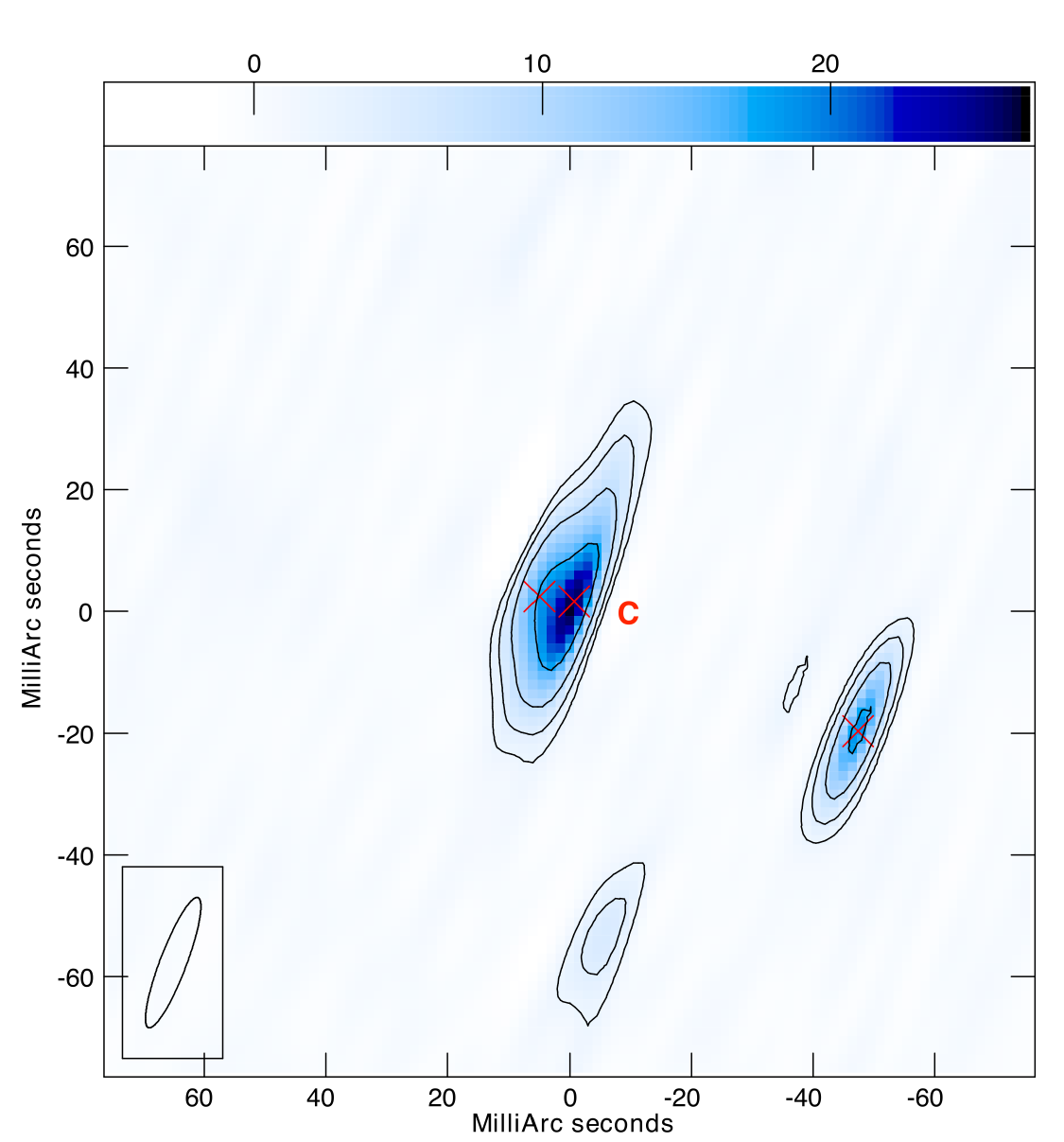}
    \includegraphics[width=0.4\textwidth]{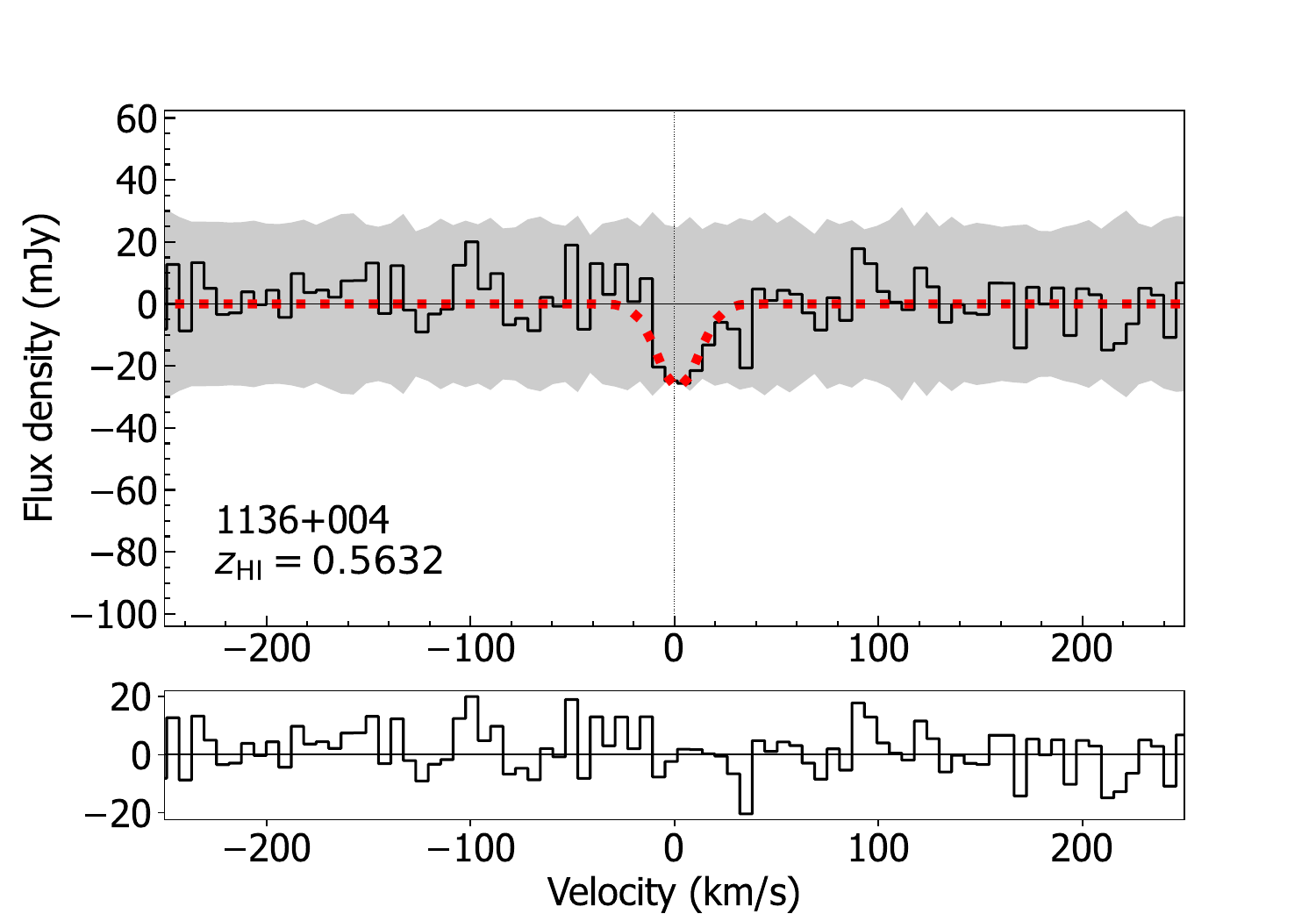} 

\caption{1136+004: Details of the images are same as that of Figure~\ref{fig:0011-023}. Contours in the left panel are at the levels $\rm 2.1 \ mJy \times (1, 2, 4, 8..)$. The peak flux density in the image is 24.0 mJy/bm.}\label{fig:1136+004}
\end{figure*}

\begin{figure*}[!h]

\centering

\includegraphics[width=0.4\textwidth]{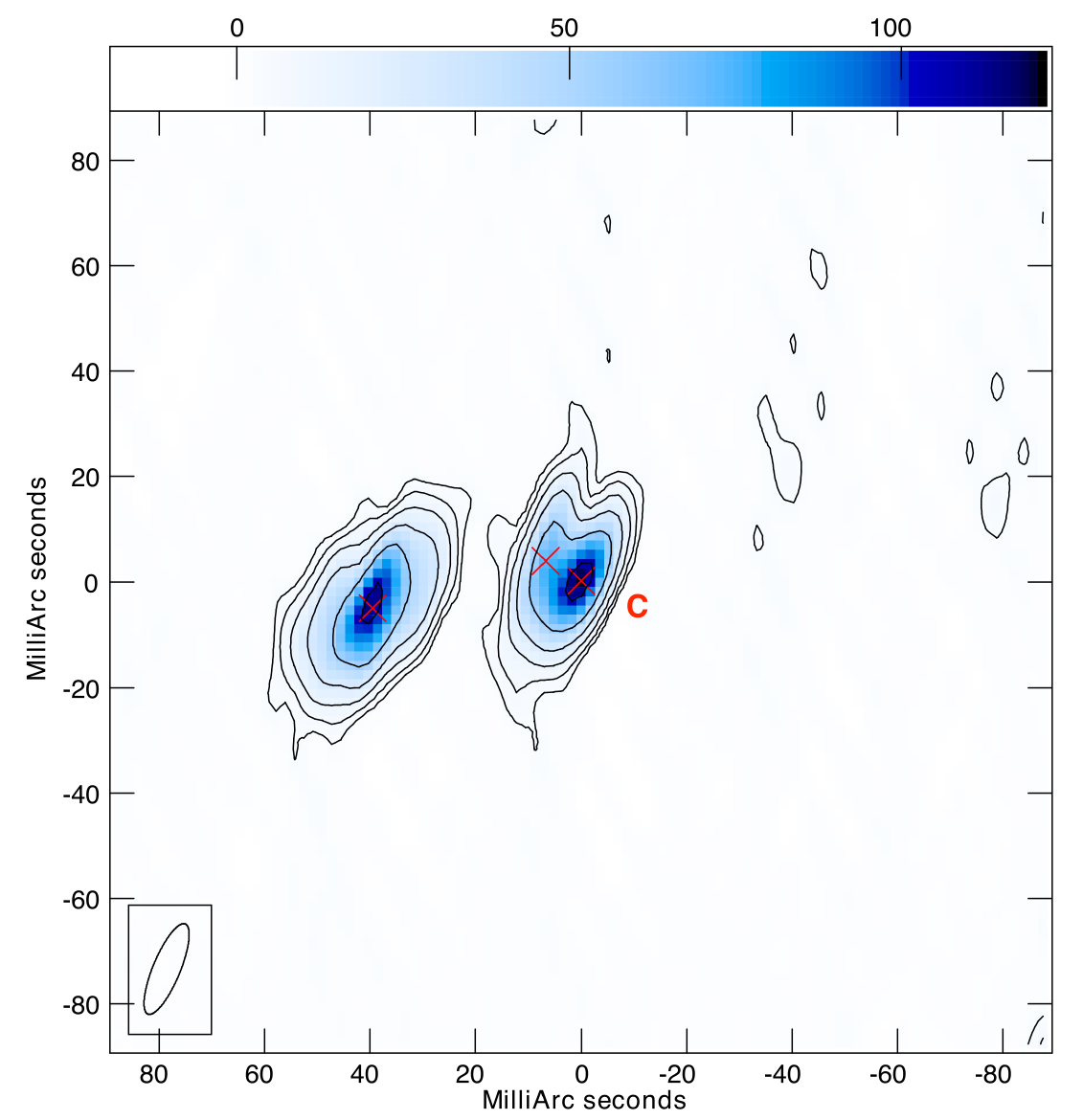}
    \includegraphics[width=0.4\textwidth]{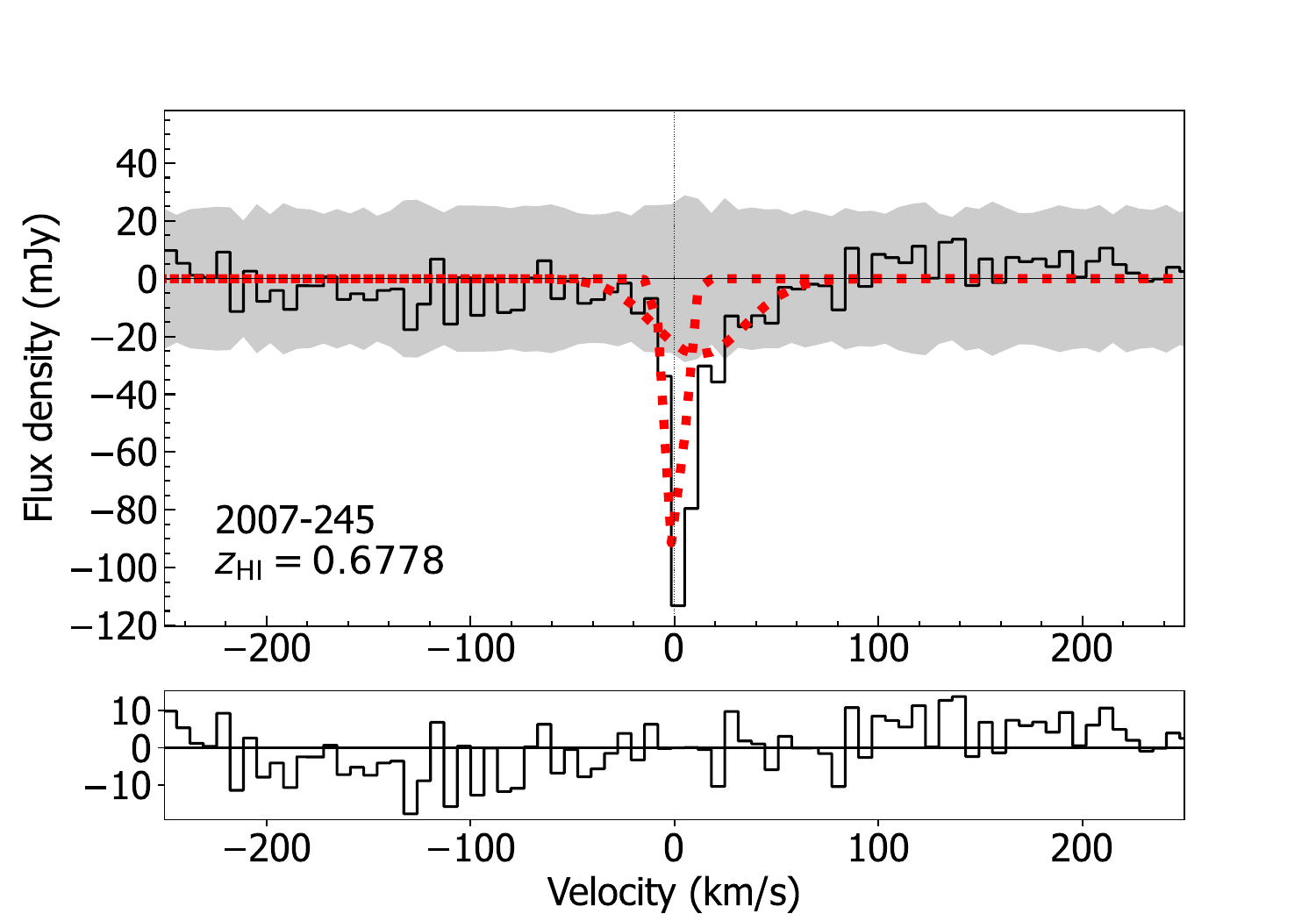}
\includegraphics[width=0.4\textwidth]{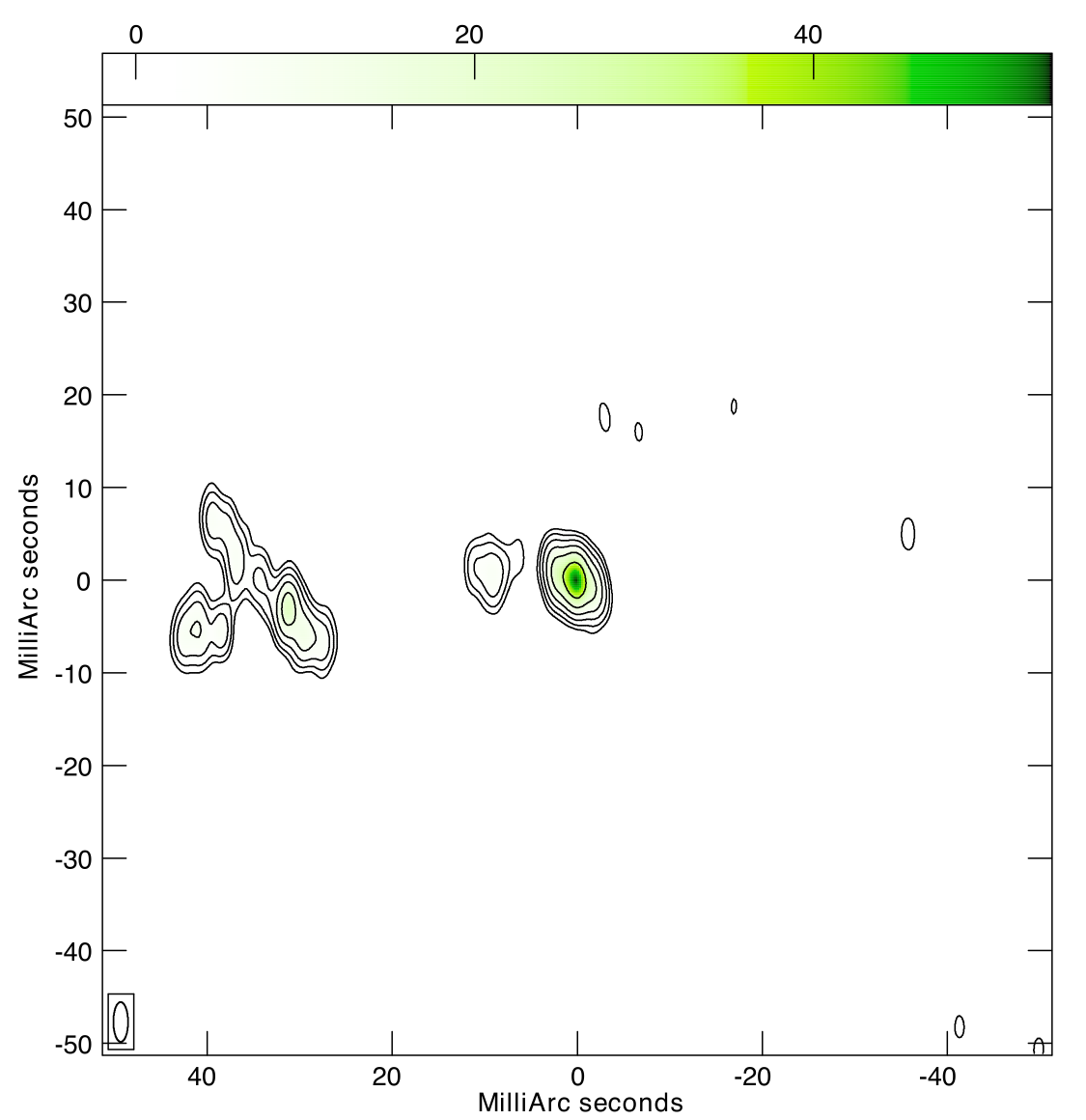}
    \includegraphics[width=0.4\textwidth]{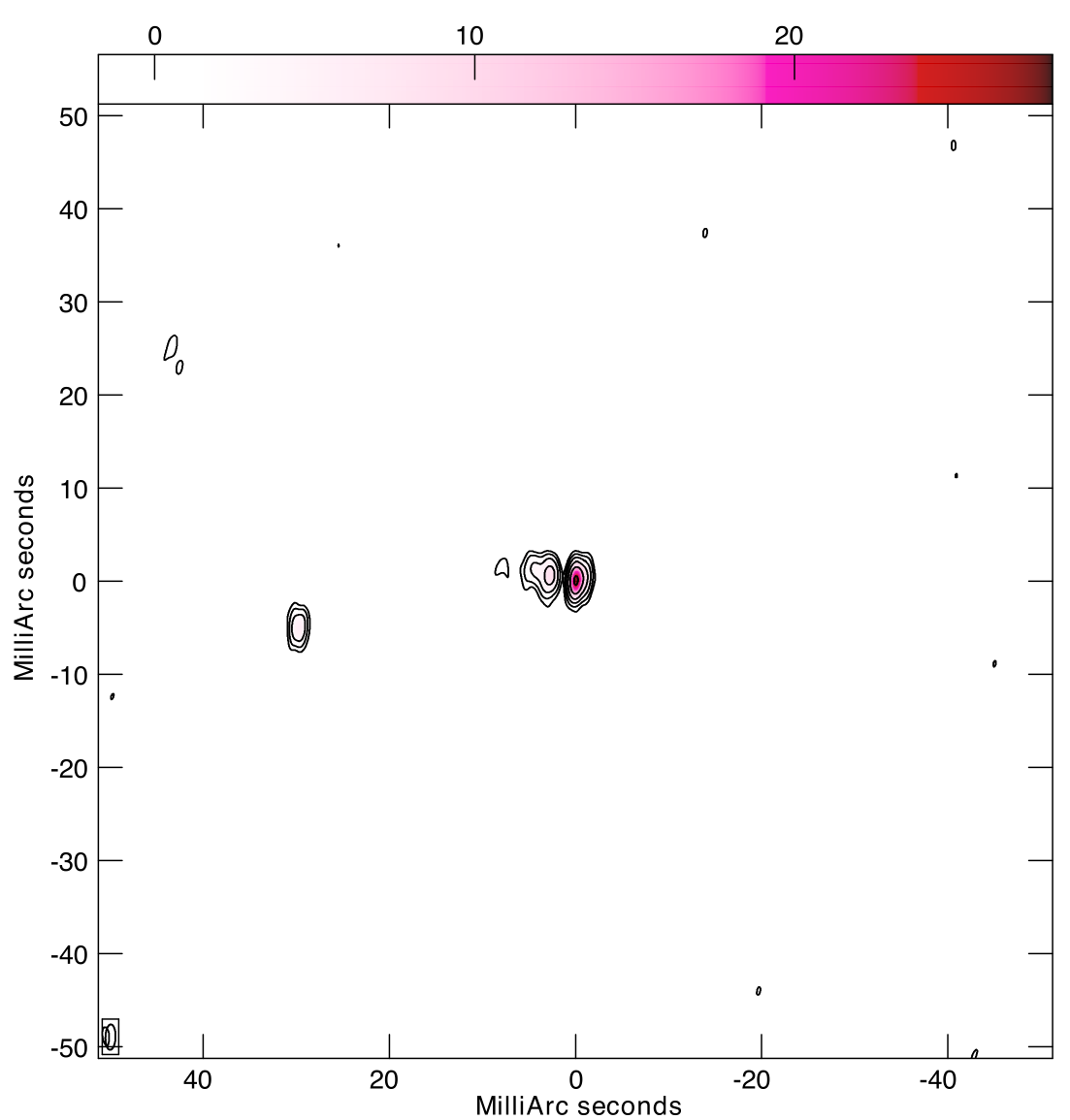}
\caption{2007-245: Details of the top left and top right images are same as that of Figure~\ref{fig:0011-023}. Contours in the top-left panel are at the levels  $\rm 3.3 \ mJy \times (1, 2, 4, 8..)$. The peak flux density in the image is 69.9 mJy/bm. The left and right panels in the bottom are the 4.4 GHz and 7.6 GHz continuum images. The data for these images were taken from astrogeo. The contours in the 4.4 GHz image are at $\rm 1.0 \ mJy \times (1, 2, 4, 8..) $, and those in the 7.6 GHz image are at $\rm 0.8 \ mJy \times (1, 2, 4, 8..) $. The peak flux densities in the two images are 41 mJy/bm and 31 mJy/bm, respectively. }\label{fig:2007-245}
\end{figure*}

\section {Discussion}

\subsection{Source characteristics}
Among the twelve sources, half have core jet (CJ) morphology, four have TSJ morphology, one has a complex structure, and one is unresolved. Even the single source that is unresolved in the L band shows a resolved core-double jet structure in the 8.7 GHz VLBI image (see Figure~\ref{fig:0518-245}).

In seven of the twelve targets (excluding 0023+010 where the identification of the core is uncertain), namely 0141-231, 0518-245, 1136+004, 1701-294, 2007-245, 2233-015, and 2236-251, the core fraction is higher than the fractional peak absorption of the \hi 21-cm line. This means that the core has sufficient flux density to cause the detected absorption, and thus, a bulk of the absorption could arise against the core. However, this does not rule out the gas covering an emission larger than that of the core. 
Our results are consistent with a hypothesis that for absorbers with $\rm FW20 < 250 \ km \ s^{-1}$, a significant fraction of the absorption could arise at the scales of $\rm few \times 100-1000$ pc, the spatial extents of the core.

It is interesting to note that nine out of the 12 targets have peaked SED (spectral energy distribution) shapes, with spectral peaks at $\rm \lesssim 1 \ GHz$ \citep[see][]{yoon2024, kerrison2024}. The inverted spectral shapes indicate that the sources are either young or undeveloped, and the plasma is still dense where either synchrotron or free-free self-absorption is occurring. However, the sources are not as compact as the high-frequency peakers \citep[e.g.][]{orienti2006} that are mostly unresolved in VLBI maps and whose SEDs peak at frequencies above a few GHz. In general, the  detection fraction of AGN-associated \hi 21-cm absorption is expected to be relatively higher in compact and peaked spectrum sources, due to a high gas covering factor \citep[e.g.][]{gupta2006, curran2013b}.

\subsection{Projected size of the radio source}

\begin{figure*}[!h]
\includegraphics[width=0.49\textwidth]{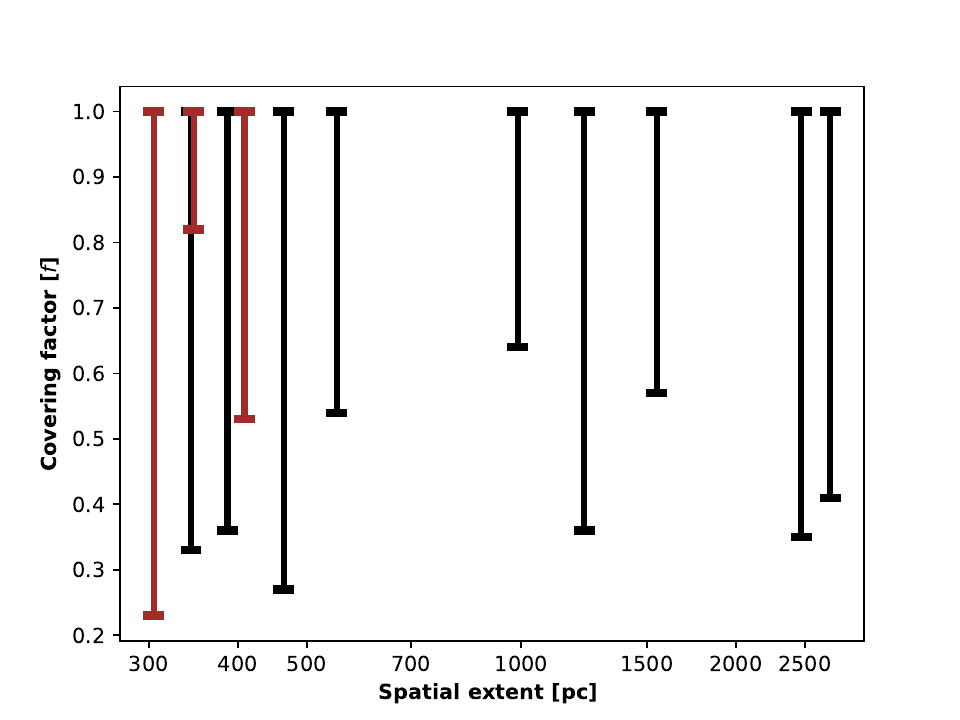}
    \includegraphics[width=0.49\textwidth]{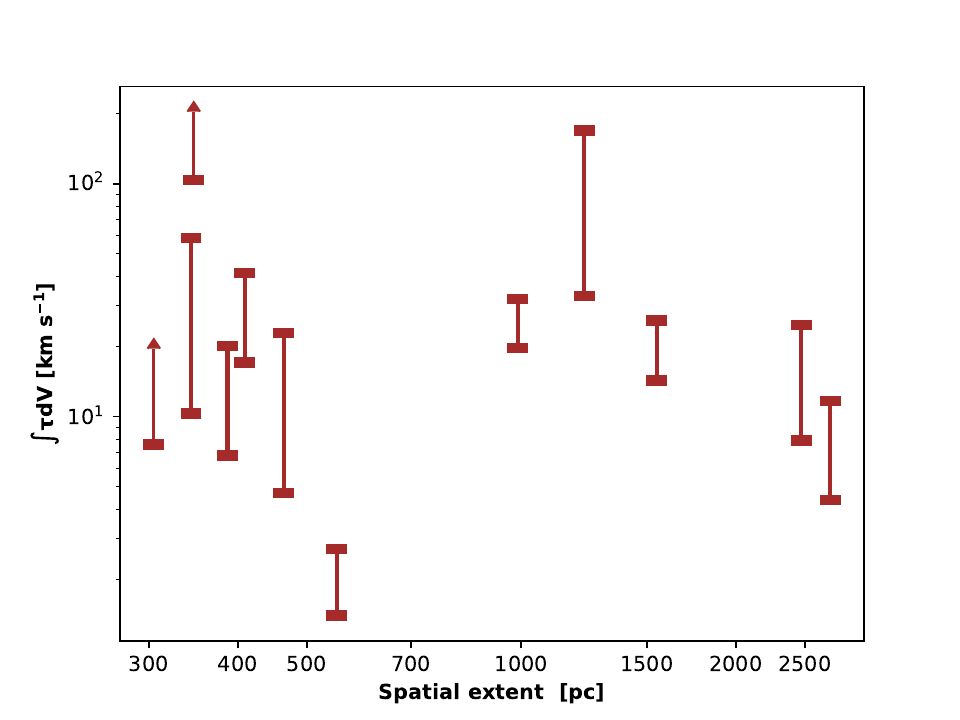}
        
  \caption{Left panel: the estimated limits on the covering factor plotted as a function of the spatial extent of the radio source at the \hi redshift. The three sources marked in Red represent 0903+010, 0920+161 and 1002-195, for which we estimate that a large fraction of the emission in the VLBA image is likely covered by gas.
  Right panel: the limits on \hi velocity integrated optical depths plotted as a function of the projected radio continuum size. For two targets, 0903+010 and 0920+161, we could not estimate the upper limits on the optical depth.  }\label{fig:opd_pc_f}
\end{figure*}

In Figure~\ref{fig:opd_pc_f} we plot the estimated limits on the covering factor and VOD for the 12 targets, as a function of the spatial extent of the radio source estimated at the redshift of the absorber. In the left panel of the figure, the three markers in Red represent
targets 0903+010, 0920+161 and 1002-195, with projected sizes 347 pc, 305 pc and 409 pc, respectively. For these three sources a large fraction of the entire emission detected in the VLBA image, including the core, is likely occulted by \hi gas (see left panel of Figure~\ref{fig:opd_pc_f}). For 0903+010, we deduce that at least $\approx 73\%$ of the peak absorption cannot arise against large-scale undetected emission, and should arise against emission detected in the VLBA image. These three targets are also among the sources with smallest spatial extensions, $\lesssim 400$ pc.

In the panel on the right, we do not see any relation between the optical depth limits and the source size. However, source 0903+010 has the second highest estimate of optical depth in our sample, with a lower limit on VOD of $\rm 103.9 \ km \ s^{-1}$. Overall, the results hint at a scenario in which sources with projected size $\rm \lesssim 400 \ pc$ could have 
most of the emission in the central few $\times 100$ pc covered by gas, and thus yield a high \hi 21-cm absorption strength.


\subsection{Highest optical depths}

In Figure~\ref{fig:opd_z} we plot the VOD limits for the 12 targets, together with the optical depths of the associated and intervening systems that were searched at redshifts $0.4 < z < 1.0$ with smaller interferometers, in the literature. Among the estimates, we can see that the values cross $\rm 100 \ km \ s^{-1}$ for two targets; $\rm 169.0 \ km \ s^{-1}$ for 0023+010, and $\rm 103.9 \ km \ s^{-1}$ for 0903+010, respectively. In the case of 0023+010, the VOD is an upper limit; the true estimate needs to be confirmed using further spectroscopic VLBI observations. However, for 0903+010, the VOD is a lower limit; the true value could be higher than this value. We note that FLASH detection towards 0903+010 was reported earlier in \citet[][]{su2022}.

Objects with a VOD greater than $\rm 100 \ km \ s^{-1}$, detected in surveys with smaller interferometers, are very rare in the literature. This can be seen in the column density distribution of absorbers at $z > 0.1$, reported by \citet[][]{curran2024}. Only two objects, MRC 0531-237 and 0903+010, have $\rm N_{HI} > 10^{22} \ cm^{-1}$, assuming $\rm T_{s} = 100 \ K$, corresponding to  $\rm VOD \sim 56 \ km \ s^{-1}$.
NGC 3079 ($z = 0.003689$, VOD $\rm \approx 135 \ km \ s^{-1}$; \citealp[][]{gallimore1999} ) was a known local system until a few years ago, with MRC 0531-237 ($z = 0.851$, VOD $\rm \approx 144 \ km \ s^{-1}$; \citealp[][]{aditya2024} ) and 0903+010 recently added. Note that MRC 0531-237 is part of our sample, the LBA results of which will be discussed in our subsequent article (see Section~\ref{sec:this_work}).

Gas disks in local galaxies have been shown to contain \hi clouds with column densities reaching up to $\rm 10^{23} \ cm^{-2}$ \citep[][]{braun2012}. 
In AGN host galaxies, due to accretion of gas to the nuclear regions, we expect to find relatively higher \hi column densities compared to galactic disks. Since FLASH is a blind survey, we would detect absorption from \hi clouds that are AGN
associated, and from intervening clouds that are located in foreground galactic disks. In view of the above, in our survey we can expect to find a reasonable fraction of absorbers with column densities $\rm 10^{23} \ cm^{-2}$ or higher.

The \hi column density is related to the VOD and spin temperature ($\rm T_{s}$) as:
\begin{equation}
     {\rm N_{HI}} = 1.823 \times 10^{18} \times {\rm T_{s}} \times \int \tau dv
\end{equation}
 where VOD $\equiv \int \tau dv$. Assuming a spin temperature of 100 K, for the two highest estimates of VOD in our sample, the column densities will be $\rm 1.9 \times 10^{22} \ cm^{-2}$ and $\rm 2.8 \times 10^{22} \ cm^{-2}$, for 0023+010 and 0903+010 respectively. Even for the highest VODs, clearly the column densities are significantly lower than $\rm \approx 10^{23} \ cm^{-2}$.


Here, we note that the assumed spin temperature of 100 K is only a fiducial value, for comparison with $\rm N_{HI}$ estimates from the literature \citep[e.g.][]{morganti2018}. In fact, a major hindrance to estimating the \hi column density in AGN surroundings is the lack of available measurements of the spin temperature. Simulations predict that the spin temperature of the \hi gas in the AGN vicinities could be anywhere between a few $\rm \times 100 \ K$ to 8000 K \citep[e.g.][]{maloney1996}. Due to highly complex environments, it is also difficult to justify the assumption of a uniform spin temperature for gas in the AGN environments. In general, for both associated and intervening systems, the inferred spin temperature will be a column-density-weighted harmonic mean of the temperatures of different \hi phases along the sightline \citep[e.g.][]{kanekar2014}. In this sense, it is likely that the spin temperatures for 0023+010 and 0903+010 could be higher than 100 K, by up to an order of magnitude.

\subsection{Optical depth distribution}

\begin{figure*}[!h]

    \includegraphics[width=0.8\textwidth]{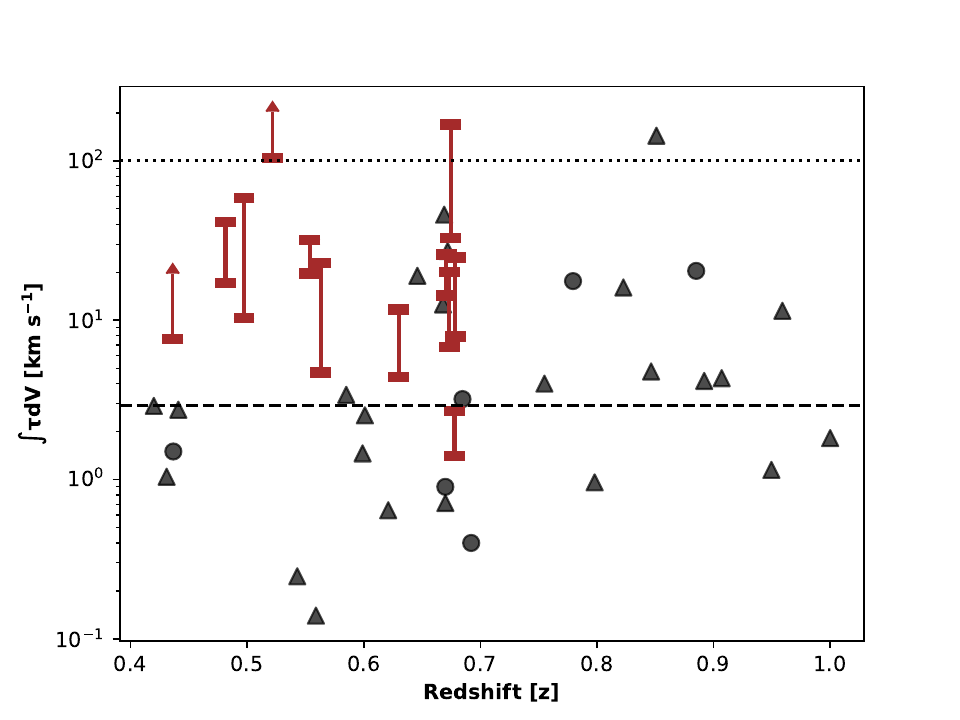}
  \caption{Optical depth distribution of \hi 21-cm absorbers at $0.4<z<1.0$. Red bars represent the estimated limits to the \hi 21-cm velocity integrated optical depth, of the FLASH detections. The black triangles are AGN-associated \hi 21-cm absorbers in the literature, while the circles are intervening \hi 21-cm absorbers. The dashed horizontal line represents the median velocity integrated optical depth of absorbers from the literature, and the dotted horizontal line represent a VOD of $\rm 100 \ km \ s^{-1}$. The literature references for the \hi 21-cm absorbers are: \citet[][]{aditya2016, aditya2018b, aditya2019, aditya2024, brown73, carilli1993, chengalur1999, darling04, lane2001,  moss2017, salter2010, vermeulen03, yan2016} }\label{fig:opd_z}
\end{figure*}

In Figure~\ref{fig:opd_z}, it is clear that for most of the 12 targets, the upper limits are significantly higher than the previous estimates of VOD (the lower limits) estimated from ASKAP observations. Excluding two targets for which the lower limits on the covering factor could not be estimated, for the remaining ten targets, the difference between the upper and lower limit on VOD is in the range of $\rm 1.3- 119.4 \ km \ s^{-1}$. The median ratio between the upper and lower limits on the VOD is 2.8. On the other hand, the median VOD for associated and intervening absorbers reported in the literature, at redshifts $0.4 < z < 1.0$, is $\rm 2.9 \ km \ s^{-1}$, represented by the dashed horizontal line in the figure. Note that the literature values have not been corrected for covering factors. The results indicate that upon correction of the gas covering factors based on VLBI measurements, the distribution of \hi VODs could increase nearly three times, with the median VOD of the sample reaching $\rm \approx 8.1 \ km \ s^{-1}$. However, we strongly note that the above estimates are based on a small sample of just twelve targets and that the optical depth distributions of associated and intervening absorbers could be different. 

In the literature, for the source 4C 12.5 \citet[][]{morganti2013} have reported a highest VOD of $\rm 241.4 \ km  \ s^{-1}$ at a spatial resolution $32 \times 11 $ pc, while a VOD of just $\rm 14.26 \ km \ s^{-1}$ was reported in observation with the Westerbork Synthesis Radio Telscope (WSRT) \citep[][]{morganti2005}, implying a factor $\approx 17$ increase in optical depth upon correction for the covering factor. For 4C 31.04 \citet[][]{murthy2024} reported VODs of $\rm 13.7 \ km \ s^{-1}$ and $\rm 7.7 \ km \ s^{-1}$ at the VLBI and WSRT scales respectively, which implies a factor of 1.8 increase. Note that the VOD of $\rm 13.7 \ km \ s^{-1}$ is an average value over a scale of $\approx 140$ pc. For 3C 84, \citet[][]{morganti2023} report that the high-resolution VOD could be much higher than a value of $\rm 1.3 \ km \ s^{-1}$ measured in VLA observations, as the gas appears to cover a faint background emission detected at VLBI resolutions. However, they could not retrieve any absorption in their VLBI observations and so could not report an estimate. We strongly note here that all these observations are for $z \lesssim 0.1$ objects, where the VLBI spatial resolution is a $\rm few \times 10$ pc. These cannot be directly compared with our observations due to a significant difference in the spatial resolutions.

Estimating the \hi VOD distributions at various redshift intervals is critical to understanding the redshift evolution in gas properties, both in AGN hosts and in normal galaxies, and to understanding the effects of AGN radiation on the surrounding neutral hydrogen gas \citep[see e.g.][]{curran2008, aditya2016, aditya2018b, curran2019}. In terms of associated \hi studies, \citet[][]{aditya2016, aditya2018b} have found a significant ($3\sigma$) variation in the \hi 21-cm absorption strength associated with flat-spectrum radio sources, as a function of the redshift. The sources in their subsample at $z > 1$ showed lower VODs compared to those at $z < 1$. The decline in VODs was attributed to a redshift evolution in gas properties or to relatively higher AGN (ultraviolet and radio) luminosities at high redshifts (due to the Malmquist effect in a flux-limited sample) \citep[see e.g.][]{curran2008}, which could significantly reduce absorption strength \citep[see][]{aditya2016, aditya2018b}. Studies by \citet[][]{curran2008, curran2017, curran2019, grasha2019} found a decrease in \hi column densities as a function of redshift, which was mainly attributed to high AGN ultraviolet luminosities at high redshifts. However, \citet[][]{murthy2022} found low \hi detection rates even in sources with low UV and radio luminosities, suggesting that UV luminosity alone cannot explain the decrease in absorption strength with redshift.

In the case of intervening absorbers, earlier studies find a steep decline in the \hi 21-cm absorption strength, particularly at redshifts above $z \approx 1$. \citet[][]{curran2006, curran2012b} suggest geometric effects due to the expanding Universe, leading to relatively lower gas covering factors in $z > 1$ systems, as the dominant reason for the decline in the absorption strength. However, \citet[][]{kanekar2003a, kanekar2014} suggest that the high spin temperatures at $z > 1$ are the likely cause of the decline.  

In studies of AGN-associated systems, the radio sources have not been corrected for the gas covering factor in any of the above-mentioned studies. It remains to test whether the decline in the \hi 21-cm absorption strength with redshift persists after the VLBI correction to the optical depths. In the studies of the intervening absorbers, since the number of objects in the samples was $< 40$, it is problematic to draw statistically significant statistics. The suggestions of \citet[][]{curran2006} and \citet[][]{kanekar2014} need to be rigorously tested using larger unbiased surveys, to identify the dominant cause, if any.

\section{Conclusions}
We have conducted VLBA L band observations towards 12 targets that have been detected with \hi 21-cm absorption in the FLASH radio survey. We have made continuum images of the radio sources, identified the core and jet components, and classified the source morphology. We describe a methodology to test whether the emission from the radio core or the total emission in the VLBA image has sufficient flux density to cause the entire \hi 21-cm absorption detected in the ASKAP spectrum. We used this method to estimate the lower and upper limits on the gas covering factor and the VOD for each target.
The VOD estimated solely using the ASKAP spectra and continuum flux density assumes a uniform covering factor, and hence, this provided a lower limit on the VOD. We find that for seven of our targets, the radio core has sufficient flux density to cause the entire detected \hi 21-cm absorption, which supports the hypothesis that for absorbers with a narrow line width a significant fraction of the absorption could arise against the core. Note that the radio cores identified in our sources have a typical scales of $\rm few \times 100 - 1000$ pc.  For three compact sources in our sample, with scales in the range $\rm 305-409 \ pc$, we find that a large fraction of the entire emission in the VLBA image is probably occulted by gas. For 0903+010, we find that $\approx 73\%$ of the peak absorption detected in the ASKAP spectrum could arise against the emission detected in the VLBA map. Further, for 0903+010 we estimate a lower limit on the VOD of $\rm \approx 104 \ km \ s^{-1}$, and for 0023+020 we estimate an upper limit of $\rm \approx 169 \ km \ s^{-1}$. 
These absorbers could have high \hi column densities or spin temperatures exceeding 100 K, or both.
Finally, we find that after correction for the covering factor, the distribution of VODs could increase by a factor of $\approx 3$, although we caution here that these estimates are based on a small sample of 12 targets, and correspond to the spatial scales of $\rm few \times (100 - 1000)$ pc.

\section{Future Work}

We plan to report the LBA L-band observations and results of the remaining 21 targets from the FLASH pilot surveys \citep[][]{yoon2024} in a subsequent publication. Following this, we plan to expand the sample using VLBA and LBA observations of 31 new targets that have been detected with \hi 21-cm absorption in the FLASH full survey. In addition to the L-band observations, we aim to conduct 5 GHz VLBA and LBA observations for all targets. These high-frequency observations will allow us to 1) estimate the spectral indices of the various milliarcsecond scale components, helping us to identify the radio core and assess the morphology with better accuracy, and 2) yield accurate astrometric positions of the radio sources. A comparison of radio and optical astrometric positions \citep[e.g.][]{dey2019} will allow us to classify the absorbers as associated with AGN or as intervening systems. In associated systems, we expect the radio position to coincide with the nucleus of the optical galaxy, while for intervening systems we expect a radio-optical offset, termed the `impact parameter’. Overall, these observations will allow us to estimate the distribution of \hi 21-cm optical depths for associated and intervening absorbers at intermediate redshifts, $0.4 < z < 1.0$, and allow us to identify and study potential targets with high \hi column densities. 

\begin{appendix}
\section{Estimation of limits on the \hi covering factor and optical depth}
\label{sec:appendix}
In the following, we describe a method to test if one or more continuum components in the VLBA image have sufficient flux density to cause the entire \hi 21-cm absorption that is detected in the ASKAP spectrum, and to estimate lower and upper limits to the covering factor and optical depth, respectively.

We use the parameters described in Section~\ref{params}. In addition, we use the parameters the covering factor ($\rm f$), the measured/apparent optical depth ($\tau_{\rm app}$) and the true optical depth ($\tau_{\rm true}$), as defined in Section ~\ref{sec:intro}.

The flux density of the \hi 21-cm absorption profile at the line peak, $\rm S_{HI}$, is related to the integrated flux density of the background source measured in the ASKAP beam ($\rm S_{ASKAP}$) and $\tau_{\rm app}$ as:

\begin{equation}    
 {\rm S_{HI} = S_{ASKAP}} \cdot {\rm e}^{-\tau_{\rm app}} \ - \ {\rm S_{ASKAP}}  \label{eqn3}
\end{equation}  \citep[e.g.][]{draine2011}

Note that $\rm S_{HI}$ is negative since the continuum flux density ($\rm S_{ASKAP}$) is subtracted on the right-hand side (RHS). Here, we assume that the gas obscures the complete radio emission. However, the gas could be obscuring only a fraction of the region of radio emission that we see.
Then, the flux density of the occulted emission is $ f \cdot \rm S_{ASKAP}$ \citep[e.g.][]{briggs1983}.

In such a case, equation~\ref{eqn3} translates to
\begin{equation}
{{\rm S_{HI}} = f  \cdot {\rm S_{ASKAP}} \cdot e}^{-\tau _{\rm true}} \ - \ { f \cdot \rm S_{ASKAP}} \label{eqn4}
\end{equation}
where $ f \cdot \rm S_{ASKAP}$ is the flux density of the emission obscured by gas. Note that $f = 1$, the upper limit to the covering factor, yields $ \tau_{\rm true} = \tau_{\rm app}$, a lower limit to the true optical depth.

Equation~\ref{eqn4} can be written as:
\begin{equation}
 {\rm S_{HI}} / ({ f \cdot \rm S_{ASKAP}}) = {\rm e}^{-\tau_{\rm true}} - 1.0  \label{eqn5}
\end{equation}

The absolute value of RHS is always $< 1$ for an absorbing medium with finite optical depth $\tau_{\rm true}$.
This condition imposes a lower limit on the flux density of the occulted emission. 
\begin{equation}
{f}\cdot {\rm S_{ASKAP}} > \lvert {\rm S_{HI}} \rvert \label{eqn6} 
\end{equation}

The above equation implies that the gas has to cover an emission with a flux density larger than $\rm \lvert S_{HI} \rvert$ to cause the entire \hi 21-cm absorption. This yields a lower limit to the covering factor, $ f \rm > \lvert S_{HI} \rvert /S_{ASKAP}$.
However, note that using $f \rm = \lvert S_{HI} \rvert / S_{ASKAP}$ in equation~\ref{eqn4} will yield $ \tau_{\rm true} = \infty$, i.e. the absorbing medium is completely opaque. So, while the above relation imposes a constraint on the fractional emission, it does not yield a realistic upper limit to $\tau_{\rm true}$.

Therefore, we used the VLBA continuum image to estimate the flux density of a minimum emission that should be occulted by the absorbing medium (${\rm S_{min}} = f \rm \cdot S_{ASKAP} $), which will yield a realistic upper limit of $\tau_{\rm true}$. So, $\rm S_{min}$ should satisfy the relation
\begin{equation}
{\rm S_{min} > \lvert S_{HI} \rvert } \label{eqn7}
\end{equation}
However, not that $\rm S_{min}$ is measured in the 1.4 GHz band (the VLBA observations) while $\rm \lvert S_{HI} \rvert$ at $\sim 700$ MHz (in ASKAP observations), and so a direct comparison cannot be made. Instead, we compare their ratios w.r.t the total flux density of the source measured at the respective frequency bands. That is,
\begin{equation}
    {\rm S_{min} / S_{RACS} > \lvert S_{HI} \rvert / S_{ASKAP} }      \label{eqn8} 
\end{equation}
Note that $\rm S_{RACS}$ is the L-band flux density of the source measured using ASKAP, in RACS-mid survey, while $\rm S_{ASKAP}$ is the flux density of the source measured in the FLASH survey, again with ASKAP, at $\rm \sim 700 \ MHz$. 
\newline
$\rm \lvert S_{HI} \rvert / S_{ASKAP}$ is abbreviated as PF, ratio of the absolute value of the peak absorption w.r.t. the total source flux density. So, the above relation is written as 
\begin{equation}
    {\rm S_{min} / S_{RACS} > PF } \label{eqn9}
\end{equation}
The emission in the VLBA image is divided into the core and other radio lobes. So, $\rm S_{min}$ can be assigned iteratively the flux density of one or more components to test whether $\rm S_{min}$ satisfies the above relation. If so, then it implies that the assigned component/s has sufficient flux density to cause the \hi 21-cm absorption. 

After a sufficient $\rm S_{min}$ has been estimated, the ratio $\rm S_{min} / S_{RACS}$ could be used as a lower limit to the covering factor, that is, $ f \rm \geq S_{min} / S_{RACS}$, which in turn, substituting in Equation~\ref{eqn4}, yields a finite upper limit to the optical depth.

\end{appendix}




\begin{acknowledgement}
This scientific work uses data obtained from Inyarrimanha IlgariBundara / the Murchison Radio-astronomy Observatory. We acknowledge the Wajarri Yamaji People as the Traditional Owners and native title holders of the Observatory site. CSIRO’s ASKAP radio telescope is part of the Australia Telescope National Facility (\url{https://ror.org/05qajvd42}). Operation of ASKAP is funded by the Australian Government with support from the National Collaborative Research Infrastructure Strategy. ASKAP uses the resources of the Pawsey Supercomputing Research Centre. Establishment of ASKAP, Inyarrimanha Ilgari Bundara, the CSIRO Murchison Radio-astronomy Observatory, and the Pawsey Supercomputing
Research Centre are initiatives of the Australian Government, with support from the Government of Western Australia and the Science and Industry Endowment Fund. This research was supported by the Australian Research Council Centre of Excellence for All Sky Astrophysics in 3 Dimensions (ASTRO 3D), through project number
CE170100013. Support for the operation of the MWA is provided
by the Australian Government (NCRIS), under a contract to Curtin University administered by Astronomy Australia Limited. MG acknowledges support from IDIA and was partially supported by the Australian Government through the Australian Research Council’s Discovery Projects funding scheme (DP210102103).
TA acknowledges the support of the Xinjiang Tianchi Talent Program.
RZS is supported by the National
Science Foundation of China (grant 12473019), the National SKA
Program of China (grant No. 2022SKA0120102).
HY is supported by the National Research Foundation of Korea (NRF) grant funded by the Korea government (MSIT) (RS-2025-00516062).

\end{acknowledgement}




\paragraph{\textbf{Data Availability Statement:}}
All data from FLASH observations are available online at the CSIRO ASKAP Science Data Archive (CASDA; \url{https://research.csiro.au/casda/}), with the project code AS109.
The VLBA data are available at the National Radio Astronomical Observatory (NRAO; \url{https://data.nrao.edu/portal/}) data archive, with project code VLBA/24A-059, BJ124. The astrogeo data are available at \url{http://doi.org/10.25966/kyy8-yp57}.


\bibliography{vlbi}

\end{document}